\documentclass[aps,prd,twocolumn,floatfix,superscriptaddress,balancelastpage,nofootinbib,amsmath,showpacs]{revtex4}
\usepackage{graphicx}
\usepackage{epstopdf}
\usepackage{epsfig}
\usepackage{bm}
\usepackage{dcolumn}
\usepackage{natbib}
\usepackage{amsmath}
\usepackage{amssymb}

\newcommand{\be}{\begin{equation}}
\newcommand{\ee}{\end{equation}}
\newcommand{\bea}{\begin{eqnarray}}
\newcommand{\eea}{\end{eqnarray}}

\def\lsim{\mathrel{\raise.3ex\hbox{$<$\kern-.75em\lower1ex\hbox{$\sim$}}}}
\def\gsim{\mathrel{\raise.3ex\hbox{$>$\kern-.75em\lower1ex\hbox{$\sim$}}}}

\begin{document}

\hspace*{130mm}{\large \tt FERMILAB-PUB-13-052-A}

\title{Two Emission Mechanisms in the \emph{Fermi} Bubbles: A Possible Signal of Annihilating Dark Matter} 

\author{Dan Hooper}
\affiliation{Fermi National Accelerator Laboratory, 
             Theoretical Astrophysics Group, Batavia, IL 60510} 
\affiliation{University of Chicago, 
             Department of Astronomy and Astrophysics, Chicago, IL 60637} 

\author{Tracy R. Slatyer}
\affiliation{School of Natural Sciences, Institute for Advanced Study, Princeton, NJ 08540.}

\begin{abstract}
\noindent 

We study the variation of the spectrum of the \emph{Fermi} Bubbles with Galactic latitude. Far from the Galactic plane ($|b|\gsim 30^{\circ}$), the observed gamma-ray emission is nearly invariant with latitude, and is consistent with arising from inverse Compton scattering of the interstellar radiation field by cosmic-ray electrons with an approximately power-law spectrum. The same electrons in the presence of microgauss-scale magnetic fields can also generate the the observed microwave ``haze''. At lower latitudes ($|b|\lsim 20^{\circ}$), in contrast, the spectrum of the emission correlated with the Bubbles possesses a pronounced spectral feature peaking at $\sim$1-4 GeV (in $E^2 dN/dE$) which cannot be generated by any realistic spectrum of electrons. Instead, we conclude that a second (non-inverse-Compton) emission mechanism must be responsible for the bulk of the low-energy, low-latitude emission. This second component is spectrally similar to the excess GeV emission previously reported from the Galactic Center (GC), and also appears spatially consistent with a luminosity per volume falling approximately as  $r^{-2.4}$, where $r$ is the distance from the GC. Consequently, we argue that the spectral feature visible in the low-latitude Bubbles is most likely the extended counterpart of the GC excess, now detected out to at least $\sim$2-3 kpc from the GC. The spectrum and angular distribution of the signal is broadly consistent with that predicted from $\sim$10 GeV dark matter particles annihilating to leptons, or from $\sim$50 GeV dark matter particles annihilating to quarks, following a distribution similar to, but slightly steeper than, the canonical Navarro-Frenk-White (NFW) profile. We also consider millisecond pulsars as a possible astrophysical explanation for the signal, as observed millisecond pulsars possess a spectral cutoff at approximately the required energy. Any such scenario would require a large population of unresolved millisecond pulsars extending at least 2-3 kpc from the GC.
 
\end{abstract}


\maketitle

\section{Introduction}

Data from the \emph{Fermi} Gamma-Ray Space Telescope have revealed a pair of large gamma-ray lobes extending approximately $50^\circ$ north and south of the Galactic Center \cite{Su:2010qj}. These lobes, known as the \emph{Fermi} Bubbles, are visible in gamma-rays between $\sim$1-100 GeV and have a markedly harder spectrum ($dN/dE \propto E^{-2}$) than the gamma-ray emission associated with the Galactic Disk.

The Bubbles were originally studied as a possible gamma-ray counterpart to the \emph{WMAP} haze \cite{Dobler:2009xz}, a spectrally hard microwave excess in the inner Galaxy most clearly visible in \emph{WMAP}'s 23 and 33 GHz frequency bands. The haze was first discovered in 2003 \cite{Finkbeiner:2003im}, and has been studied over the past decade as a possible signature of a new hard electron population in the inner Galaxy \cite{Hooper:2007kb, Hooper:2007gi, Dobler:2011rd}, producing microwave synchrotron radiation in the Galactic magnetic field. Recently, the existence of the microwave haze has been confirmed by the \emph{Planck} experiment~\cite{:2012fb}, whose data indicate a strong degree of spatial coincidence between the microwave haze and the gamma-ray Bubbles, further supporting the hypothesis that these signals have a common origin. Perhaps the simplest possibility is that the gamma rays arise from inverse Compton scattering (ICS) by the same hard electron population that produces the haze via synchrotron.

The question of the origin and nature of the \emph{Fermi} Bubbles has been the subject of much debate. One key question is whether these gamma-rays are produced by a hadronic~\cite{Crocker:2010dg,Zubovas:2011py,Zubovas:2012bn} or leptonic~\cite{Guo:2011eg,Guo:2011ip,Lang:2011zr,Cheng:2011xd,Yang:2012fy} mechanism, i.e. whether they arise from the scattering of energetic protons on the gas of the interstellar medium, or from the ICS of photons from the interstellar radiation field by energetic electrons. An example of a hadronic scenario was proposed by Aharonian and Crocker \cite{Crocker:2010dg}, in which the Bubbles are billion-year-old reservoirs of energetic protons, which were injected as a result of star formation in the Galactic Center and are confined within the Bubbles by magnetic fields. Leptonic scenarios have garnered somewhat more attention in the literature, and provide a straightforward link with the spatially correlated emission observed in the microwave and radio~\cite{Carretti:2013sc} (hadronic scenarios will also generate synchrotron emission through the electrons produced in charged pion decays, but since these electrons will diffuse after being produced, the connection between their spectrum and spatial distribution and that of the gamma-rays is not as straightforward as in the leptonic scenario). In such leptonic models, electrons are accelerated by a shock or a series of shocks and/or by Fermi acceleration in turbulent magnetic fields behind the shock front \cite{Mertsch:2011es}; the shock(s) may be fueled by accretion onto the supermassive black hole at the Galactic Center, by starburst activity, or by some other mechanism.

In this article, we reexamine the \emph{Fermi} Bubbles and the variation of their spectrum with Galactic latitude. Far from the Galactic plane ($|b|\gsim 30^{\circ}$), the observed gamma-ray spectrum is nearly invariant with latitude and fairly flat over the energy range observed by \emph{Fermi}. This spectrum can be well explained by inverse Compton scattering of cosmic microwave background (CMB), infrared, and starlight photons by a population of GeV-TeV electrons with an approximately power-law spectrum ($dN_e/dE_e \sim E_e^{-3}$). Furthermore, we find that this same population of cosmic ray electrons leads to synchrotron emission of the same amplitude as the observed microwave haze, if microgauss-scale magnetic fields are present in the high-latitude regions of the Bubbles. The success of this simple and self-consistent picture provides strong support for a leptonic origin of the high-latitude emission from the \emph{Fermi} Bubbles. 

At latitudes closer to the disk, however, a leptonic origin of \emph{all} the emission associated with the Bubbles does not appear possible. The gamma-ray spectrum of the \emph{Fermi} Bubbles at latitudes within approximately $20^{\circ}$ of the Galactic plane possesses a peak at energies of a few GeV, and cannot be generated by inverse Compton scattering of starlight, infrared, or cosmic microwave background radiation by any realistic steady-state electron population. Furthermore, no realistic spectrum of cosmic ray protons is capable of accounting for the gamma-ray spectrum observed at these low latitudes.

Gamma-ray emission with a similar spectrum has been previously identified from the region surrounding the Galactic Center (GC)~\cite{Hooper:2010mq,Hooper:2011ti, Goodenough:2009gk, Boyarsky:2010dr, Abazajian:2012pn}. Proposed origins for this excess include annihilating dark matter~\cite{Hooper:2010mq,Hooper:2011ti, Goodenough:2009gk,Abazajian:2012pn}, a population of millisecond pulsars~\cite{Hooper:2010mq,Hooper:2011ti,Abazajian:2010zy,Wharton:2011dv,Abazajian:2012pn}, or cosmic ray interactions with gas~\cite{Hooper:2010mq,Hooper:2011ti,Linden:2012iv,Linden:2012bp,Abazajian:2012pn}. In this paper, we show that the non-inverse Compton component of the emission from the \emph{Fermi} Bubbles identified in this study is spectrally and morphologically consistent with being the extended counterpart of this GC excess, revealing that this emission is not confined to the GC, but extends out to at least $\sim$2-3 kiloparsecs from the Galactic plane. The morphology of this signal is consistent with originating from annihilating dark matter distributed according to a generalized Navarro-Frenk-White (NFW) profile with an inner slope of $\rho \propto r^{-1.2}$, where $r$ is the distance to the GC (the GC excess, in isolation, favors a power-law slope in the range of $1.2$-$1.4$ \cite{Hooper:2010mq,Hooper:2011ti,Abazajian:2012pn}). The spectral shape of this signal can be accommodated by dark matter particles of mass $\sim$10 GeV annihilating to leptons, or by $\sim$50 GeV particles annihilating to quarks. In either case, the normalization of the observed signal requires an annihilation cross section on the order of $\sigma v\sim(6-8)\times 10^{-27}$ cm$^3$/s, similar to that expected of a thermal relic of the Big Bang. 

The remainder of this paper is structured as follows. In Sec.~\ref{analysis} we describe the analysis used to extract the spectra from various regions of the \emph{Fermi} Bubbles. In Sec.~\ref{high} we demonstrate that the high-latitude regions of the Bubbles can be accounted for with an approximately power-law spectrum of GeV-TeV electrons, which can simultaneously produce the observed microwave haze as synchrotron. In Sec.~\ref{low} we turn our attention to the low-latitude regions of the \emph{Fermi} Bubbles, and show that their spectra cannot be accounted for by inverse Compton scattering. Instead, an additional mechanism is required, capable of producing a spectrum which peaks strongly at energies of a few GeV. In Sec.~\ref{compare} we compare this signal with that previously observed from the Galactic Center. In Sec.~\ref{morphology}, we study the residuals remaining when the (best-fit) known backgrounds are subtracted, and demonstrate how the few-GeV peak appears in these residuals. In Sec.~\ref{nfw}, we show that the emission observed at low-latitudes is better described by a spherically symmetric, NFW-like morphology than by a flat-brightness distribution confined to the regions of the \emph{Fermi} Bubbles. In Sec.~\ref{interpretation} we discuss possible interpretations of this signal, including annihilating dark matter and a population of gamma-ray pulsars. We summarize our results and draw conclusions in Sec.~\ref{conclusions}. This paper also includes five appendices, which describe various cross checks of our results and other supplementary material.

\section{Extracting the Latitude-Dependent Spectrum of the Gamma-Ray Bubbles}
\label{analysis}

In our analysis, we employ the publicly available \texttt{Pass 7\_Version 6} data release from \emph{Fermi}, with 4.5 years of photon data.\footnote{The dataset we employ may be downloaded from \texttt{http://heasarc.gsfc.nasa.gov/FTP/fermi/data/lat/weekly}.} We apply a standard zenith angle cut to exclude the Earth limb, rejecting events with zenith angles greater than $100^\circ$. We also employ the recommended diffuse analysis cuts on the data quality, nominal science configuration and rocking angle: \texttt{DATA$\_$QUAL==1}, \texttt{LAT$\_$CONFIG==1}, \texttt{ABS(ROCK$\_$ANGLE)$<$52}. Throughout, we use the class of events designated \texttt{ULTRACLEAN}, but have confirmed that employing the \texttt{SOURCE} or \texttt{CLEAN} classes does not alter our conclusions.

We generate skymaps for 30 log-spaced energy bins spanning the range from 0.3 GeV to 300 GeV, binning the photons on an \texttt{NSIDE=256} HEALPix grid, and smooth all maps to $2^\circ$ FWHM (full width at half maximum). As in Ref.~\cite{Su:2010qj}, we use front-converting events only (which have better inherent angular resolution) at energies below 1 GeV. We follow the prescription from Ref.~\cite{Su:2010qj} for point source subtraction, using the \emph{Fermi} 1-year source catalog and masking the very brightest sources. 

\begin{figure}[!]
\includegraphics[width=0.45\textwidth ]{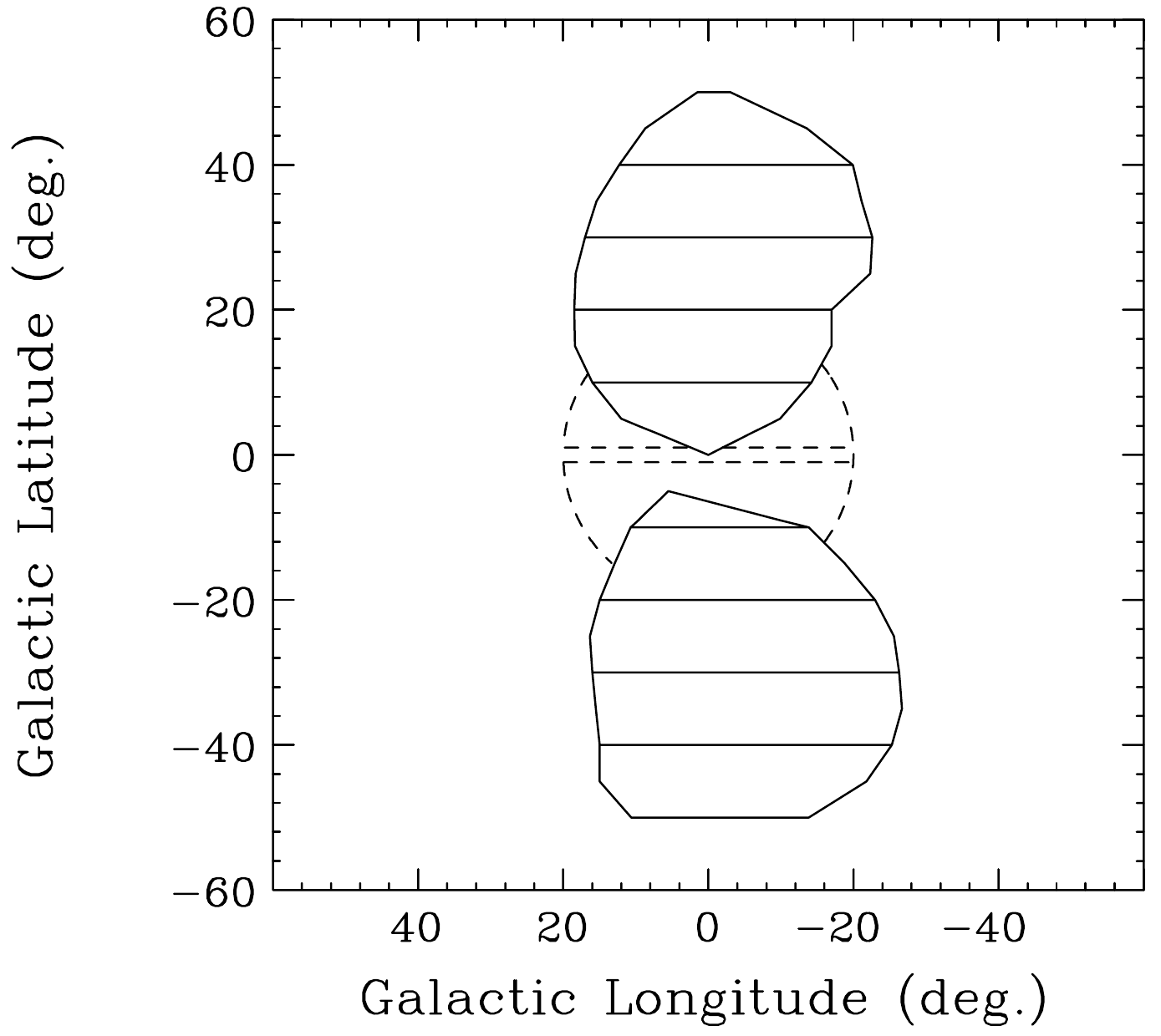}
\caption{The regions of the sky considered in our analysis. The \emph{Fermi} Bubbles themselves are broken into five pairs of regions according to Galactic latitude ($|b|<10^{\circ}$, $10^{\circ}-20^{\circ}$, $20^{\circ}-30^{\circ}$, $30^{\circ}-40^{\circ}$, and $40^{\circ}-50^{\circ}$). Also shown as dashed lines is the inner complement region to the Bubbles, as described in Appendix \ref{app:complement}.}
\label{sky}
\end{figure}

For each energy bin, we fit the skymap as a linear combination of templates, maximizing the Poisson likelihood. The Gaussian errors on the fit coefficients are computed from the likelihood by $\Delta \ln \mathcal{L} = 1/2$, and do \emph{not} take into account the systematic error in the event that the linear combination of templates is a poor description of the data.  Further details of the fitting procedure may be found in Ref.~\cite{Su:2010qj} and in Appendix B of Ref.~\cite{Dobler:2009xz}. We employ several different template combinations to test the robustness of our results to the foreground model. 

In the Galactic disk, there is a substantial population of unsubtracted point sources, as well as bright diffuse emission; consequently, we mask the inner disk. Throughout our study, we will show results for masks with $|b| < 1^\circ$, $|b| < 2^\circ$ and $|b| < 5^\circ$, to test the dependence of our results on this parameter.

To determine the spectrum of the \emph{Fermi} Bubbles as a function of latitude, we divide the standard spatial template for the Bubbles (as defined in Ref.~\cite{Su:2010qj}) into sub-regions by (absolute) latitude: $|b| < 10^\circ$, $10^\circ < |b| < 20^\circ$, $20^\circ < |b| < 30^\circ$, $30^\circ < |b| < 40^\circ$, and $40^\circ < |b| < 50^\circ$ (see Fig.~\ref{sky}). We smooth all templates to the scale of the maps. We fit separately for the spectrum in each of these latitude bands, varying the degree of masking of the Galactic Disk, and with a range of template-based models for the known backgrounds. All fits include an isotropic offset, to absorb residual cosmic-ray contamination and isotropic diffuse emission. The two primary possibilities we consider for the additional templates are:
\begin{itemize}
\item \emph{Diffuse model}: We take the \emph{Fermi} diffuse model from version \texttt{P6V11} of the \emph{Fermi} Science Tools, smooth it to match the maps we are using, interpolate to the appropriate energies, and perform the fit using only this template (in addition to the latitude-sliced Bubbles templates and the isotropic template, which are universal to all subtraction methods). This version of the diffuse model has been adjusted to fit the data assuming \emph{no} contribution from the Bubbles; consequently it may absorb some of the Bubbles-correlated emission at the cost of oversubtraction in neighboring regions. It was also designed primarily to model the emission at energies $\lesssim 50$ GeV, and is not recommended for use at very low latitudes, $|b| < 3^\circ$. However, since our signal extends to quite high latitudes and the energies of greatest interest are at $\lesssim 50$ GeV, these latter caveats do not pose severe problems for our study.
\item \emph{Low-energy template}: We employ the Schlegel-Finkbeiner-Davis (SFD) dust map \cite{Schlegel:1997yv} as a template for emission from cosmic-ray protons scattering on the gas (see Refs.~\cite{Dobler:2009xz, Su:2010qj} for a discussion). We take the \emph{Fermi} data at 0.5-1.0 GeV (where the Bubbles are less pronounced~\cite{Su:2010qj}) and subtract the SFD dust map to obtain an approximate template for emission from inverse Compton scattering by cosmic-ray electrons; this is the dominant contribution to the diffuse background after the dust/gas-correlated emission has been removed. We then fit the higher-energy data using this template, the SFD dust map, and a flat template for the large soft-spectrum structure known as \emph{Loop I}. This method avoids the use of complicated models and minimizes the use of external maps, but by construction cannot probe the Bubbles spectrum at energies around or below 1 GeV, and does not take into account spectral variation in the various emission components with position in the Galaxy.
\end{itemize}

\begin{figure*}
\includegraphics[width=.32\textwidth]{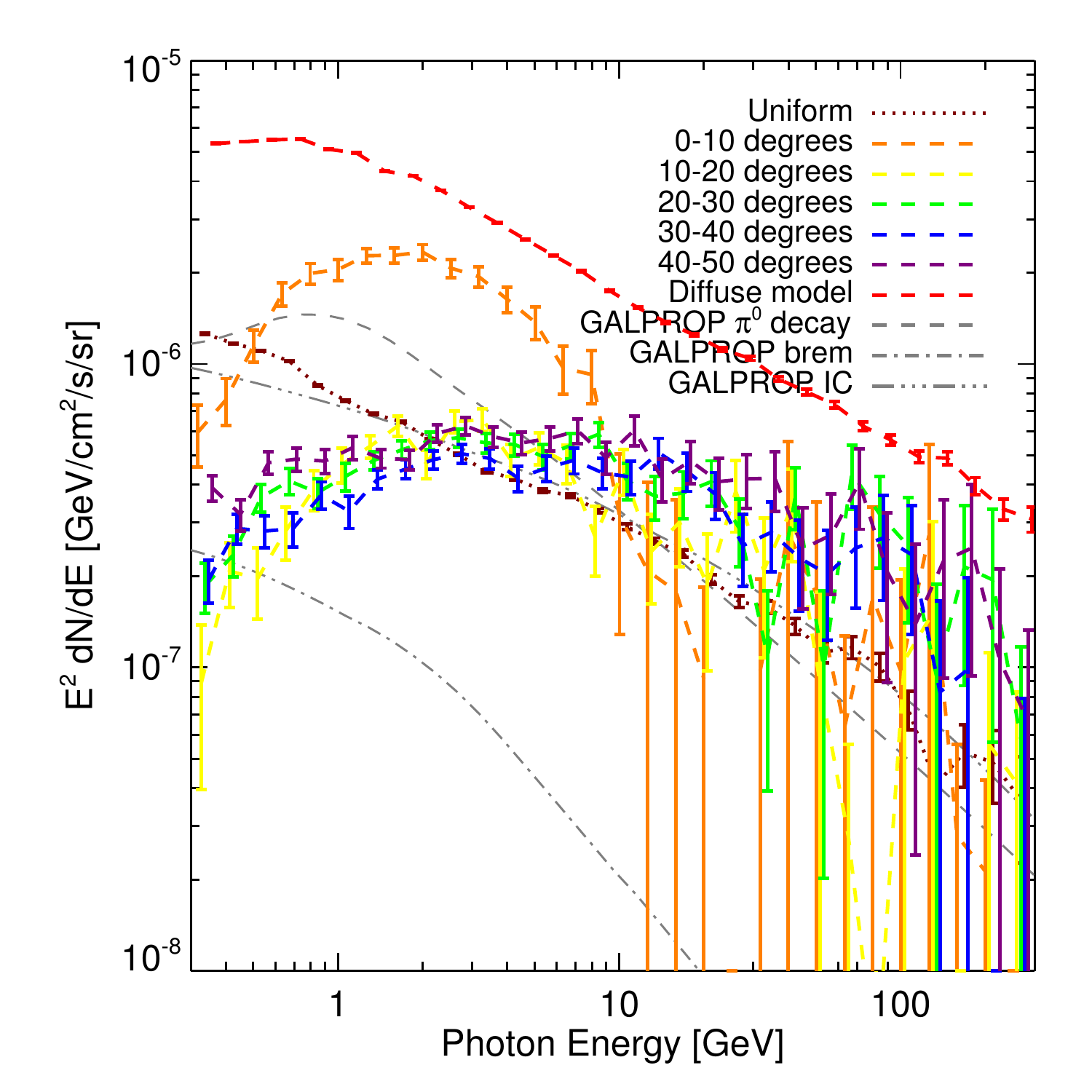}
\includegraphics[width=.32\textwidth]{plots/fermi_bubble_specFeb13_v1_4_0_25_26_27_28_29_diffuse_diskcut30_bcut1_southonly}
\includegraphics[width=.32\textwidth]{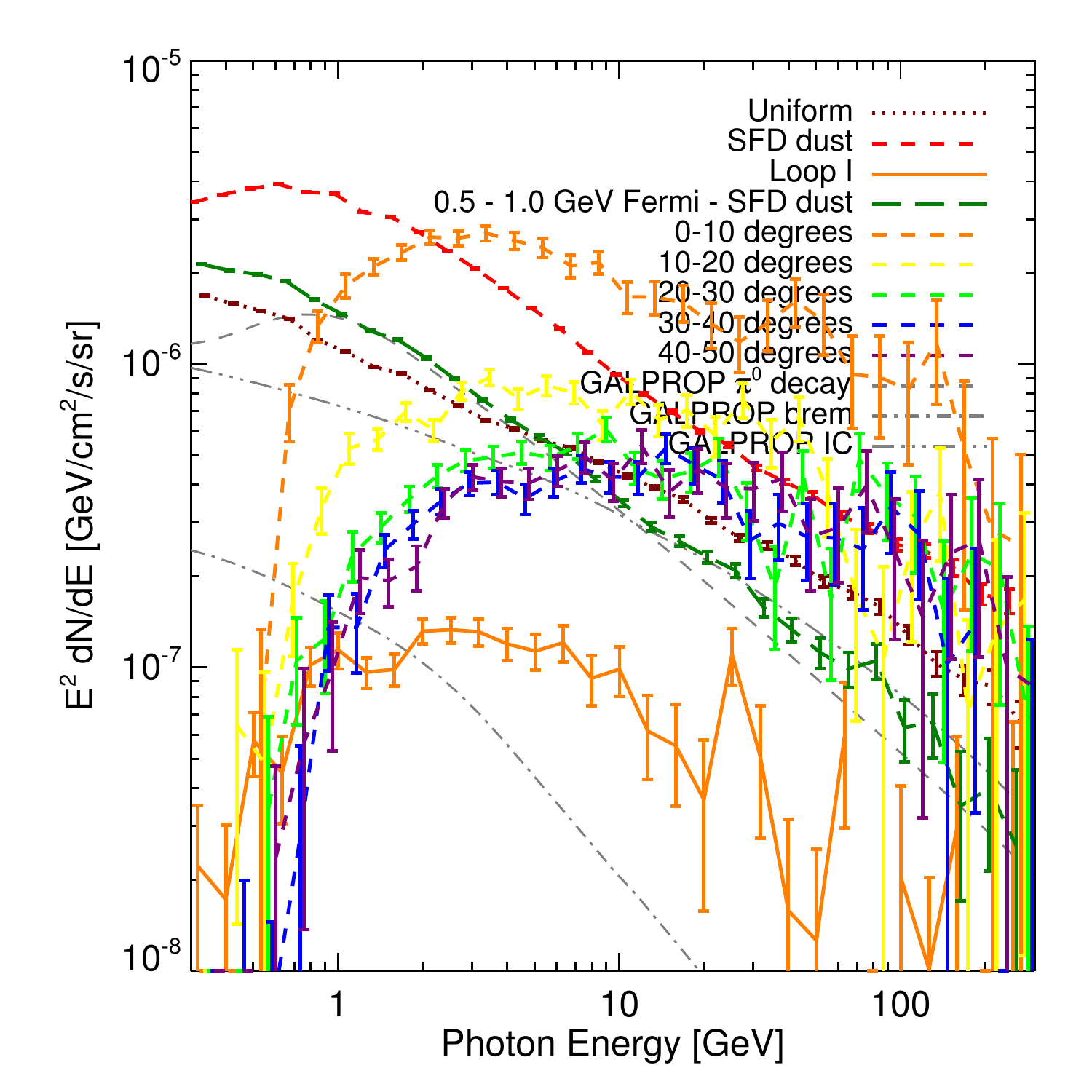}
\caption{\label{fig:alltemplatesspec}
The spectra of the various fit components, including five separate latitude-sliced templates for the Bubbles (see Fig.~\ref{sky}), for the two foreground models we employ (see text). The Galactic Disk is masked for $|b| < 1^\circ$ in each case. The \emph{left} and \emph{center} panels employ the ``diffuse model'' fit, for the entire sky in the \emph{left} panel and the southern hemisphere in the \emph{center} panel. The \emph{right} panel employs the ``low-energy template'' fit over the entire sky (see text for the details of the fitting procedures).}
\end{figure*}

\begin{figure*}
\includegraphics[width=.75\textwidth]{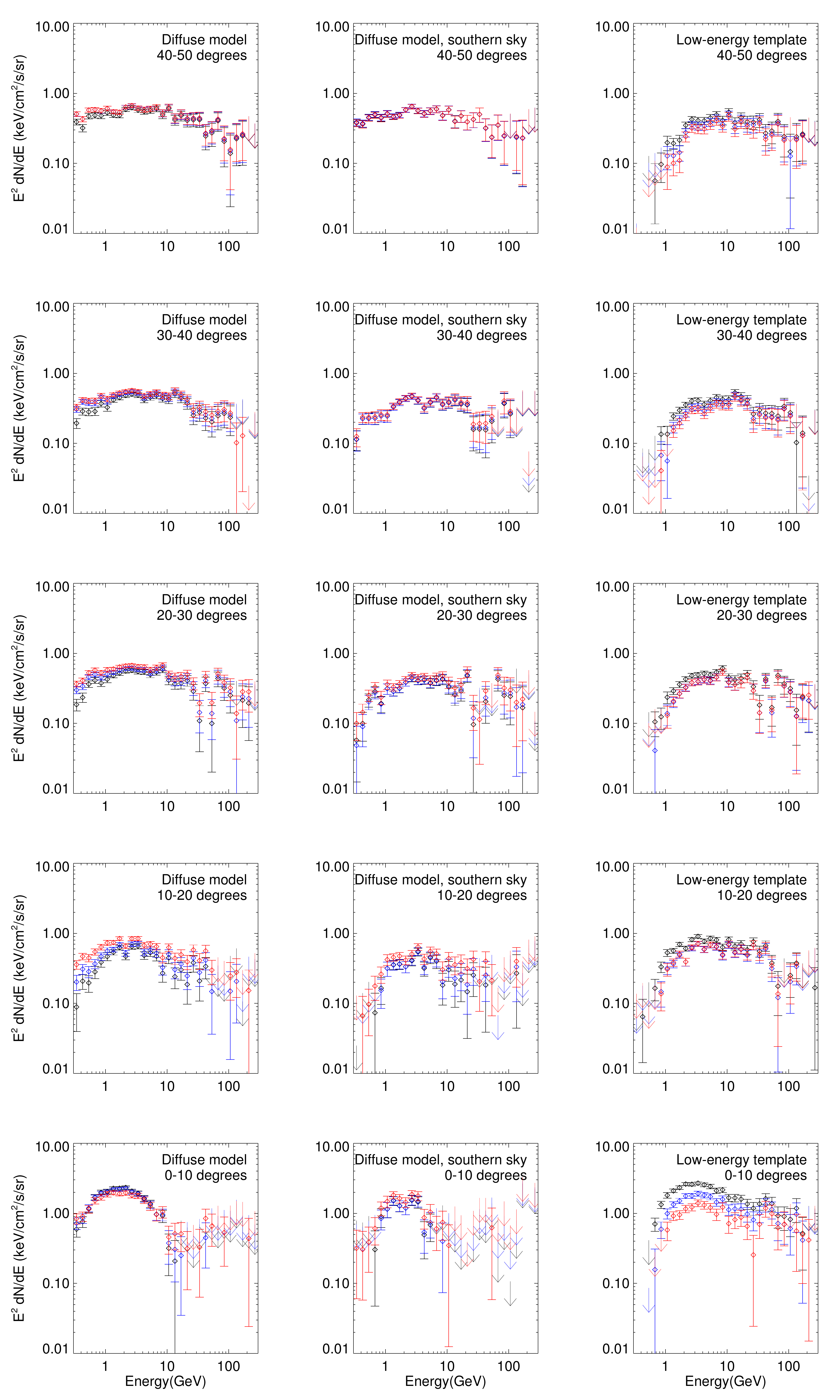}
\caption{\label{fig:specbylatitude}
The spectrum extracted for the gamma-ray Bubbles in ten-degree latitude bands: in order from the top row, $40^{\circ} <|b| < 50^\circ$, $30^\circ < |b| < 40^\circ$, $20^\circ < |b| < 30^\circ$, $10^\circ < |b| < 20^\circ$, $|b| < 10^\circ$. The \emph{left} and \emph{center} panels use the ``diffuse model'' template fit (see text); in the \emph{center} panels, the fit is restricted to $b < 0$ in addition to the masking. The \emph{right} panels use the ``low-energy template" approach (see text). The different colors show different choices for the latitude cut to remove the Galactic Disk: $|b| < 1^\circ$ (black), $|b| < 2^\circ$ (blue), $|b| < 5^\circ$ (red). Where the $1 \sigma$ error bars overlap with zero, we instead plot downward-pointing arrows corresponding to the $3 \sigma$ upper limits on the emission.}
\end{figure*}

Each of these templates has been discussed in greater depth in Ref.~\cite{Su:2010qj}; we refer the reader to that work for further details. We also employ the same normalization convention as Ref.~\cite{Su:2010qj}; the coefficients of the SFD dust map, 0.5-1.0 GeV map, and \emph{Fermi} diffuse model are multiplied by the average value of these templates within the entire region defined by the Bubbles (and outside of the mask). The other templates are flat (in projected intensity) within the regions that they are non-zero (over a given latitude range within the Bubbles, for example). 

We employ the ``diffuse model'' fit for our primary results, and use the ``low-energy template'' fit as a check for possible systematic errors introduced by our use of the \emph{Fermi} diffuse model. We have also tested the ``simple disk'' model as employed in Ref.~\cite{Su:2010qj}, where the ICS emission is described by a simple geometric diffused disk template. As this template proved inadequate for modelling the data close to the Galactic plane, and the results obtained using it were found to depend strongly on the degree of masking of the disk, we have relegated discussion of this model to Appendix~\ref{app:simpledisk}.

In Fig.~\ref{fig:alltemplatesspec} we show the extracted spectrum for each of the fitted templates, masking only the region within one degree of the Galactic plane, $|b| < 1^\circ$. We show results found using the ``diffuse model'' and the ``low-energy template''. In the center frame, we show the fit restricting to the southern sky ($b < 0$), which we might expect to be less contaminated by bright features such as Loop I. As expected, the error bars are larger in this case due to lower statistics, but the results are not otherwise significantly altered. In Fig.~\ref{fig:specbylatitude}, we show the spectra extracted for the gamma-ray Bubbles, and the dependence on the degree of masking of the disk, in each range of Galactic latitude. For our two preferred template models, the results are largely stable to changes in the mask. 

While the gamma-ray spectra extracted using the low-energy template appear somewhat different from those derived using the diffuse model, this is natural and expected, particularly at low energies, since part of the emission associated with the Bubbles is included in the low-energy template itself. Additionally, at high energies and low latitudes the low-energy template fit yields a significant amount of emission roughly flat in $E^2 dN/dE$, which is nearly absent in the diffuse model fit; these issues are discussed in further detail in Appendix~\ref{app:dmle}.

We note that the spectrum is almost invariant from $|b|=20^{\circ}-50^\circ$. This suggests that the electrons responsible for the observed emission in any leptonic scenario must either be accelerated \emph{in situ} or instead travel from the inner Galaxy very rapidly, avoiding significant energy losses (the distance over which TeV electrons propagate via standard diffusion without significant energy losses is considerably less than the 5 or more kpc to which this angular range corresponds). In contrast, a pronounced change in the Bubbles' spectrum is observed at lower latitudes. In an attempt to quantify the significance of this transition, we have compared the quality of the fit found using five separate latitude-sliced Bubbles templates to that found using only a single Bubbles template. Even conservatively limiting our analysis to the cleaner southern bubble, and masking within 5$^{\circ}$ of the disk, we find that the five-Bubbles-templates model is favored over the single Bubbles template at the level of approximately 16$\sigma$. However, it is important to note that this is a \emph{formal} significance, accounting only for statistical error; there is a degree of unavoidable and unaccounted-for systematic error in that neither model is a ``good fit'', in the sense of describing the sky to the level of Poisson noise.

\section{Cosmic Ray Electrons as the Source of the High-Latitude Gamma-Ray Bubbles and Synchrotron Haze}
\label{high}

\begin{figure*}[!]
\includegraphics[width=0.45\textwidth ]{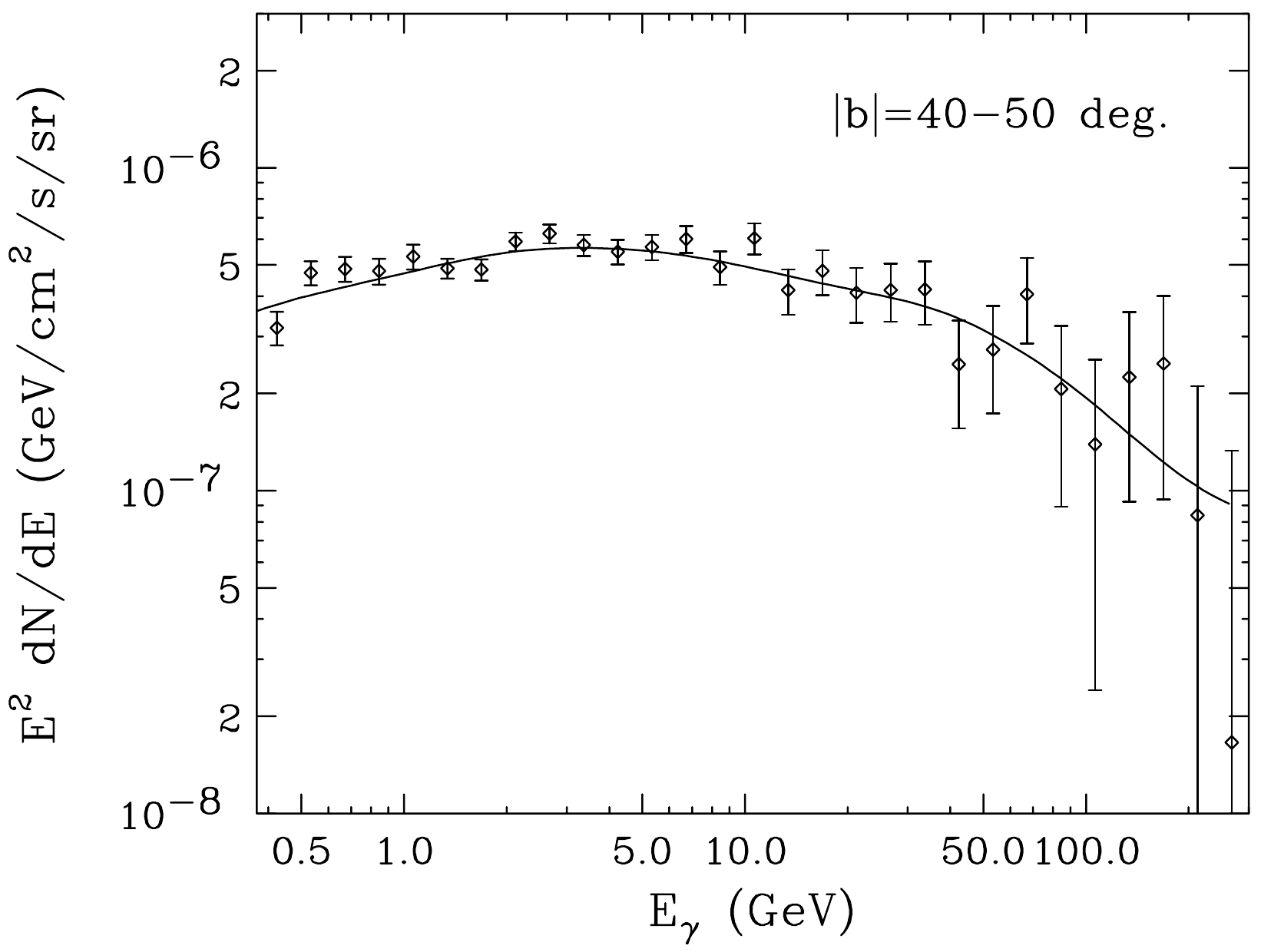}
\hspace{0.4cm}
\includegraphics[width=0.45\textwidth ]{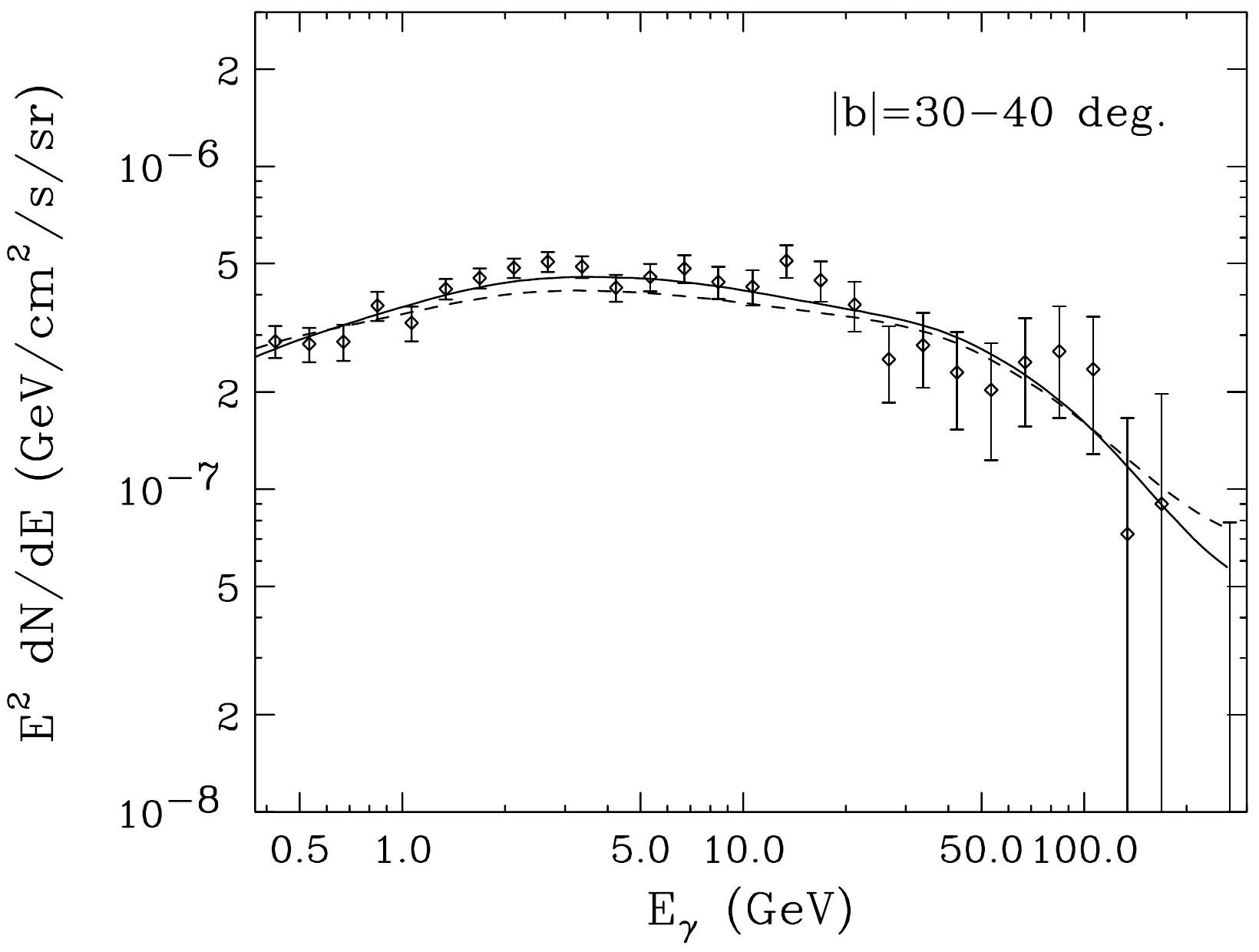}\\
\includegraphics[width=0.45\textwidth ]{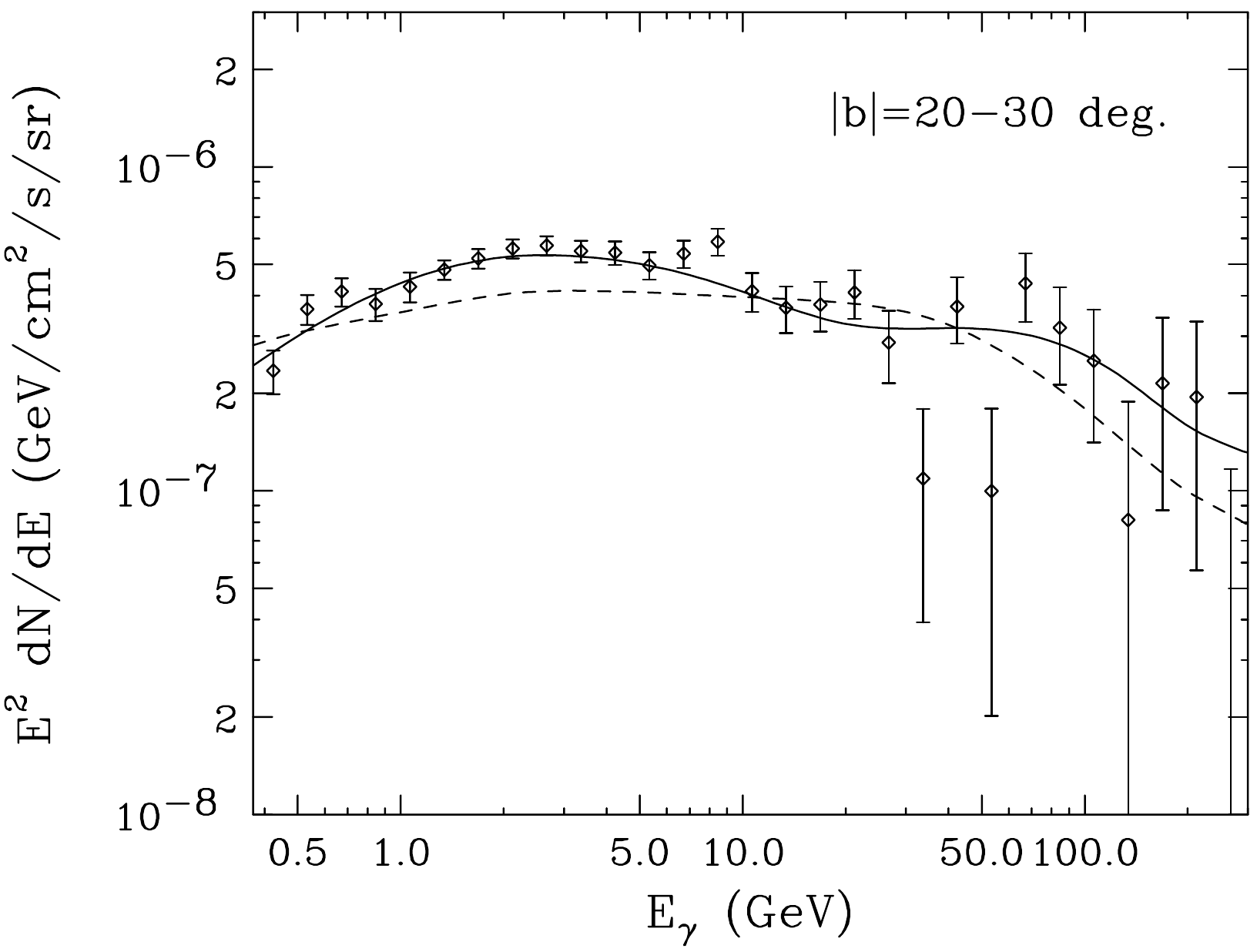}
\hspace{0.4cm}
\includegraphics[width=0.45\textwidth ]{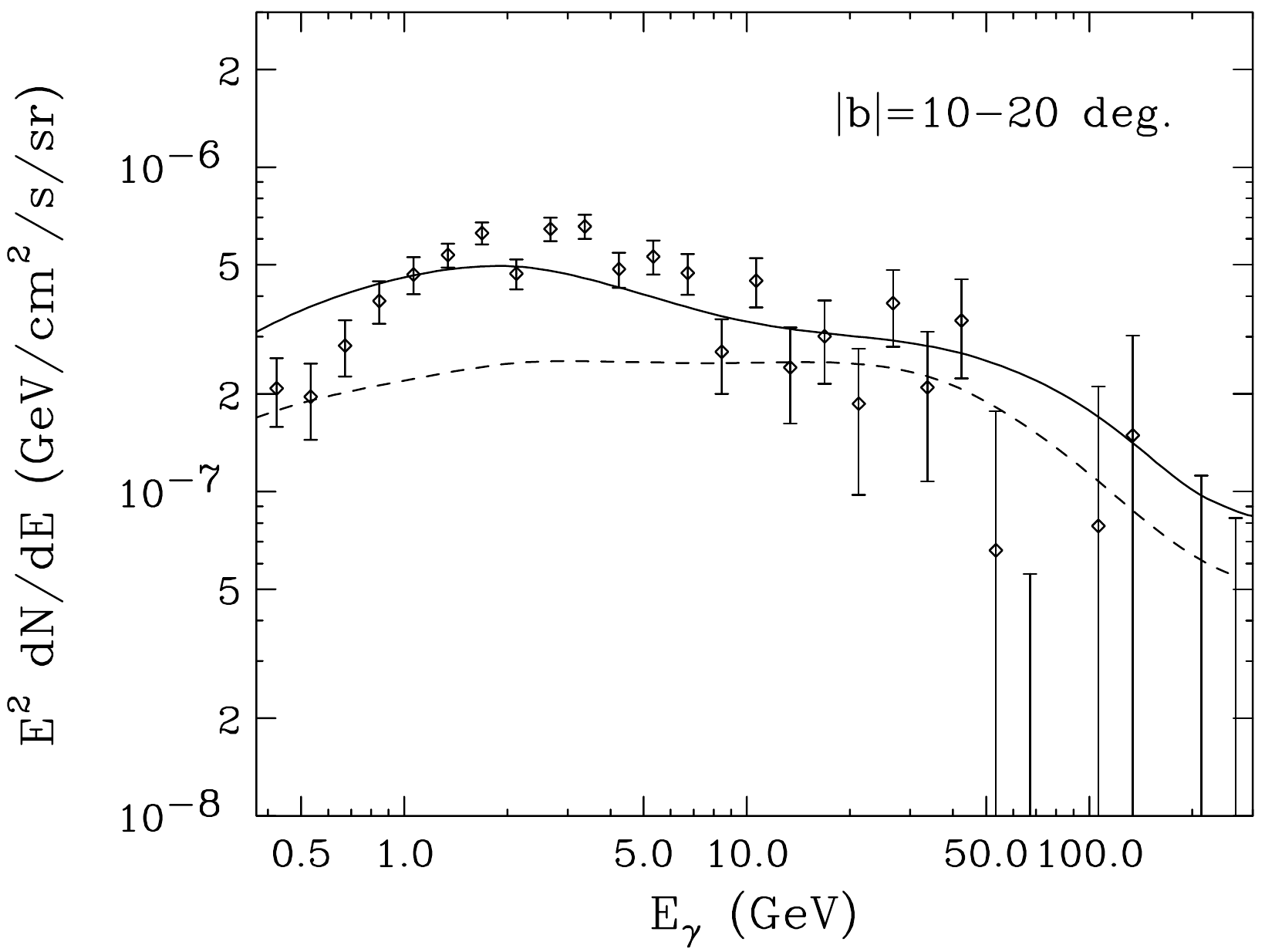}\\
\includegraphics[width=0.45\textwidth ]{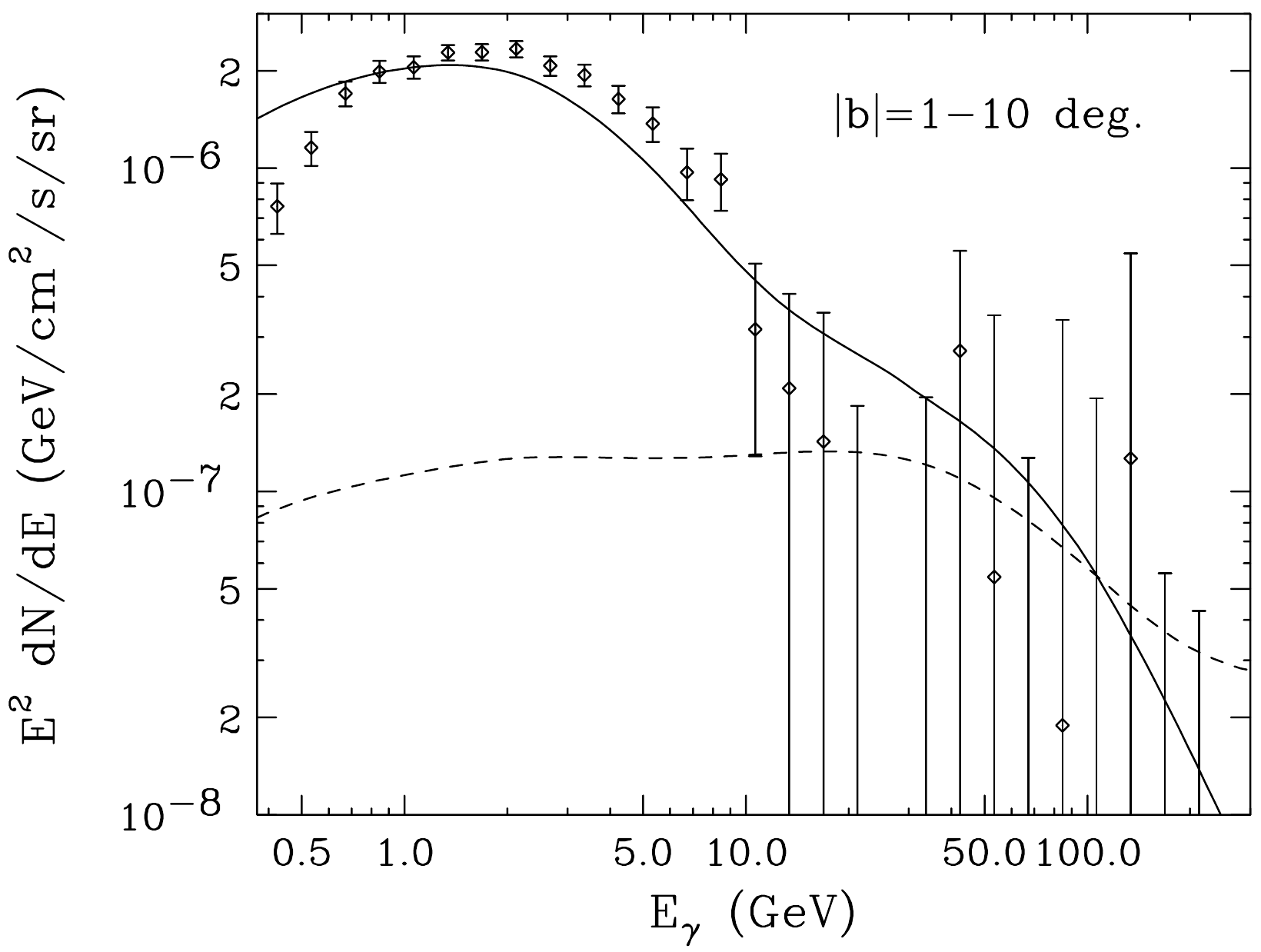}
\caption{The gamma-ray spectrum of the \emph{Fermi} Bubbles, broken into different regions by Galactic latitude (see Fig.~\ref{sky}). The solid lines denote the best-fit spectrum of inverse Compton emission, as calculated from the central values of the electron spectra shown in Fig.~\ref{elecspec}. At high latitudes, the spectrum of the \emph{Fermi} Bubbles is consistent with originating entirely from the inverse Compton scattering of GeV-TeV electrons, while at lower latitudes inverse Compton scattering alone cannot account for the observed emission. The dashed line in each frame denotes the spectrum of inverse Compton scattering that would be predicted from a spectrum of electrons the same as that required to generate the inverse Compton scattering spectrum observed in the highest latitude region ($|b|=40^{\circ}-50^{\circ}$), as discussed in Sec.~\ref{low}.}
\label{bubblesspec}
\end{figure*}

\begin{figure*}[!]
\includegraphics[width=0.45\textwidth ]{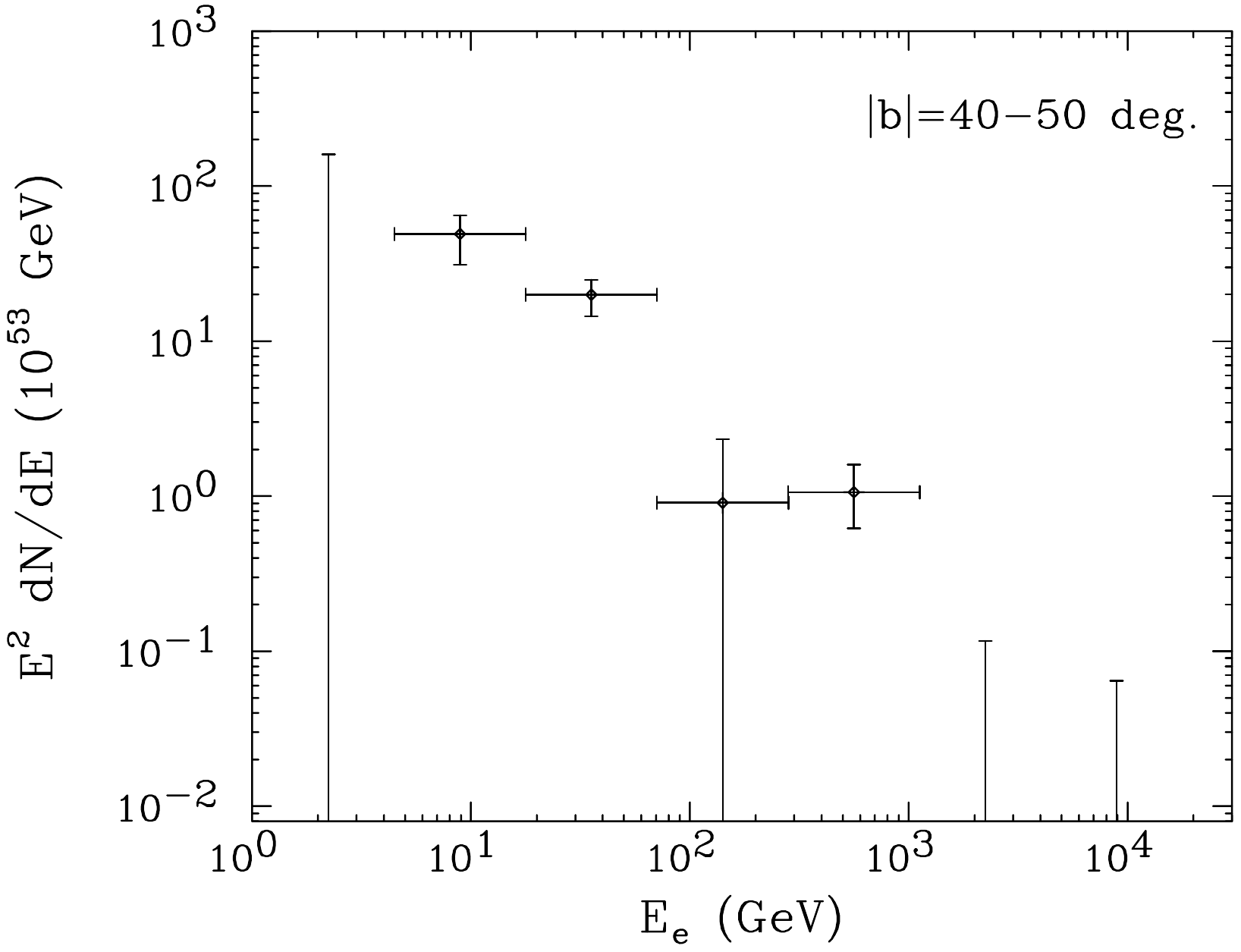}
\hspace{0.4cm}
\includegraphics[width=0.45\textwidth ]{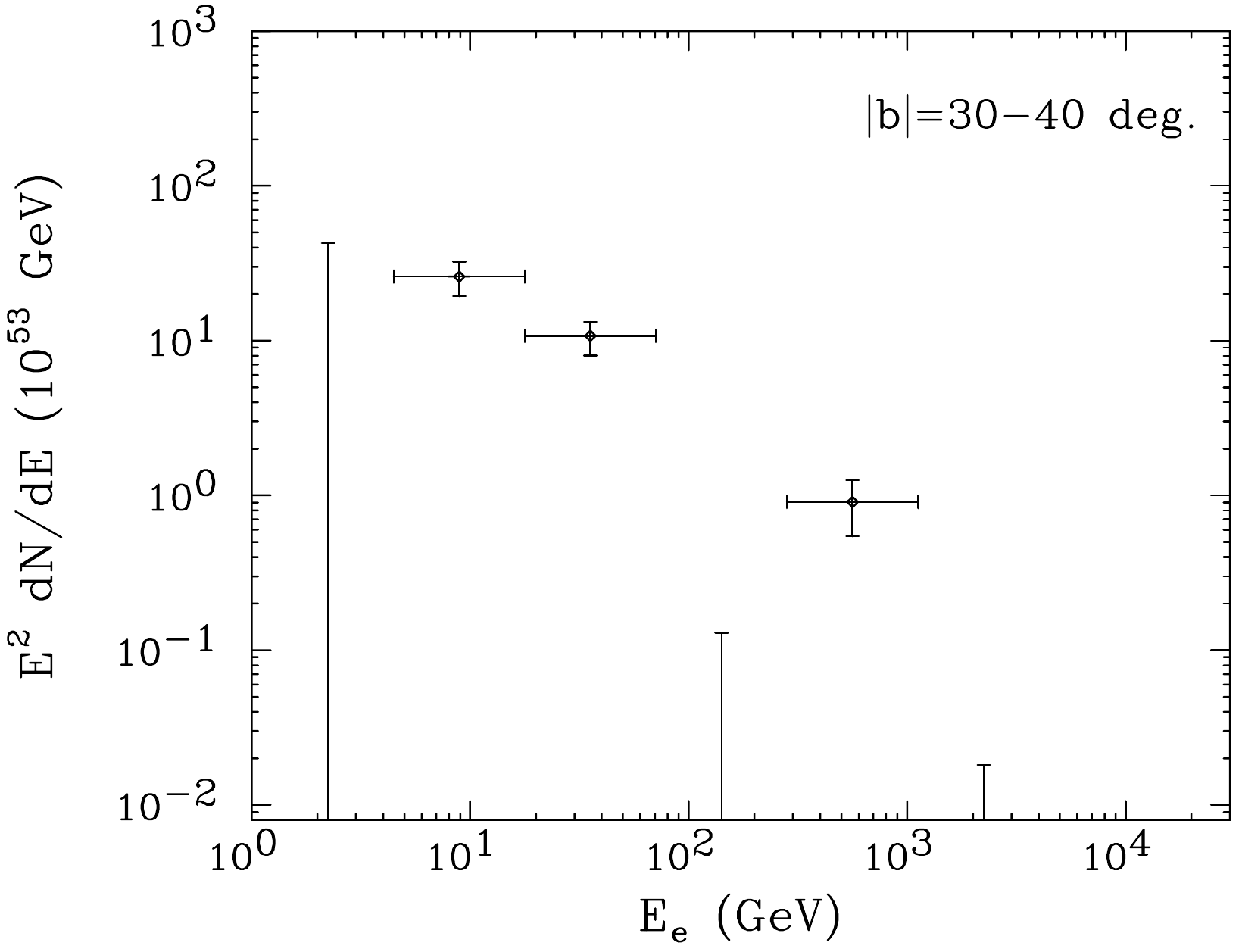}
\hspace{0.4cm}
\includegraphics[width=0.45\textwidth ]{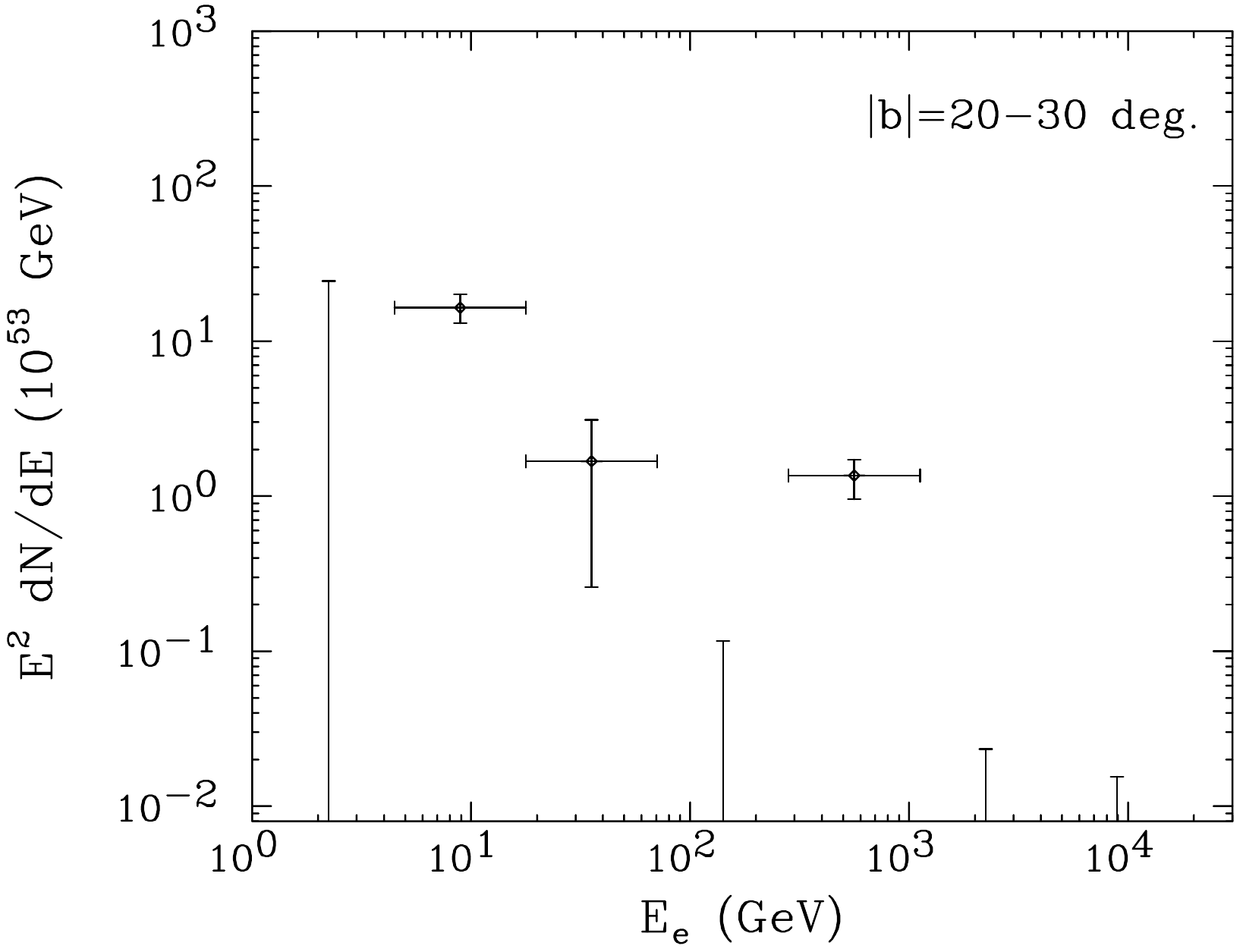}
\hspace{0.4cm}
\includegraphics[width=0.45\textwidth ]{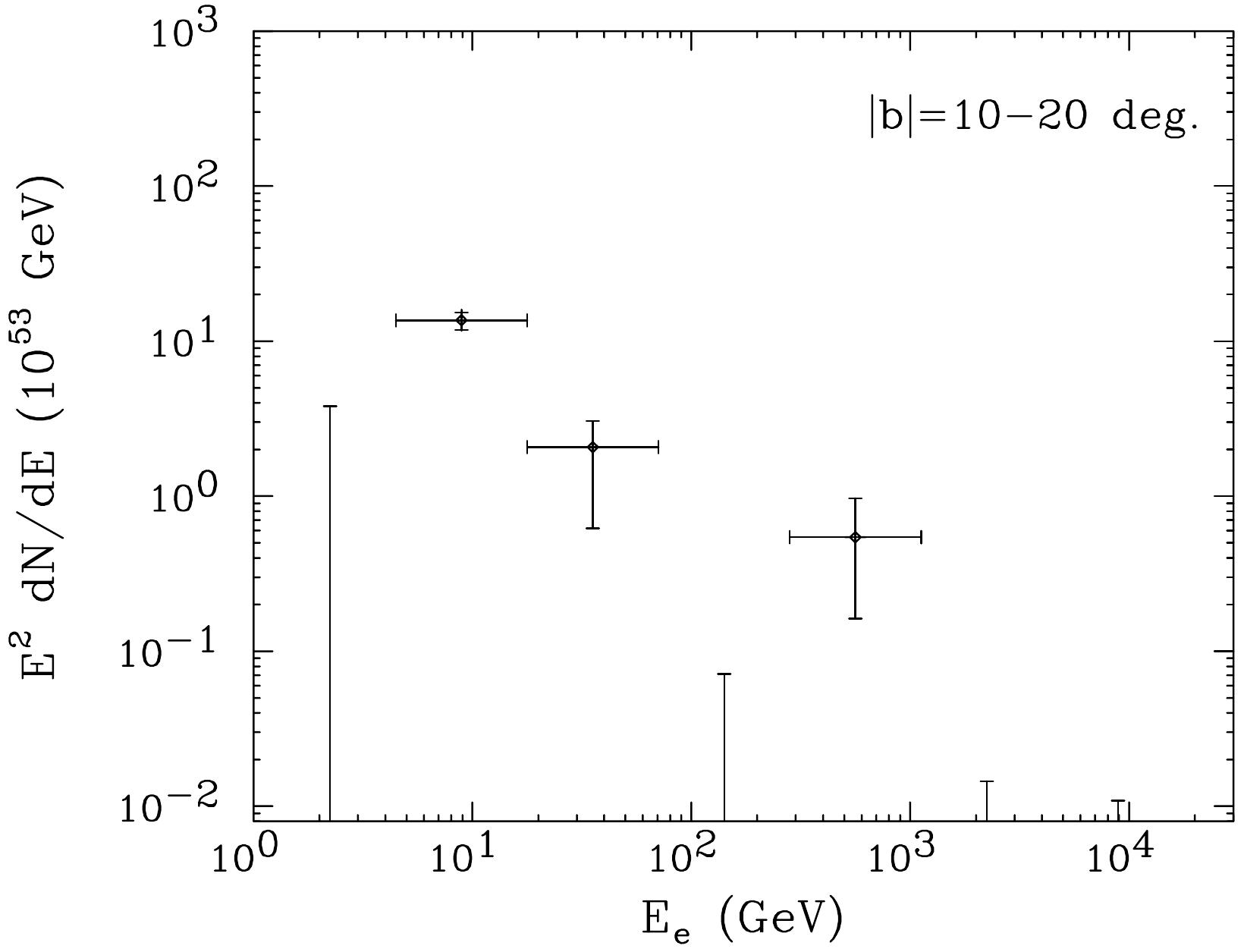}
\hspace{0.4cm}
\includegraphics[width=0.45\textwidth ]{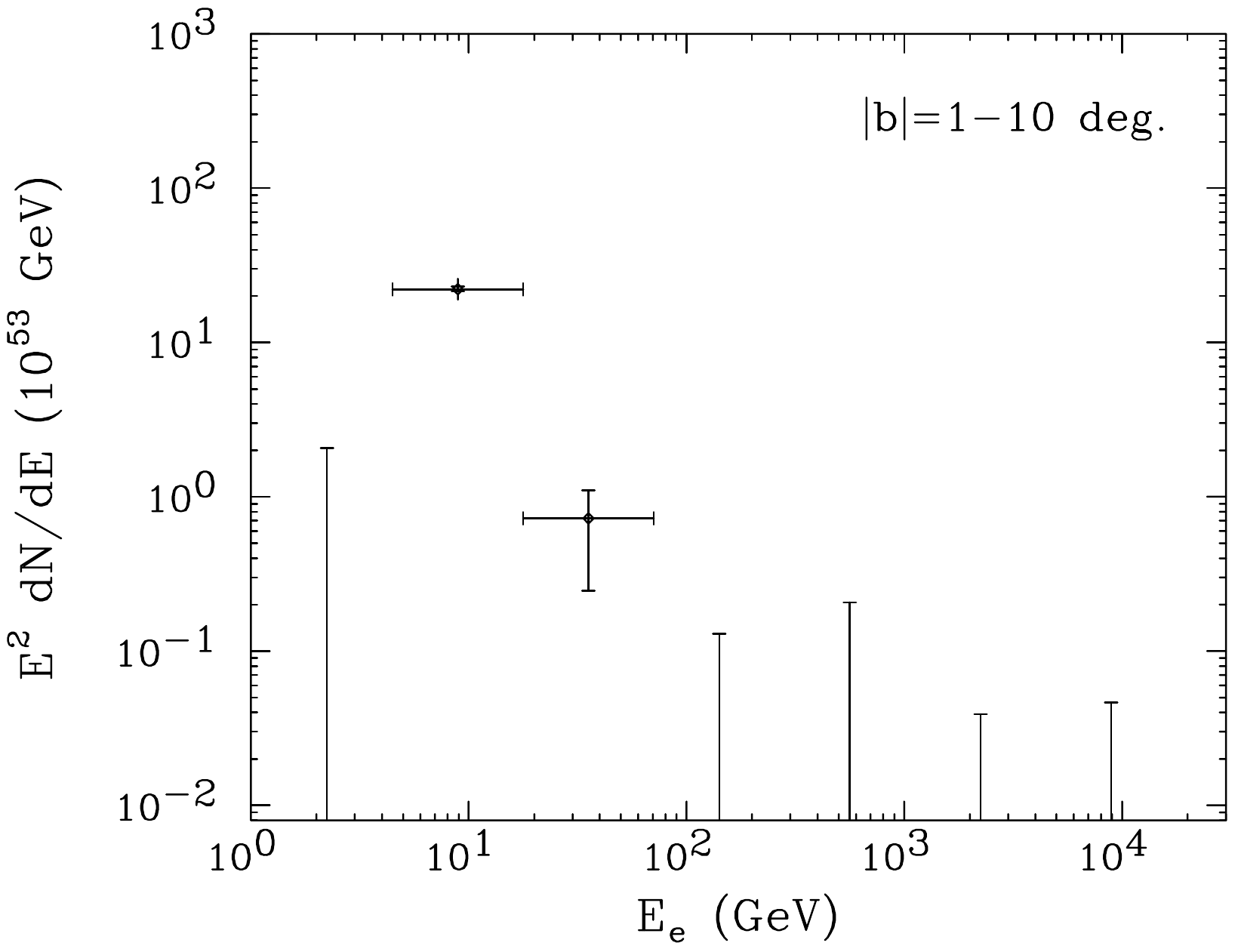}
\caption{The spectrum of electrons within the volume of the \emph{Fermi} Bubbles that is best able to fit the observed gamma-ray spectrum (see Fig.~\ref{bubblesspec}) through inverse Compton scattering alone. At high latitudes, the electron spectrum yields a good fit to the observed gamma-ray spectrum, while at low latitudes no spectrum of electrons is able to produce a gamma-ray spectrum consistent with observations.}
\label{elecspec}
\end{figure*}

Following Blumenthal and Gould~\cite{1970RvMP...42..237B}, we employ the full Klein-Nishina formula to compute the spectrum of inverse Compton emission from an arbitrary electron population. For the problem at hand, we need to consider scattering with the CMB as well as with starlight and infrared radiation. In our calculations, we adopt the interstellar radiation model of Ref.~\cite{Porter:2005qx}. At energies below $\sim$$3\times 10^{-3}$ eV, the CMB dominates the energy density, while starlight is important at higher energies.

\begin{figure*}
\includegraphics[width=.49\textwidth]{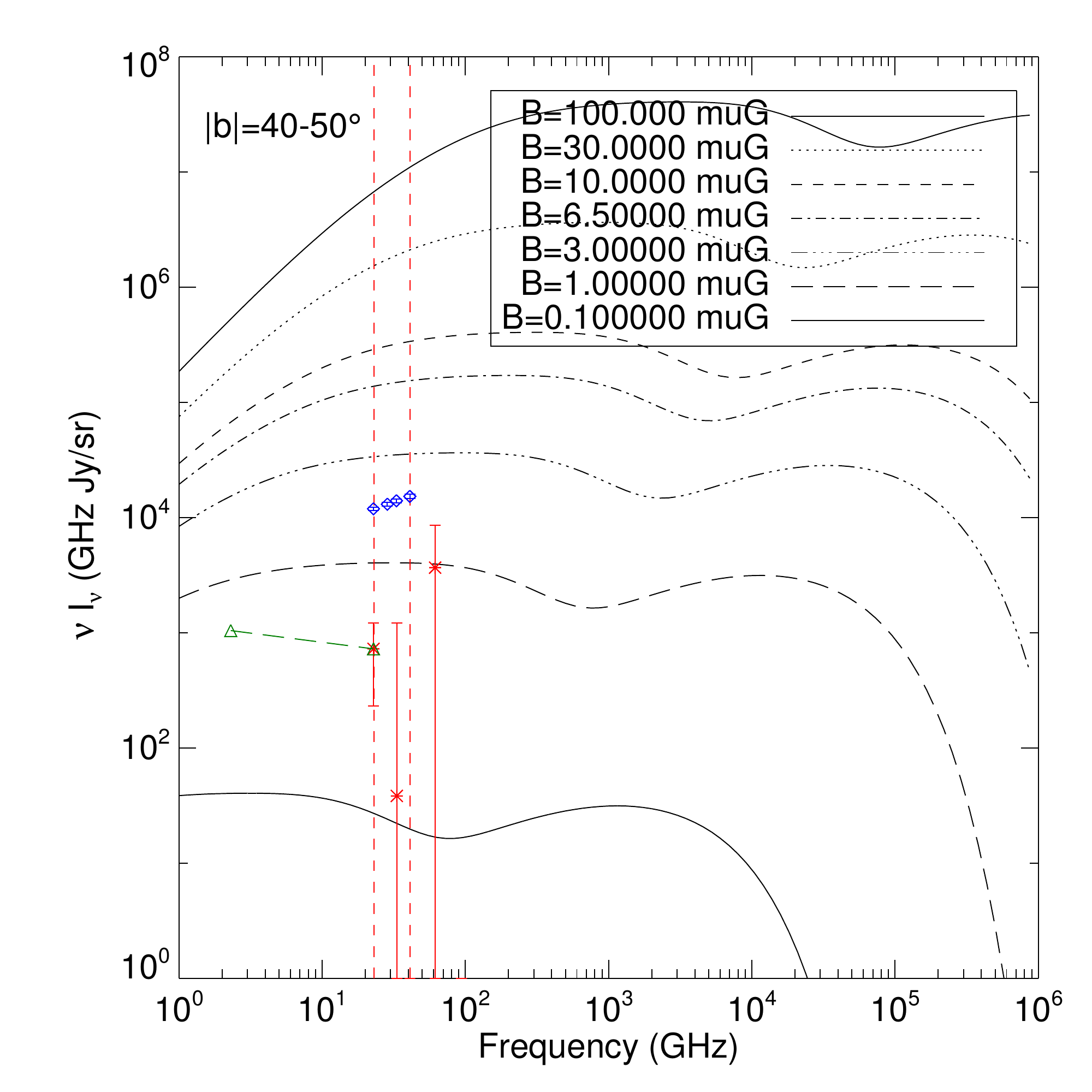}
\includegraphics[width=.49\textwidth]{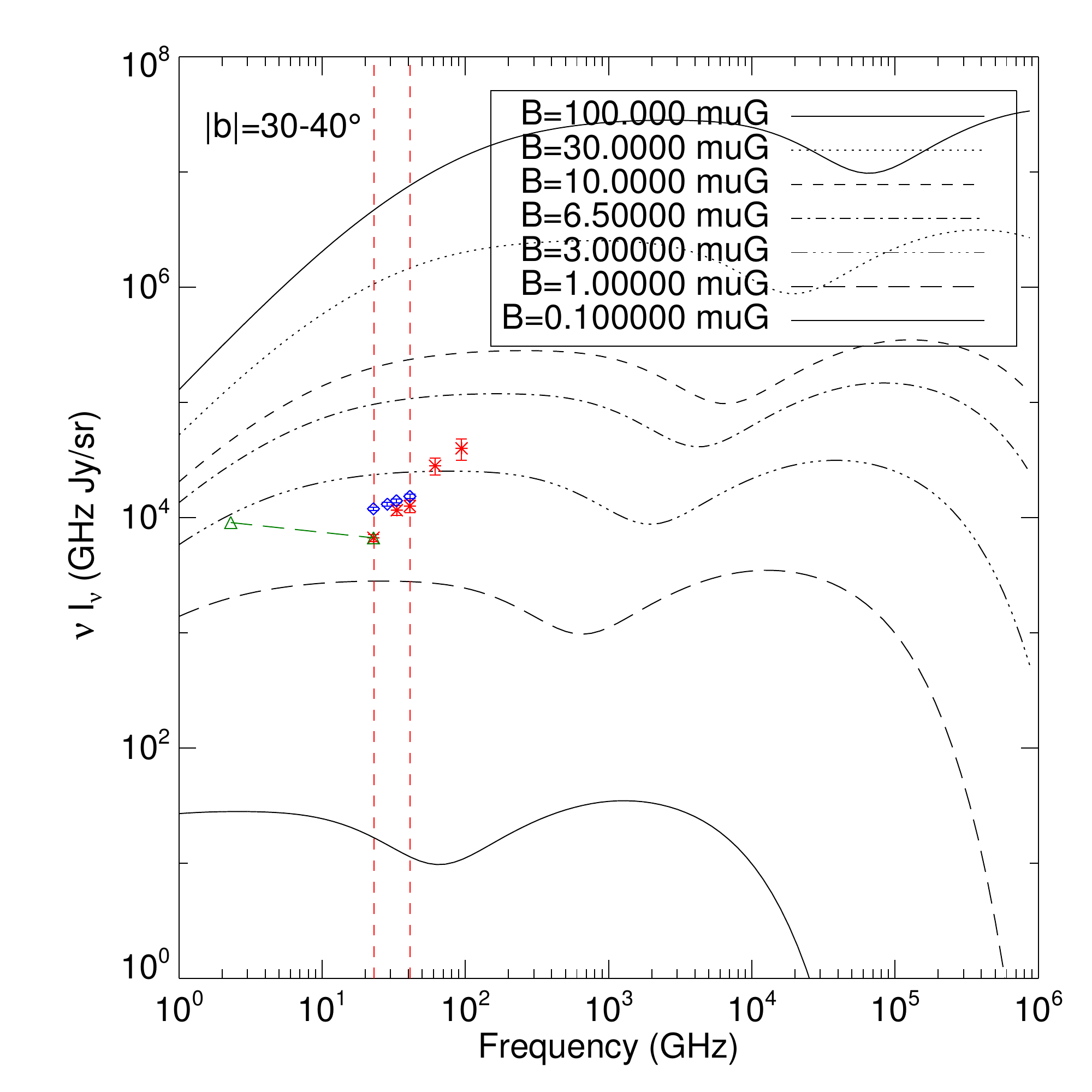}\\
\includegraphics[width=.49\textwidth]{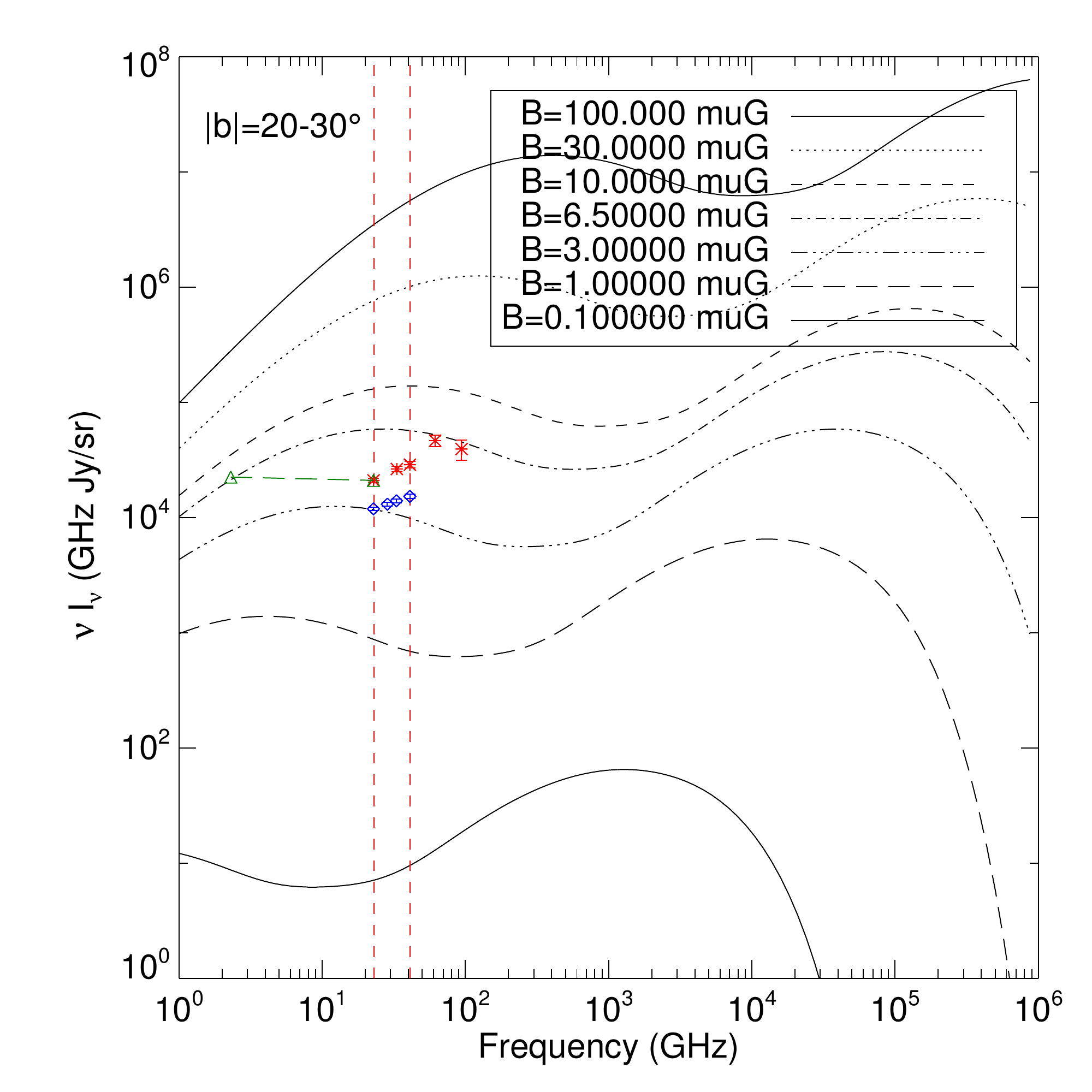} 
\includegraphics[width=.49\textwidth]{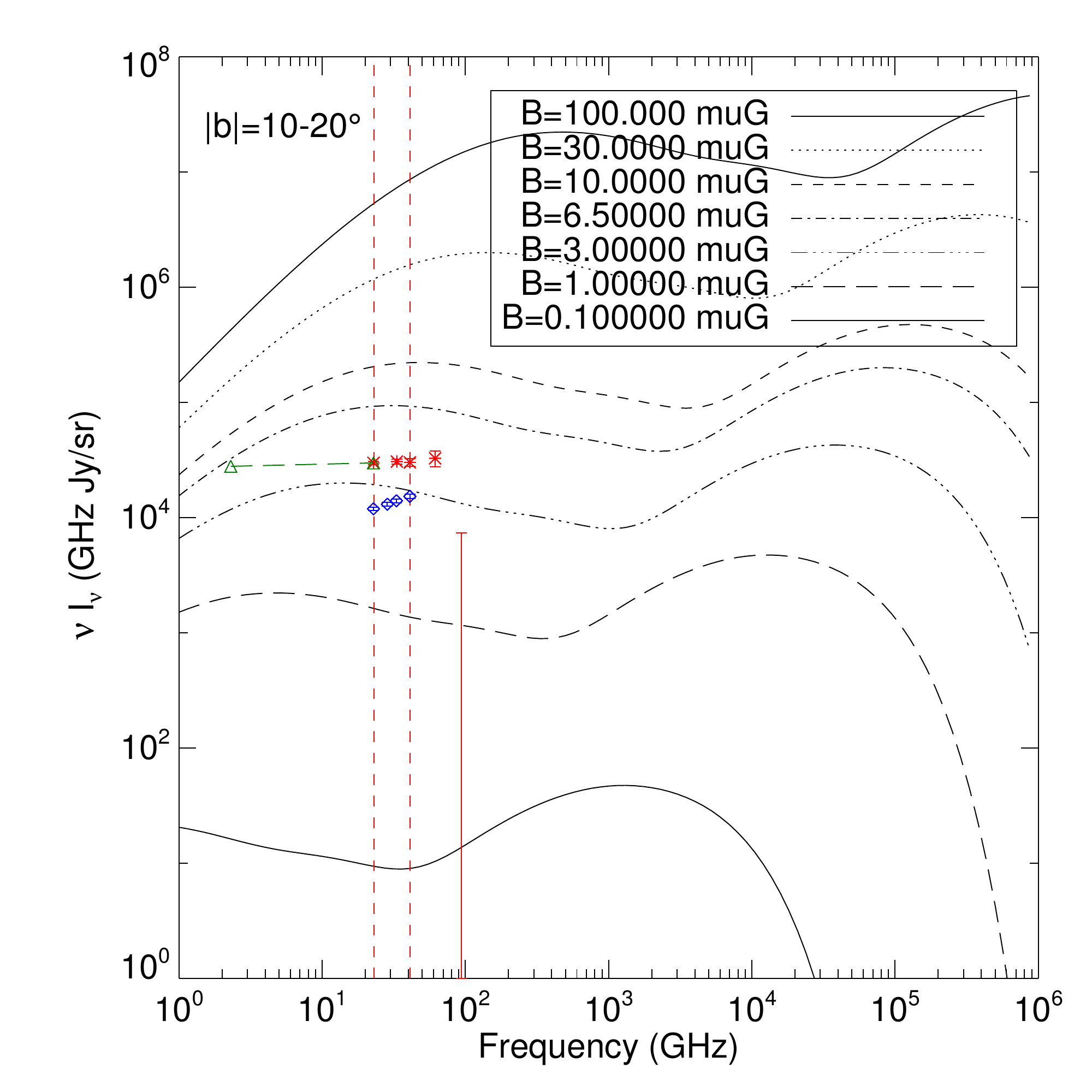}
\caption{\label{fig:hazecompare}
The spectrum of synchrotron emission predicted from the best-fit electron spectra as shown in Fig.~\ref{elecspec}, for various values of the magnetic field (see legend). Each frame represents a range of latitudes of the \emph{Fermi} Bubbles: $40^{\circ}-50^{\circ}$ away from the Galactic plane (\emph{top left}), $30^{\circ}-40^{\circ}$ away from the Galactic plane (\emph{top right}), $20^{\circ}-30^{\circ}$ away (\emph{bottom left}), and $10^{\circ}-20^{\circ}$ away (\emph{bottom right}). Red data points indicate the spectrum of the haze in the southern sky within each given latitude range, as derived from Ref.~\cite{Dobler:2011rd}; the average over $l$ is performed for $-5^\circ < l < 15^\circ$, as in Fig. 5 of Ref.~\cite{Dobler:2011rd}. At somewhat lower frequencies, the green triangles and dashed lines connecting them denote the spectral index (\emph{not} amplitude) for polarized emission between 2.3 GHz and 23 GHz, as extracted from S-PASS \cite{spass} data by Ref.~\cite{Carretti:2013sc}; the averaging is performed over $-5^\circ < l < 15^\circ$ and restricted to $|b| > 15^\circ$ to avoid depolarization associated with the Galactic plane. This comparison demonstrates that in the presence of microgauss-scale magnetic fields, the electron population responsible for the observed ICS emission can also account for the observed synchrotron haze. Also shown for comparison as blue data points is the spectrum of the microwave haze as a whole, as given by Ref.~\cite{:2012fb}. As these results are not binned by latitude, however, they should not be directly compared to the predicted synchrotron spectra.}
\end{figure*}

The gamma-ray spectra observed from various regions of the \emph{Fermi} Bubbles are shown again in Fig.~\ref{bubblesspec} (as found using the diffuse model template fit, and masking within 1 degree of the disk). To determine whether these gamma-rays could be the product of inverse Compton scattering, we take an arbitrary (binned) spectrum of electrons and compare the resulting inverse Compton emission to that shown in Fig.~\ref{bubblesspec}. In Fig.~\ref{elecspec}, we plot the electron spectrum which provides the best possible fit to the gamma-ray spectrum for each latitude range (and error bars around the best fit). The solid line in each frame of Fig.~\ref{bubblesspec} denotes the best-fit spectrum of inverse Compton photons. At high latitudes, an approximately power-law spectrum of electrons ($dN_e/dE_e \propto E_e^{-3}$) between GeV and TeV energies can produce a spectrum of gamma-rays consistent with that observed from the \emph{Fermi} Bubbles. At $|b|=40^{\circ}-50^{\circ}$, for example, the best-fit electron spectrum shown in Fig.~\ref{elecspec} provides an excellent fit ($\chi^2=21.8$ over 29-7 degrees-of-freedom) to the observed gamma-ray spectrum. 

As noted in the Introduction, the inverse Compton interpretation of the \emph{Fermi} Bubbles is supported by the observation of spatially correlated emission at both microwave~\cite{:2012fb} and radio~\cite{Carretti:2013sc} wavelengths. For a given magnetic field strength (or distribution of magnetic field strengths), we can calculate the spectrum of synchrotron emission that is predicted to result from the spectrum of electrons shown in Fig.~\ref{elecspec}. For simplicity, we take the central values for the extracted spectrum and assume an isotropic electron population; we do not extrapolate to energies above the highest bin. In Fig.~\ref{fig:hazecompare}, we show the resulting synchrotron spectra for a range of magnetic fields from 0.1-100 $\mu$G, assumed to be uniform throughout the region of the Bubbles template in question. For comparison, the magnetic field strength in the local region of the Milky Way is thought to be of order a few $\mu$G~\cite{Jansson:2012pc}. One should keep in mind, however, that magnetic fields in localized regions and filaments may be much higher than the spatially averaged value.

To compare the predicted synchrotron spectrum to observations, we employ the \emph{WMAP}7 results of Ref.~\cite{Dobler:2011rd}, and the spectral index of the polarized emission between 2.3 GHz and 23 GHz, as obtained from S-PASS \cite{spass} data and described in Ref.~\cite{Carretti:2013sc}.\footnote{We thank Greg Dobler for providing us with the numerical values corresponding to Fig.~5 of Ref.~\cite{Dobler:2011rd} and the authors of Ref.~\cite{Carretti:2013sc} for providing the spectral index map corresponding to their Fig.~S4.} We also show the recent results combining data from \emph{WMAP} and \emph{Planck}~\cite{:2012fb}, but these results were not binned by latitude so should not be directly compared to the predicted synchrotron curves.

The amplitudes of the gamma-ray Bubbles and microwave haze appear consistent with arising from a common population of cosmic ray electrons for magnetic fields on the order of a few $\mu$G. The spectrum deduced from \emph{WMAP} is somewhat harder than expected from the corresponding gamma-ray data, but this may be the result of contamination by the CMB (which would be expected to produce the appearance of spectral hardening). In contrast, the spectral index for the polarized emission found by the authors of Ref.~\cite{Carretti:2013sc} appears consistent with expectations from the gamma-rays. If the spectral hardness observed in \emph{WMAP} data is confirmed by \emph{Planck}, however, it could suggest a scenario in which most of the synchrotron emission is produced in regions with higher than average magnetic fields. For example, if magnetic fields as high as $\sim 30 \mu$G are present in even $\sim$1\% of the volume of the Bubbles, then this discrepancy could be ameliorated.

We caution that the success of this picture does not \emph{rule out} hadronic scenarios; we focus on leptonic scenarios in this work because their consistency can be checked in a straightforward and model-independent way, since the microwaves and gamma-rays originate from the same steady-state electron population. In hadronic scenarios, in contrast, the gamma-rays provide a probe of the proton CR spectrum, but the microwaves probe the electron spectrum after diffusion and cooling, and so the consistency may depend on diffusion parameters in addition to the magnetic field. Consequently, we leave a careful study of consistency in the hadronic case for future work. Later in this article, we treat the inverse-Compton-like emission present in the Bubbles as a background to be subtracted; this ``background'' is an approximately flat spectrum in $E^2 dN/dE$ with a downturn around 50 GeV, and while it is modeled as inverse Compton emission, instead treating it as emission from a hadronic cascade should not impact our results in any significant way.

\section{Evidence of Non-Inverse Compton Emission At Low Latitudes}
\label{low}

In the previous section, we demonstrated that the spectrum observed from the high latitude ($|b|\gsim 30^{\circ}$) regions of the \emph{Fermi} Bubbles can be accounted for by inverse Compton scattering of an approximately power-law spectrum of cosmic ray electrons. The same electrons, in the presence of $\mu$G-scale magnetic fields, can also account for the observed amplitude of the \emph{WMAP} haze. At lower Galactic latitudes, however, we find that inverse Compton scattering alone cannot account for the observed gamma-ray emission. In particular, we find that the gamma-ray spectrum climbs rapidly between 0.3 and 1 GeV at low-latitudes (see Fig.~\ref{bubblesspec}), and this rise cannot be accounted for by the inverse Compton scattering of any physically realistic spectrum of electrons. Quantitatively, for the choice of binning used in Fig.~\ref{elecspec}, we find that an entirely inverse-Compton origin for the gamma-rays observed from the $|b|=1^{\circ}-10^{\circ}$ or $|b|=10^{\circ}-20^{\circ}$ regions of the \emph{Fermi} Bubbles yields best-fits of $\chi^2=100.4$ and 95.7, respectively, each over 29-7 degrees-of-freedom. From this, we are forced to conclude that a non-negligible fraction of the gamma-ray emission observed from the low-latitude regions of the \emph{Fermi} Bubbles is not the result of inverse Compton scattering.\footnote{The conclusion that no spectrum of electrons can produce the low-latitude gamma-ray emission could, in principle, be evaded if we had adopted a much more finely binned electron spectrum in our analysis. In particular, a cosmic ray electron spectrum that is described by a delta function at 16 GeV provides a good fit to the low-latitude feature. As such a feature is not expected from the perspective of cosmic ray acceleration, nor realistic in light of non-negligible energy loss processes, we do not consider this possibility further.} 

\begin{figure*}[!]
\includegraphics[width=0.45\textwidth ]{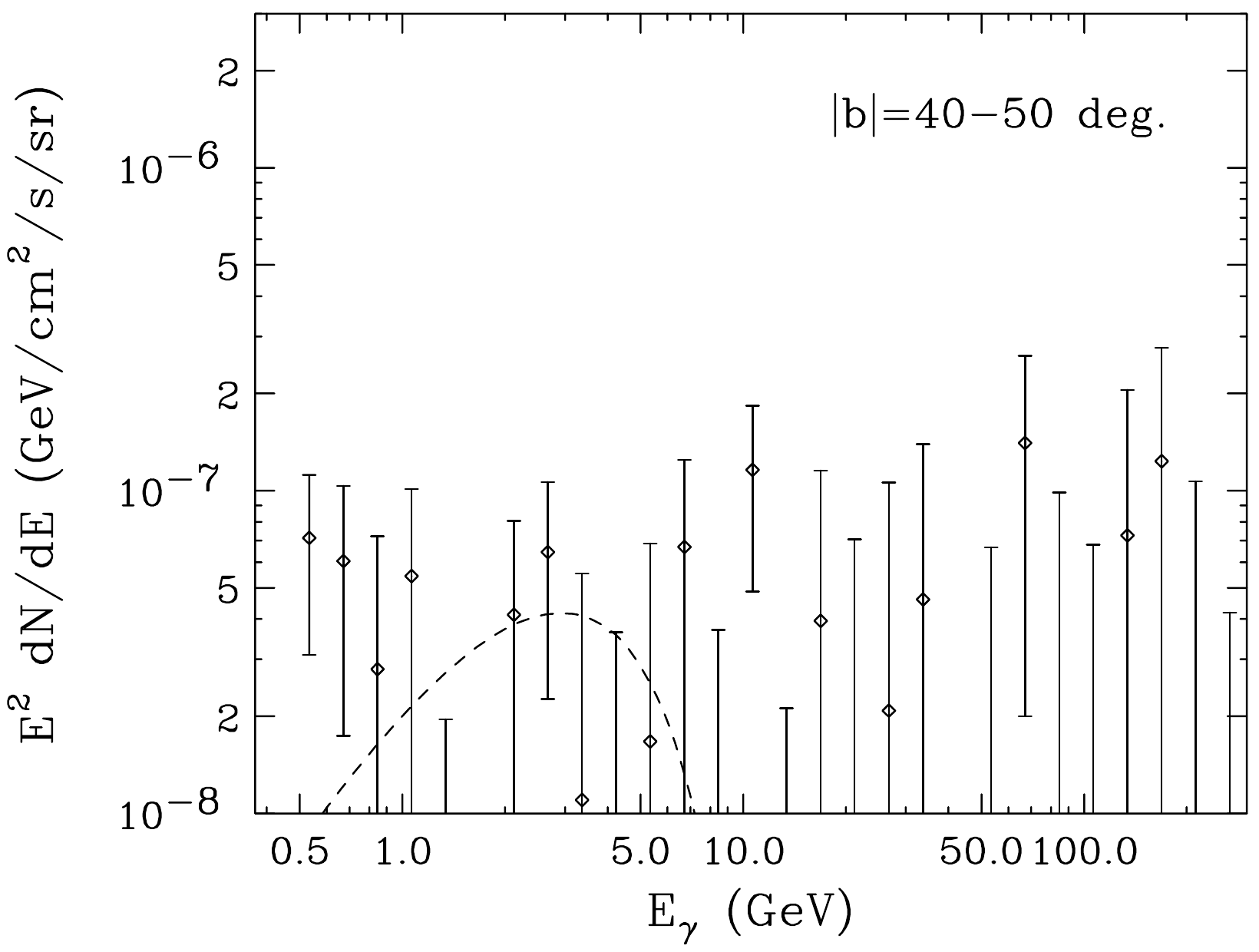}
\hspace{0.4cm}
\includegraphics[width=0.45\textwidth ]{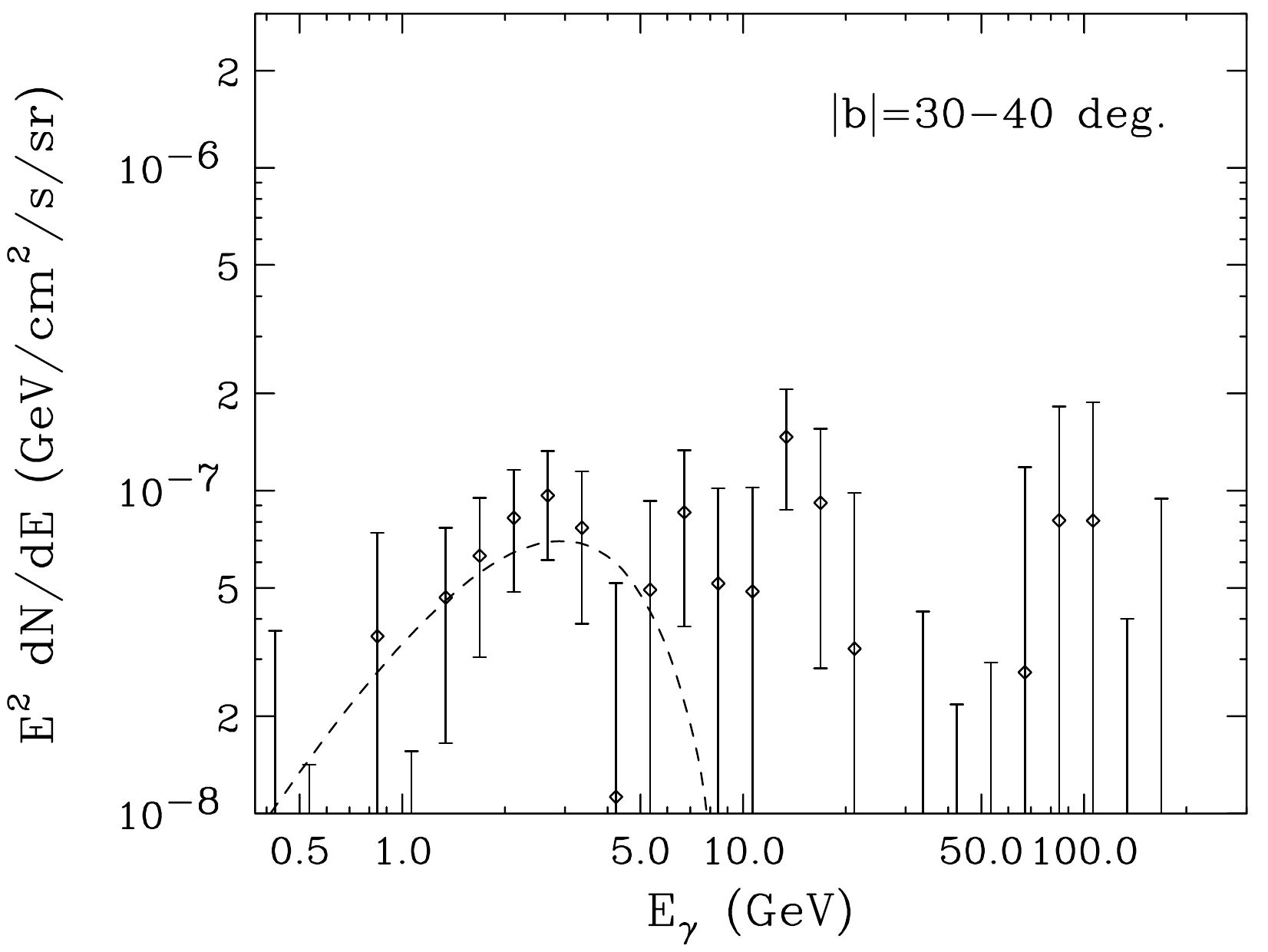}\\
\includegraphics[width=0.45\textwidth ]{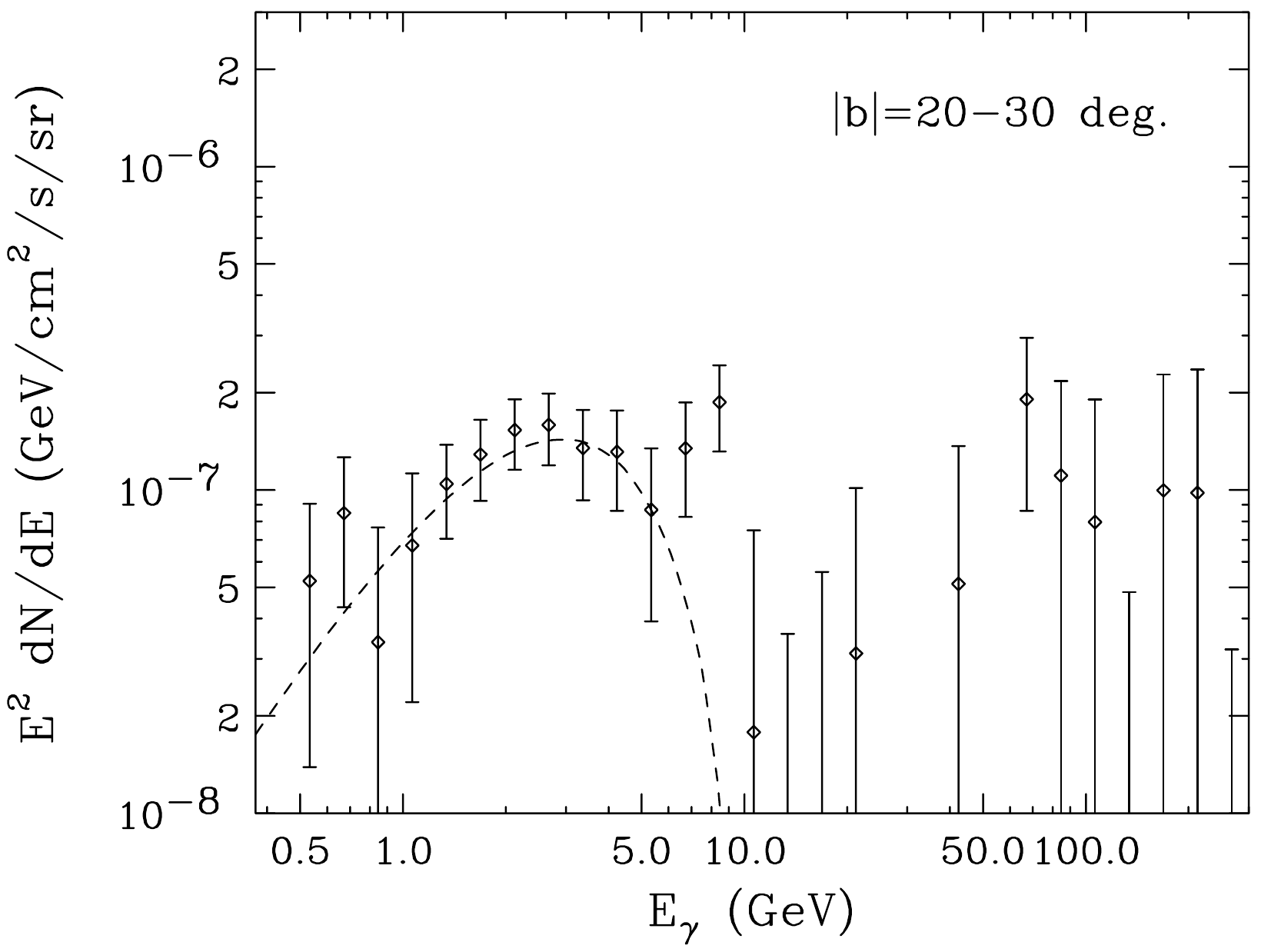}
\hspace{0.4cm}
\includegraphics[width=0.45\textwidth ]{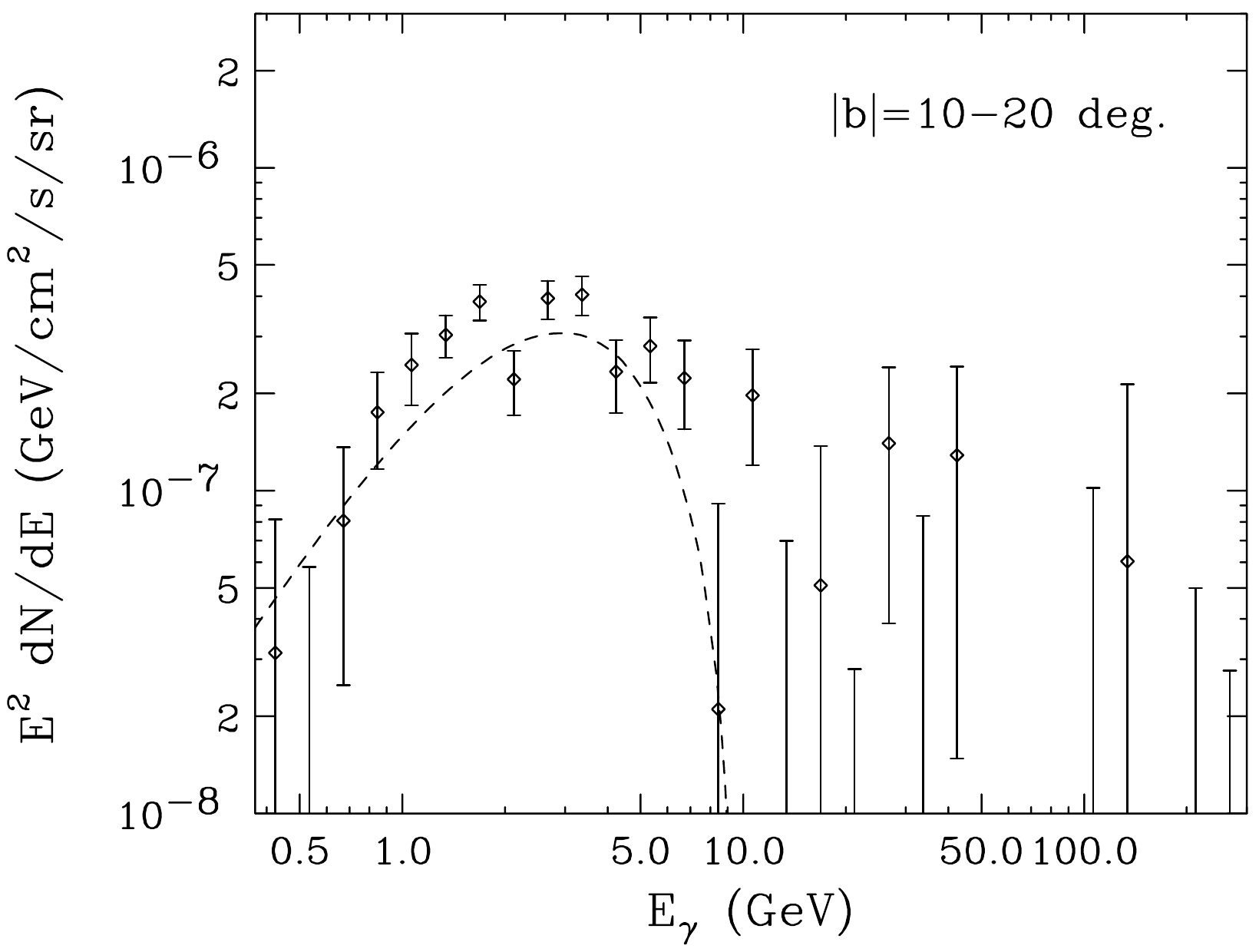}\\
\includegraphics[width=0.45\textwidth ]{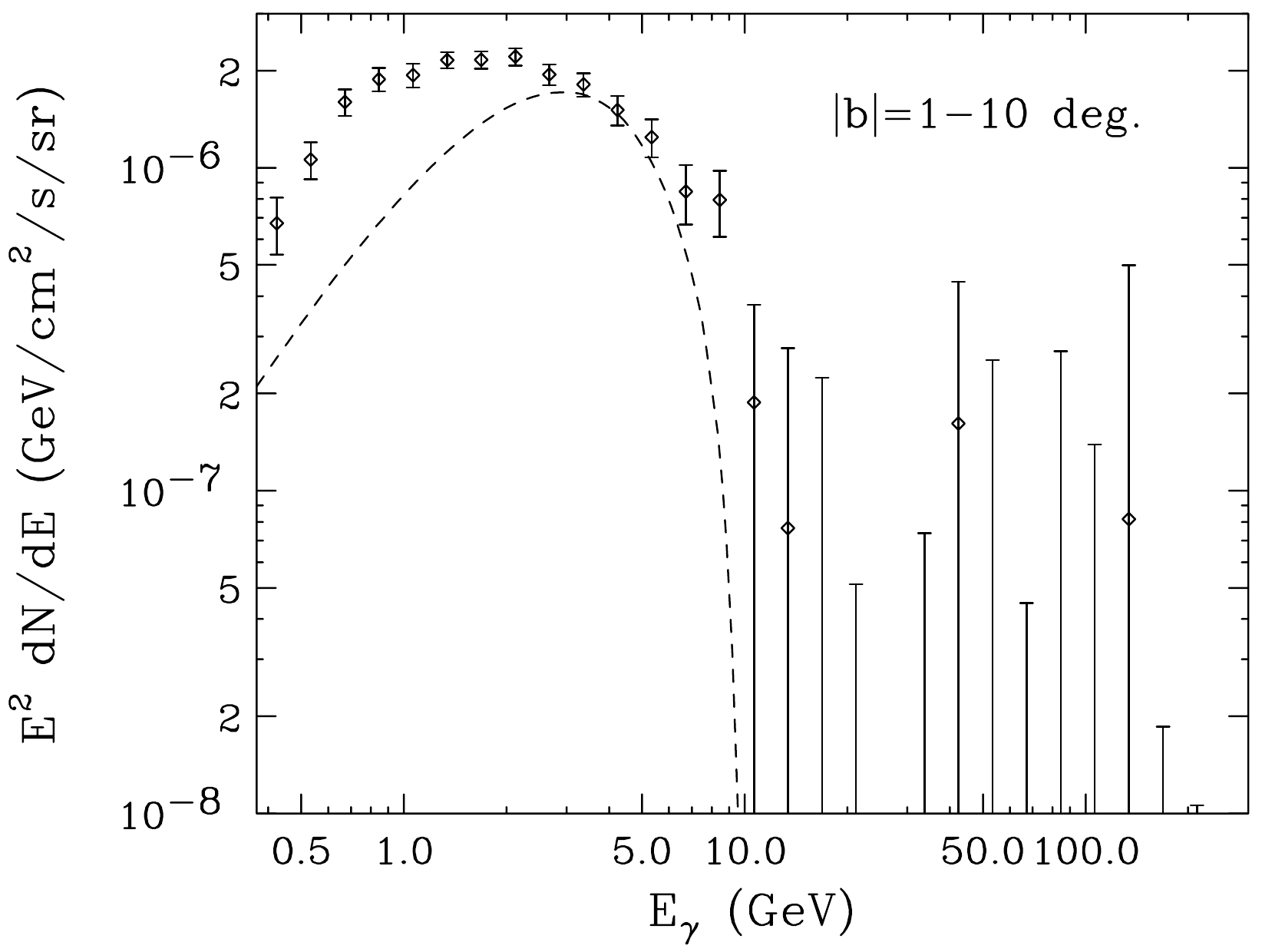}
\caption{The gamma-ray spectrum of the \emph{Fermi} Bubbles after subtracting a contribution from inverse Compton emission, derived using the electron spectrum (up to normalization) found in our best-fit to the $|b|=40^{\circ}-50^{\circ}$ region. This illustrates the characteristics of the additional (non-inverse Compton) component of the gamma-ray emission from the \emph{Fermi} Bubbles, which is quite bright at low Galactic latitudes. We caution that these extracted spectra are subject to a number of systematic uncertainties, such as those associated with the interstellar radiation field model, and due to uncertainties and variations in the electron spectra throughout the volume of the Bubbles. These extracted spectra can, however, be taken as indicative of the broad spectral features of the non-inverse Compton component of the Bubbles emission. Shown as dashed lines is the predicted contribution of gamma-rays from the annihilations of 10 GeV dark matter particles (to $\tau^+ \tau^-$) distributed according to a generalized NFW profile with an inner slope of $\gamma=1.2$, as described in Sec.~\ref{compare}. We remind the reader that the backgrounds are largest near the disk and thus there are significant systematic uncertainties in the spectrum from the low latitude ($|b|=1^{\circ}-10^{\circ}$) region, especially at low energies.}
\label{nonIC}
\end{figure*}

If we assume that the spectral shape of electrons present throughout the volume of the \emph{Fermi} Bubbles does not change significantly with latitude, and that the gamma-ray emission from the highest latitude region ($|b|=40^{\circ}-50^{\circ}$) is dominated by the products of inverse Compton scattering, we can calculate the inverse Compton contribution from each of the lower latitude regions. In each frame of Fig.~\ref{bubblesspec}, the dashed line denotes this predicted contribution from inverse Compton scattering. Not surprisingly, this component makes up most or all of the total observed emission at high latitudes, but only a small fraction at low latitudes. Note that the shapes of these dashed lines differ with latitude only as a result of variations in the radiation field model. 

To better isolate the spectrum of the non-inverse Compton emission, we subtract this estimate of the inverse Compton contribution from the total emission of the \emph{Fermi} Bubbles in each latitude range. This residual spectrum is plotted in Fig.~\ref{nonIC}. While one should keep in mind that the error bars shown in this figure do not take into account any variations in the spectral shape of cosmic ray electrons throughout the volume of the Bubbles, this provides us with what is likely to be a reasonable estimate of the spectrum and intensity of the non-inverse Compton emission exhibited in the low-latitude regions of the \emph{Fermi} Bubbles. 

The spectra shown in Fig.~\ref{nonIC} exhibit some rather distinctive features. In particular, this spectral component peaks strongly at energies of $\sim$1-5 GeV, and has no statistically significant presence above $\sim$10 GeV. Furthermore, the intensity of this component is a very strong function of Galactic latitude, being more than an order of magnitude brighter at low latitudes than at intermediate latitudes. These spectral and morphological characteristics are quite similar to those exhibited by the gamma-ray emission previously observed from the inner few degrees surrounding the Galactic Center~\cite{Hooper:2011ti,Hooper:2010mq,Goodenough:2009gk,Abazajian:2012pn,Boyarsky:2010dr}. In the following section, we will discuss this comparison in more detail.

\section{Comparison With Gamma-Ray Emission From The Galactic Center}
\label{compare}

In previous studies of \emph{Fermi} data, multiple sets of authors have identified the presence of a bright and spatially extended gamma-ray source around the Galactic Center, peaking at energies of a few GeV~\cite{Hooper:2010mq,Hooper:2011ti, Goodenough:2009gk, Boyarsky:2010dr, Abazajian:2012pn}. In particular, Ref.~\cite{Hooper:2011ti} reports that the morphology of this source implies a luminosity per volume that scales as $r^{-2.4}$ to $r^{-2.8}$, where $r$ is the distance to the Galactic Center. Similar profiles were found to provide good fits in Refs.~\cite{Hooper:2010mq,Abazajian:2012pn,Boyarsky:2010dr}. Each of these studies also found that the spectrum of this spatially extended emission peaks strongly at energies of a few GeV, very similar to the peak found in this study in the low-latitude emission from the \emph{Fermi} Bubbles. 

This comparison strongly suggests that the non-inverse Compton emission we have identified in the low-latitude regions of the \emph{Fermi} Bubbles is in fact the more spatially extended counterpart of the gamma-ray signal previously reported from the innermost few degrees around the Galactic Center. To further explore this comparison, we plot as a dashed line in each frame of Fig.~\ref{nonIC} the contribution predicted by an annihilating dark matter model found in Ref.~\cite{Hooper:2012ft} to provide a good fit to the Galactic Center signal. In this model, 10 GeV dark matter particles annihilate to tau lepton pairs (with $\sigma v = 2 \times 10^{-27}$ cm$^3$/s to $\tau^+ \tau^-$, or $\sigma v = 6 \times 10^{-27}$ cm$^3$/s if annihilations proceed equally to $\tau^+ \tau^-$, $\mu^+ \mu^-$ and $e^+e^-$), and are distributed following a generalized Navarro-Frenk-White profile, with an inner slope of $\gamma=$1.2 (the classic NFW profile corresponds to an inner slope of $\gamma=1$), a scale radius of 20 kiloparsecs, and normalized such that the local dark matter density is 0.4 GeV/cm$^3$. Specifically, the functional form is:

\begin{equation} \rho(r) \propto \frac{r^{-\gamma}}{\left(1+\frac{r}{R_s}\right)^{3 - \gamma}}, \quad \gamma = 1.2, \, R_s = 20 \,\mathrm{kpc}. \end{equation}

\begin{figure*}
\includegraphics[width=\textwidth]{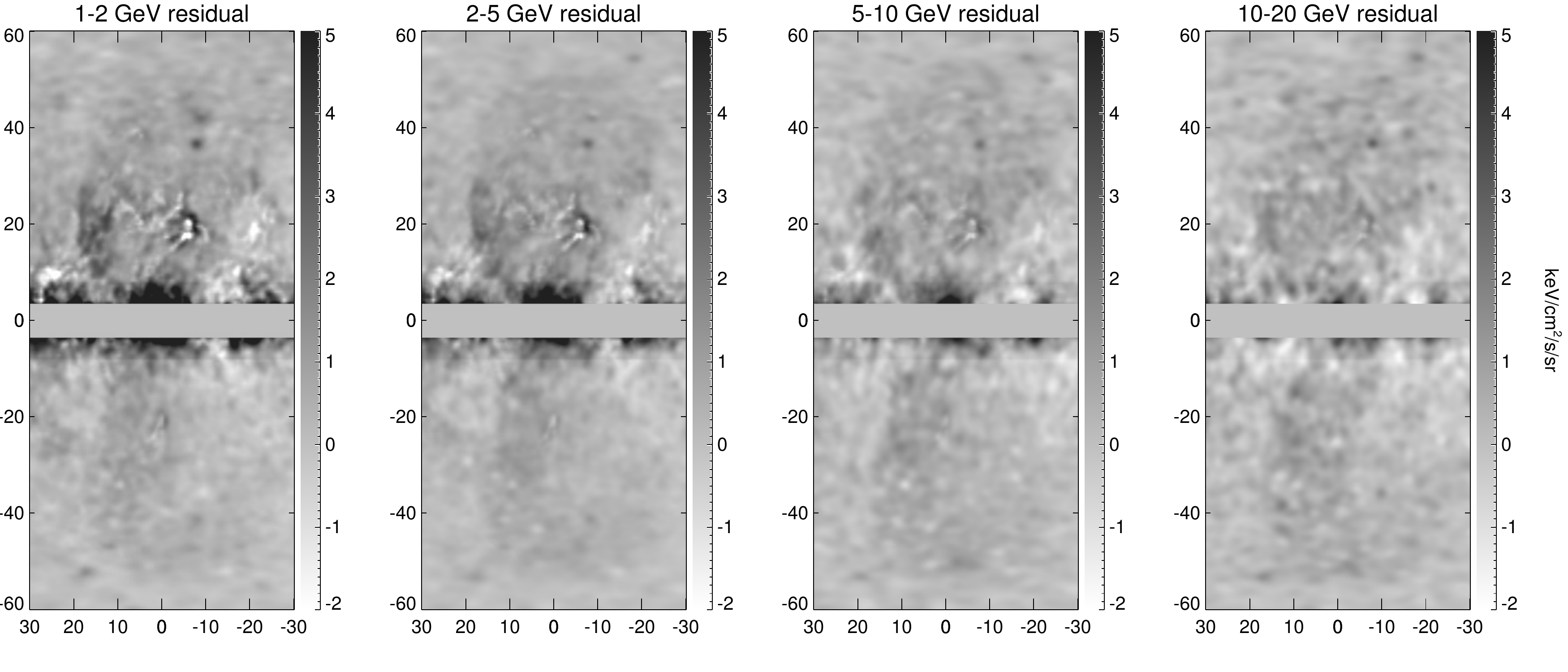}
\caption{\label{fig:residualmaps}
The residual emission after re-adding the latitude-sliced Bubbles templates with their best-fit coefficients, in $E^2 dN/dE$. Equivalently, these maps are obtained by subtracting the best-fit model for the background (in which we include all templates but the latitude-sliced Bubbles) from the data. The ``diffuse model'' fit is used, performed over regions greater than $5^\circ$ from the plane (although the mask shown in the figure is at $|b| = 3^\circ$).}
\end{figure*}

From this comparison of both the spectrum and morphology of these signals, we conclude that the gamma-ray component we have identified within the low-latitude regions of the \emph{Fermi} Bubbles is the more spatially extended continuation of the gamma-ray signal previously observed in the Galactic Center. Furthermore, we confirm that the dark matter models previously shown to be capable of accounting for the gamma-ray emission in the inner Galaxy are also capable of providing an explanation for the non-inverse Compton emission observed in the low-latitude spectrum of the \emph{Fermi} Bubbles. 

\section{Morphology of the Low-Latitude Signal}
\label{morphology}

One approach to extracting the morphology of the few-GeV spectral feature discussed in the previous sections is to examine the residual sky maps produced when the background templates (for example, the diffuse model template and isotropic background), multiplied by their best-fit coefficients, are subtracted from the data. Equivalently, these maps are produced by taking the residuals of the template fit, and re-adding the latitude-sliced Bubbles templates, weighted by their best-fit coefficients. This largely cancels out structure due to mismatches between the shape of the Bubbles templates and the actual excess, although such mismatches may still bias the fit. (This procedure has been recommended for extracting the spectrum of an excess in Ref.~\cite{:2012fb}.)

Fig.~\ref{fig:residualmaps} shows these ``residual'' maps (in average $E^2 dN/dE$) for the ``diffuse model'' fit, masked within 5 degrees of the Galactic plane, in four energy bands spanning the range from $1-20$ GeV. The high-latitude \emph{Fermi} Bubbles are visible in all bins with comparable brightness, as expected since the high-latitude spectrum is approximately flat in $E^2 dN/dE$. At low latitudes, there is a pronounced excess around the Galactic Center, not disk-like in shape, which is clearly visible in the $1-2$ and $2-5$ GeV maps, but rapidly fades at higher energies.

\begin{figure*}
\includegraphics[width=.7\textwidth]{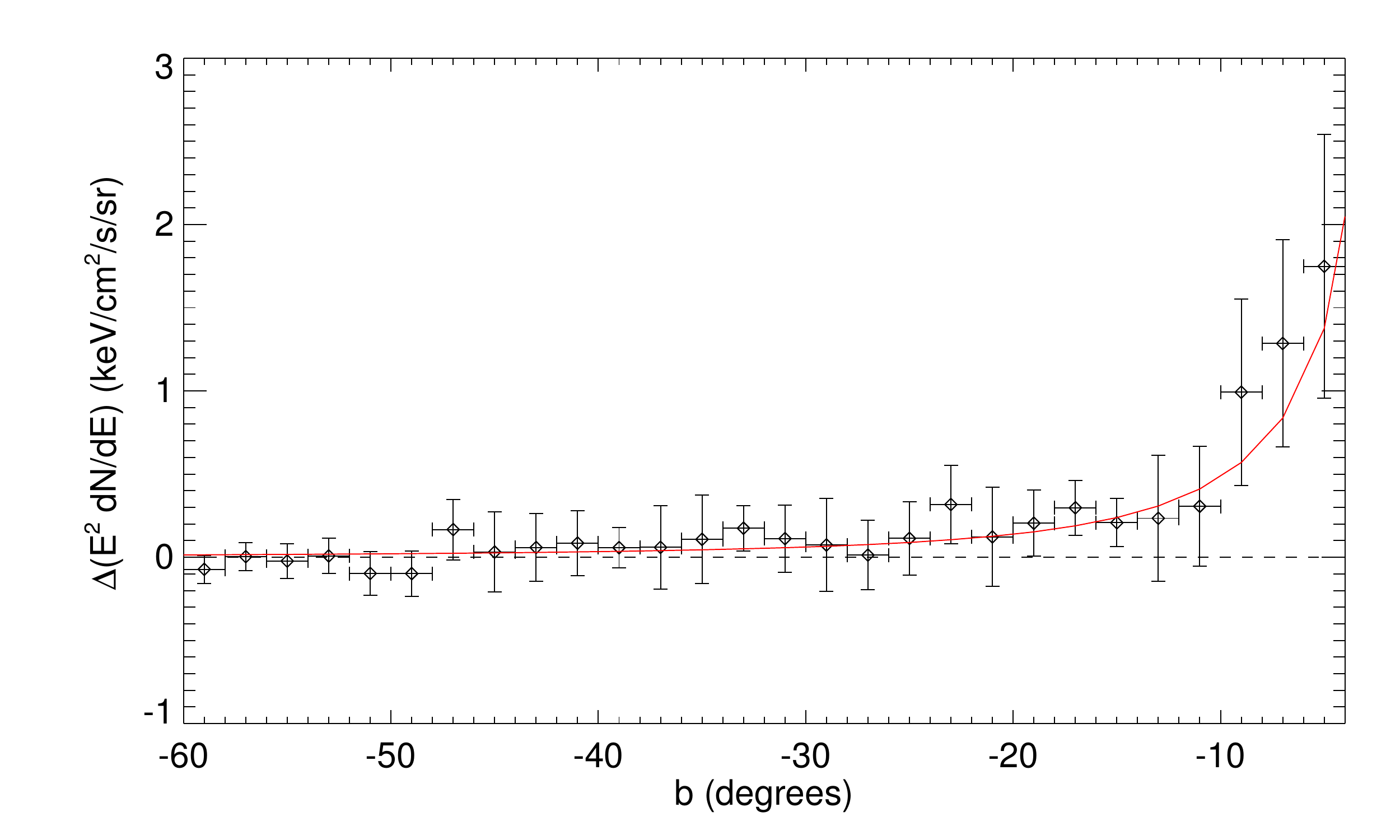}
\caption{\label{fig:diffvariance}
The difference of the residual emission maps (after re-adding the latitude-sliced Bubbles templates with their best-fit coefficients), between the 1-10 GeV bin and the 10-50 GeV bin, in $E^2 dN/dE$ averaged over $-5^{\circ} < l < 5^\circ$. The error bars describe the pixel-to-pixel scatter within each bin (standard deviation of pixel values). This analysis employs the ``diffuse model'' fit (see text), masked at $5^\circ$ from the plane. The red line shows the anticipated intensity resulting from a (squared, projected) NFW profile with inner slope of $\gamma=1.2$.}
\end{figure*}

\begin{figure*}
\includegraphics[width=\textwidth]{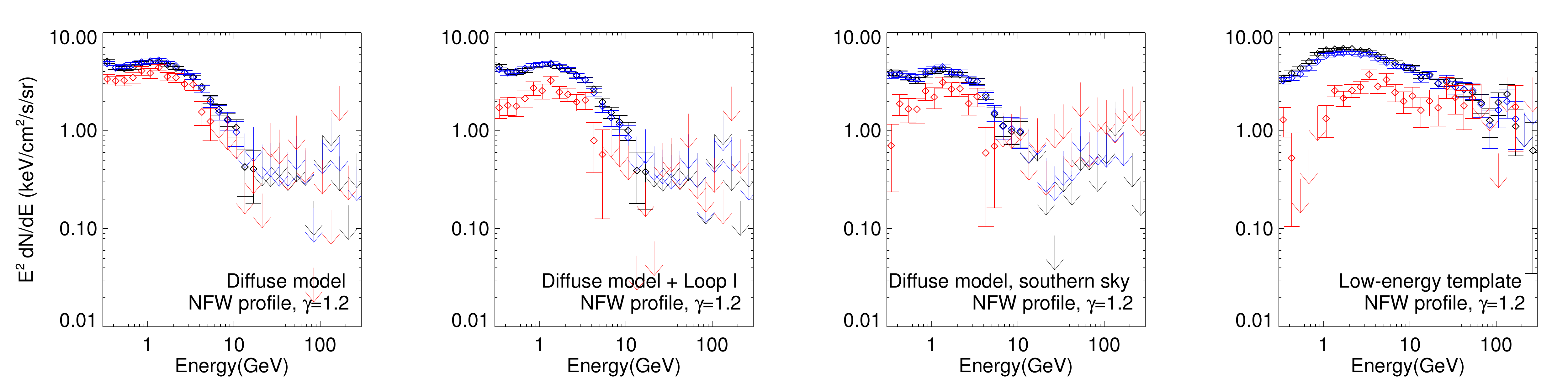}
\caption{\label{fig:nfwspec}
The spectrum of the emission fit by the NFW template ($\gamma=1.2$). The \emph{far left}, \emph{center left} and \emph{center right} panels employ the ``diffuse model'' fit, while the \emph{far right} panel uses the  ``low-energy template'' fit (see text for details). In all cases an additional (squared, projected) NFW template is added to the fit, and its extracted spectrum is plotted. The \emph{center left} panel also includes a template for Loop 1, and in the \emph{center right} panel the fit is performed using the southern hemisphere only, where there is less bright residual structure. The different colors show different choices for the latitude cut to remove the Galactic Disk: $|b| < 1^\circ$ (black), $|b| < 2^\circ$ (blue), $|b| < 5^\circ$ (red). Where the $1 \sigma$ error bars overlap with zero, we instead plot downward-pointing arrows corresponding to the $3 \sigma$ upper limits on the emission.}
\end{figure*}

\begin{figure*}
\includegraphics[width=0.75\textwidth]{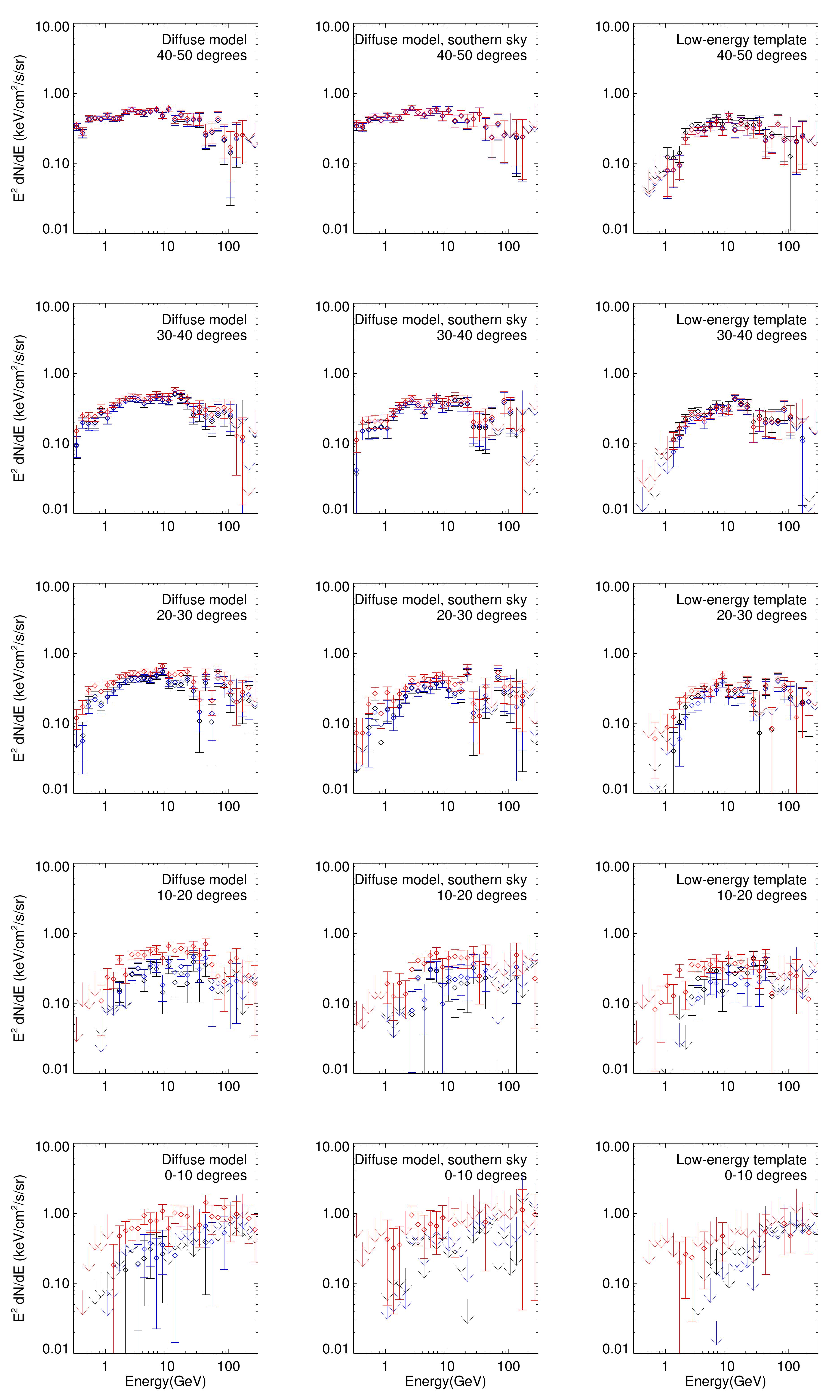}
\caption{\label{fig:specbylatitudewithnfw}
The spectrum extracted for the gamma-ray Bubbles in ten-degree latitude bands, now including an NFW template in the fit: in order from the top row, $40^\circ < |b| < 50^\circ$, $30^\circ < |b| < 40^\circ$, $20^\circ < |b| < 30^\circ$, $10^\circ < |b| < 20^\circ$, and $|b| < 10^\circ$. The \emph{left} and \emph{center} panels use the ``diffuse model'' template fit (see text); in the \emph{center} panels, the fit is restricted to $b < 0$ in addition to the masking. The \emph{right} panels use the ``low-energy template" fit (see text). In all cases an additional (squared, projected) NFW template is added to the fit, as in Fig. \ref{fig:nfwspec}. The different colors show different choices for the latitude cut to remove the Galactic Disk: $|b| < 1^\circ$ (black), $|b| < 2^\circ$ (blue), $|b| < 5^\circ$ (red). Where the $1 \sigma$ error bars overlap with zero, we instead plot downward-pointing arrows corresponding to the $3 \sigma$ upper limits on the emission.}
\end{figure*}

\begin{figure*}
\includegraphics[width=.45\textwidth]{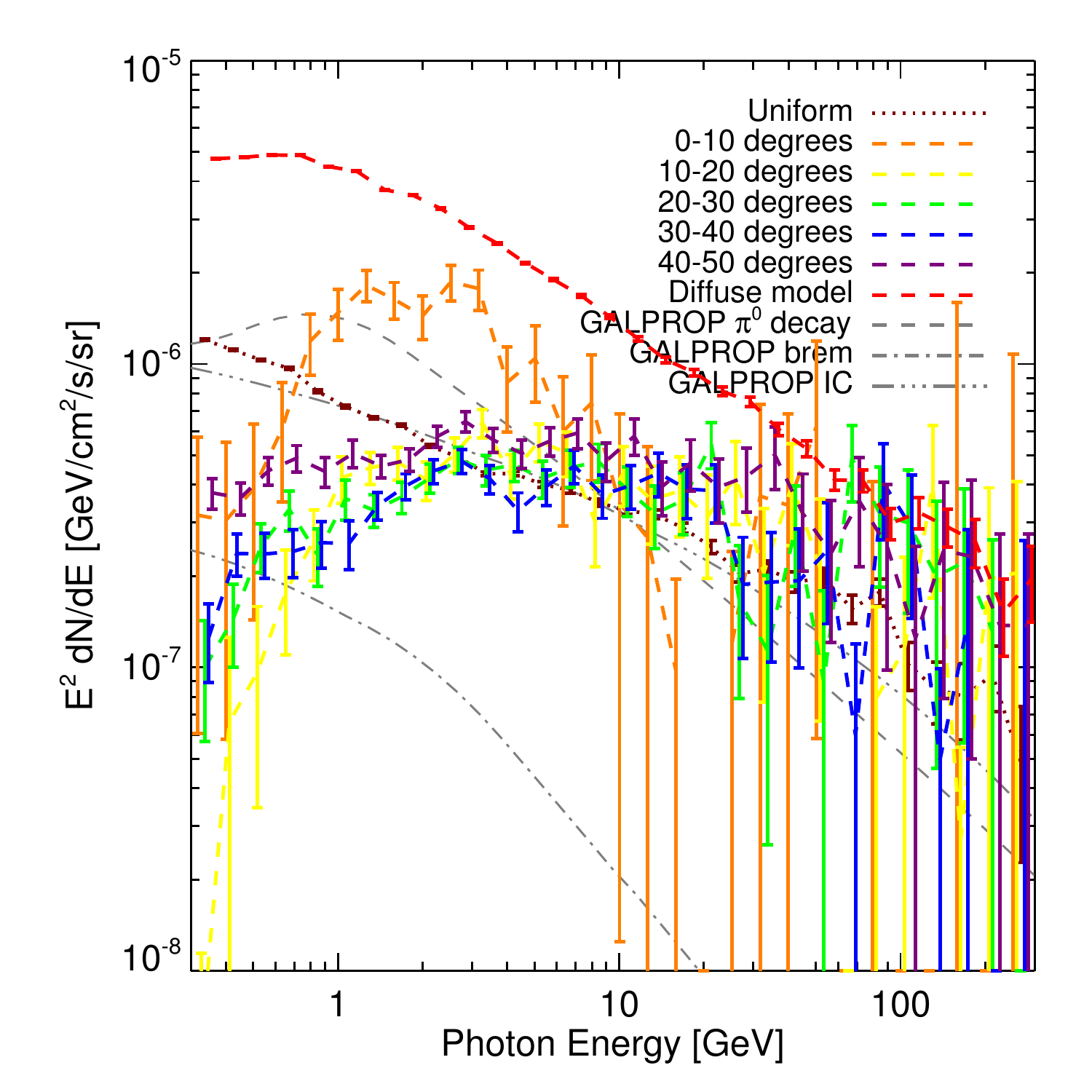}
\includegraphics[width=.45\textwidth]{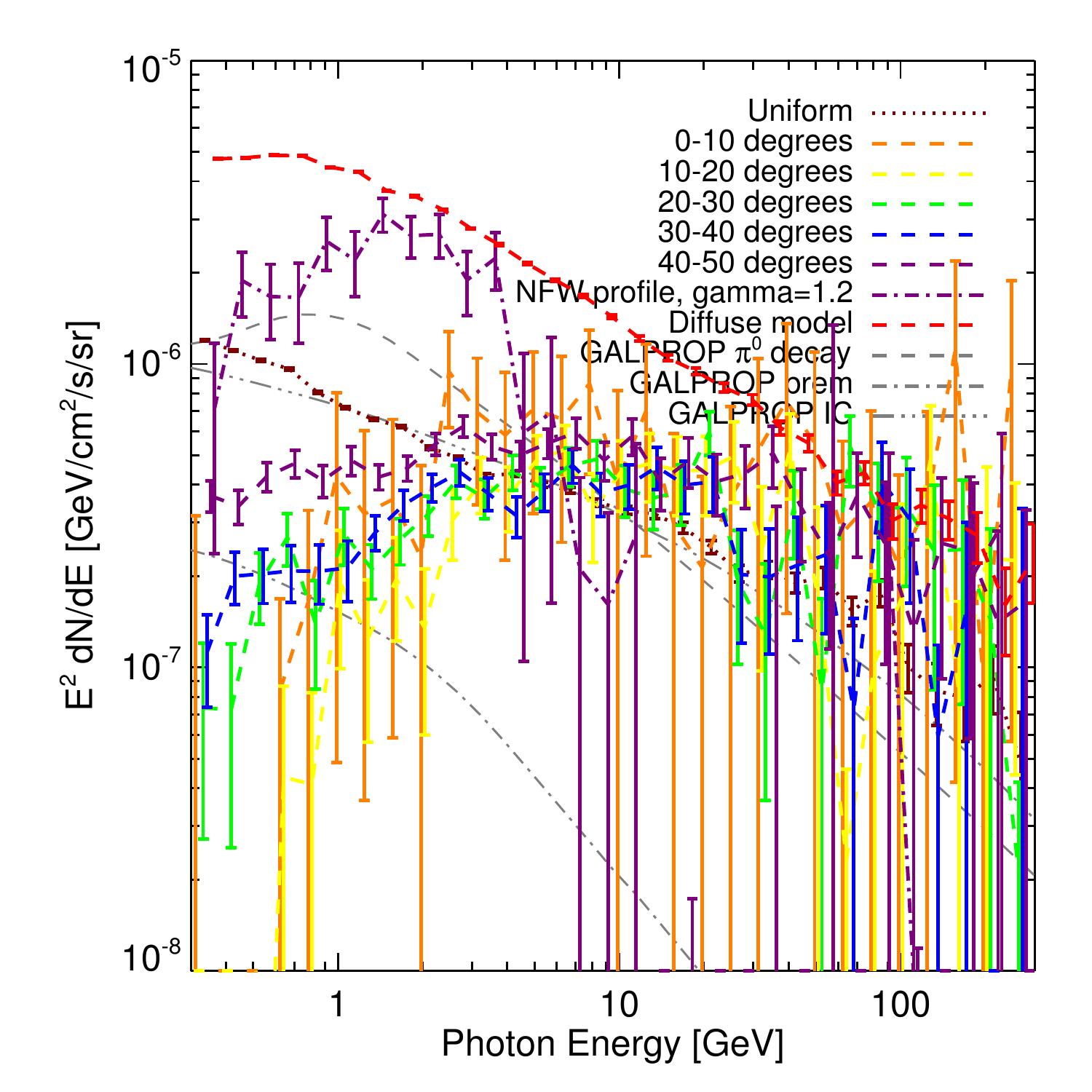}
\caption{\label{fig:normvssouthonly} 
The spectra of the various fit components, including five separate latitude-sliced templates for the Bubbles (see text), for the diffuse model foreground template, when the fit is restricted to the southern sky only. The Galactic Disk is masked for $|b| < 5^\circ$ in all cases. The \emph{left} and \emph{right} panels correspond to the case with (\emph{right}) and without (\emph{left}) an additional NFW ($\gamma=1.2$) template in the fit.}
\end{figure*}

\begin{figure*}
\includegraphics[width=.45\textwidth]{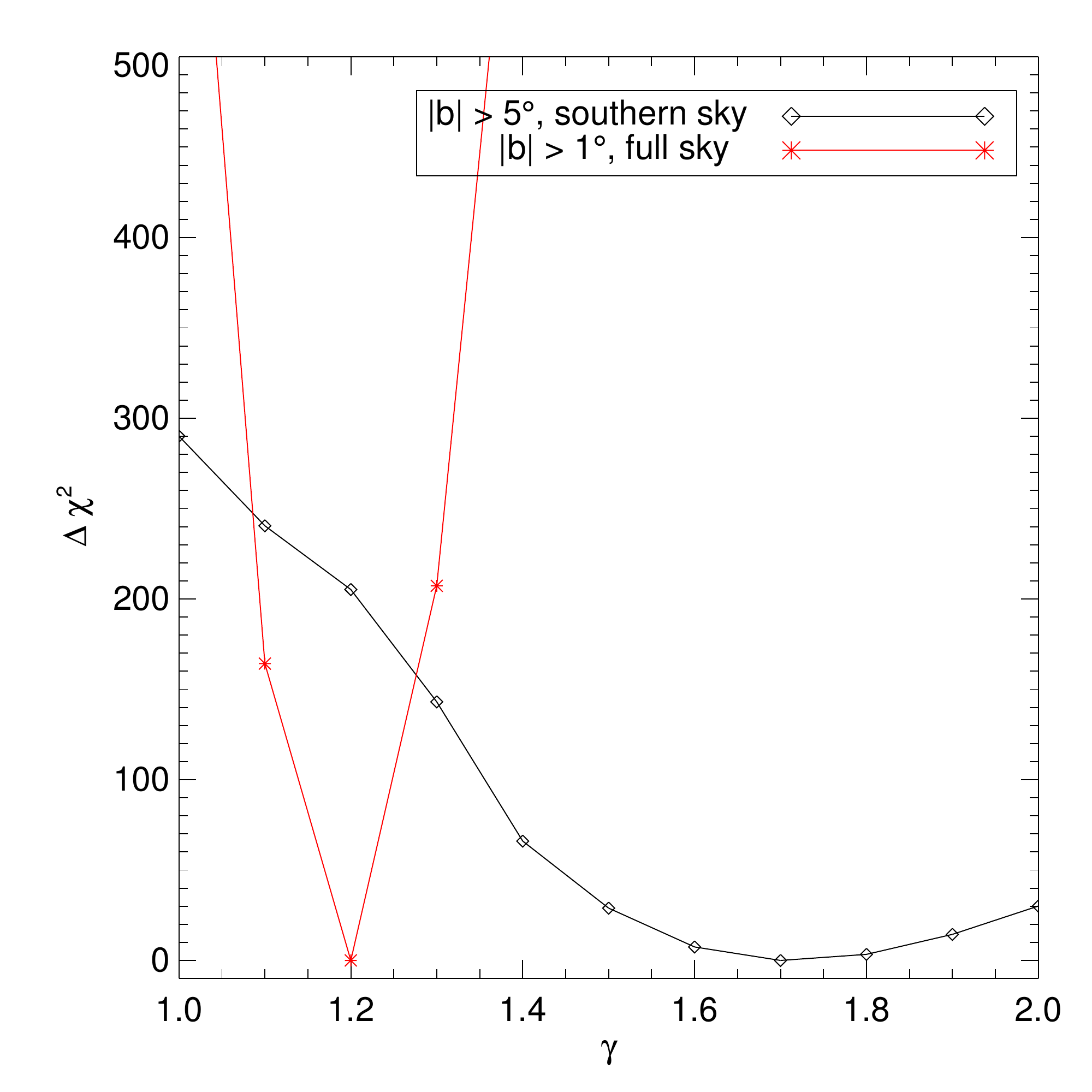}
\caption{\label{fig:chisquared}
The formal statistical $-2 \Delta \ln \mathcal{L}$ values (referred to as $\Delta \chi^2$) extracted from the likelihood fit using the diffuse model, the sliced Bubbles templates and a projected squared NFW template, for varying values of the NFW inner slope $\gamma$. The $\Delta \chi^2$ values are with respect to the tested $\gamma$ value with the highest likelihood (so the minimum is zero by definition). Red stars correspond to the full-sky fit masked for $|b| < 1^\circ$, black diamonds to the southern-sky fit masked for $|b| < 5^\circ$.}
\end{figure*}

Having generated these residual maps, we can study the emission binned in $b$, within a narrow longitude range ($|l| < 5$). As an illustrative exercise, we subtract the $10-50$ GeV (residual) map from the $1-10$ GeV (residual) map, where the excess is most pronounced (we ignore energies higher than 50 GeV because the \texttt{P6V11} \emph{Fermi} diffuse model was not designed for studies at those energies and, due to low photon statistics, little information would be gained from higher-energy photons in any case). For more details, see Appendix~\ref{app:residuals}. Since the maps are given in $E^2 dN/dE$, a zero result indicates an average spectrum of $dN/dE \propto E^{-2}$ between these two bins. The results for the southern hemisphere, where there are fewer bright sources and local features, are shown in Fig.~\ref{fig:diffvariance}. While we find that at high latitudes the spectrum is consistent with $dN/dE \propto E^{-2}$, the lower latitude emission ($|b|< 10^{\circ}-15^{\circ}$) reveals significant additional emission at low energies. The error bars shown in this figure, computed from the standard deviation of the pixel values in each $\Delta b = 2^\circ$ bin, are quite large and non-negligibly correlated, but provide a sense of the uncertainty in the rate at which the signal falls off away from the Galactic plane.

\section{Fitting the spectral bump with an NFW profile}
\label{nfw}

So far, we have focused our attention on the region of the sky occupied by the Bubbles. If annihilating dark matter is responsible for the signal observed at low latitudes, however, this signal should be distributed around the Galactic Center with approximate spherically symmetry, and will not be restricted to the Bubbles. In this section, we include an additional template in our analysis, corresponding to the signal expected from annihilating dark matter distributed according to a somewhat steepened NFW profile (with an inner slope of $\gamma=1.2$). 

In Figs.~\ref{fig:nfwspec} and~\ref{fig:specbylatitudewithnfw}, we show the spectra associated with the NFW template itself, and with the latitude-sliced Bubbles in the presence of the NFW template, respectively. The normalization of the NFW template corresponds to its brightness at $5^\circ$ from the Galactic Center.

With the inclusion of this additional NFW template, we find that the low-latitude (non-inverse Compton) emission prefers the morphology of the NFW template over being absorbed into the low-latitude portion of the Bubbles. The significant improvement in the fit is driven both by the spherical symmetry of the NFW profile, and the fact that the imposed Bubbles templates are flat in brightness rather than falling off with increasing Galactocentric radius (see also Appendix \ref{app:complement}). With the inclusion of this NFW template in the fit, the spectrum of the emission associated with the Bubbles is not only largely constant at high latitudes (as expected), but also has a similar (near-flat) spectral shape at low latitudes, at least where it can be detected (in several cases there is no significant detection of Bubbles-correlated emission in the lowest-latitude band, after the addition of the NFW template). The significant GeV-scale feature identified in Sec.~\ref{low} is no longer absorbed by these Bubbles templates, but instead is present in the spectrum of the spherically symmetric NFW template. The spectrum associated with the NFW template peaks at $\sim$1-2 GeV, and falls off above $\sim$10 GeV (there is little or no evidence for emission above 10 GeV in the spectrum of the NFW template, at least in the diffuse model fit). The total flux (integrated over solid angle) associated with this template is $\sim 1$ keV/cm$^2$/s, corresponding to a total luminosity within the solar circle of $\sim 10^{37}$ erg/s.

The regions of the sky that are most important for discriminating between the NFW template and the Bubbles templates are those near the inner Galaxy but outside the Bubbles themselves. Unfortunately, such regions are also near the Galactic plane. As a result, the extracted spectrum associated with the NFW template depends somewhat on the details of the fit, including the mask of the Galactic plane, especially below $\sim$1 GeV where the angular resolution of \emph{Fermi} is somewhat poor. In the simplest case, in which we employ the ``diffuse model'' fit and add the NFW template, the spectrum we extract below 1 GeV is considerably softer than that extracted from the Bubbles templates (see the \emph{left} frame of Fig.~\ref{fig:nfwspec}). Masking $|b| < 5^\circ$ hardens the low-energy spectrum, as does adding a separate template to absorb soft-spectrum emission from Loop I, or restricting the fit to the southern hemisphere where the structures associated with Loop I do not contaminate the signal (see \emph{left center} and \emph{right center} frames of the figure). Given these results, we find it plausible that the low-energy emission (below 1 GeV) associated with the NFW template can be attributed in large part to contamination from the disk and Loop I. 

In Fig.~\ref{fig:normvssouthonly}, we show how the spectral feature at a few GeV is reallocated from the lowest-latitude slices of the Bubbles to the NFW template, once the NFW template is added. In order to minimize background contamination, we examine the spectra masking $|b| < 5^\circ$ and restricted to the southern hemisphere. As discussed above, we find that the ``bump'' is much better fitted by the NFW template and there is no remaining significant evidence for emission correlated with the low-latitude Bubbles. Quantitatively, we find that the fit including the NFW template improves over the five-Bubbles-templates model at the level of approximately 12$\sigma$. Again, though, we reiterate that this is a formal significance derived including statistical errors \emph{only}, whereas there are significant systematics originating from the fact that the models we are using do not fully describe the data.

We have performed fits using generalized NFW profiles with a variety of inner slopes, $\gamma$. Fitting over both hemispheres and masking within 1 degree of the Galactic plane, we find that $\gamma \simeq 1.2$ provides the best fit (for this reason, we have adopted $\gamma=$1.2 as our default value). If we restrict our fit to the southern hemisphere and mask within five degrees of the Galactic plane, however, we find that a somewhat steeper distribution is preferred, $\gamma \sim 1.5-2$. This may reflect a slope which varies with distance from the GC, although systematic uncertainties prevent us from making any strong statement to that effect. We show the $\Delta \ln \mathcal{L}$ values for various choices of $\gamma$, in both these cases, in Fig. \ref{fig:chisquared}.

We note that it is not surprising that a spherically symmetric signal centered around the Galactic Center would first become apparent in an analysis of the region of the \emph{Fermi} Bubbles. By their shape, the Bubbles exclude both most of the background from the disk, and the arc structures associated with Loop I, making them reasonably well designed for extracting even a spherically symmetric excess (note the similarities between the Bubbles regions and the dark matter regions-of-interest as described in Refs.~\cite{Bringmann:2012vr,Weniger:2012tx,Profumo:2012tr}).

\section{Interpretation}
\label{interpretation}

\begin{figure*}[!]
\includegraphics[width=0.45\textwidth ]{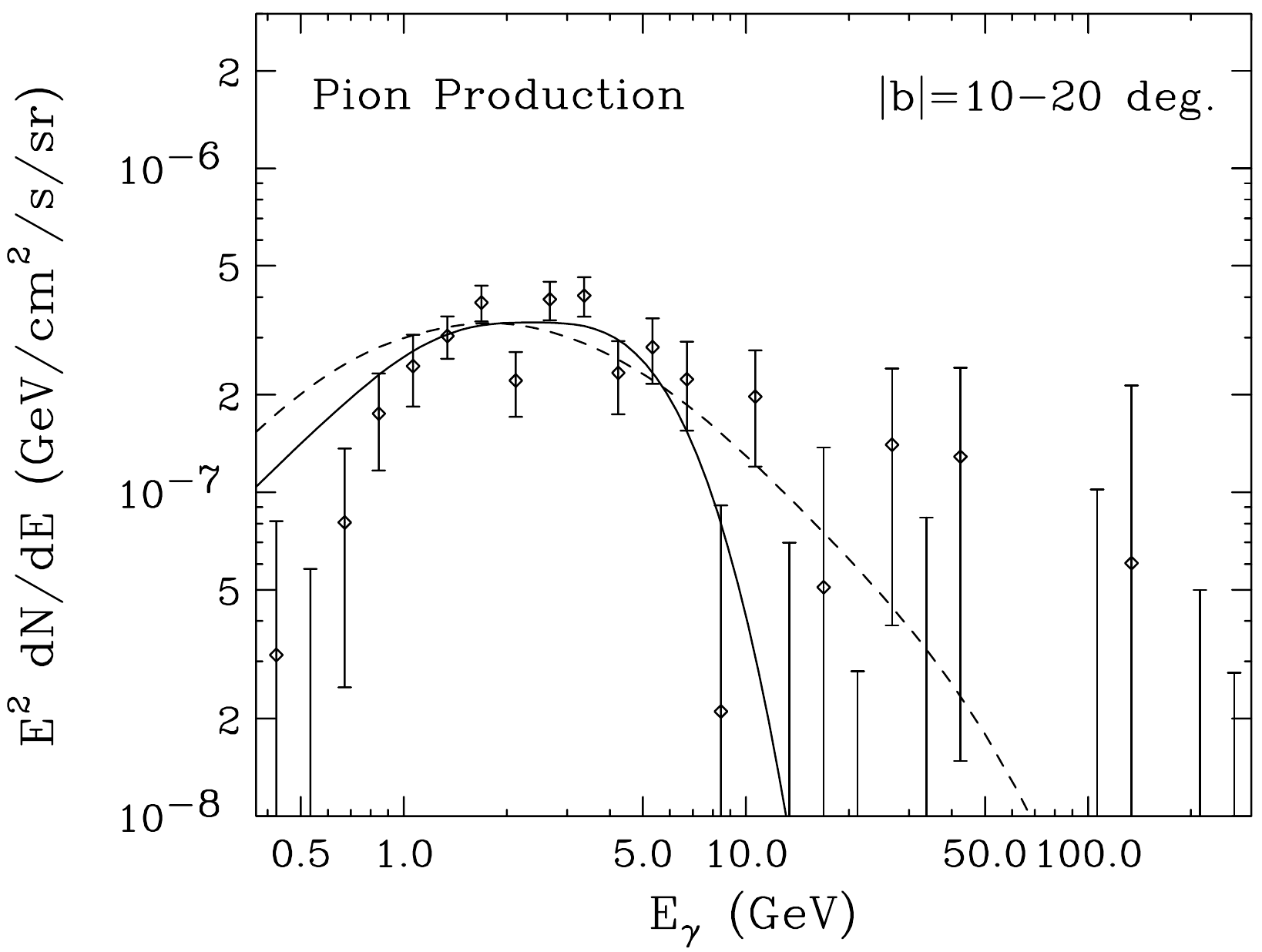}
\hspace{1.0cm}
\includegraphics[width=0.45\textwidth ]{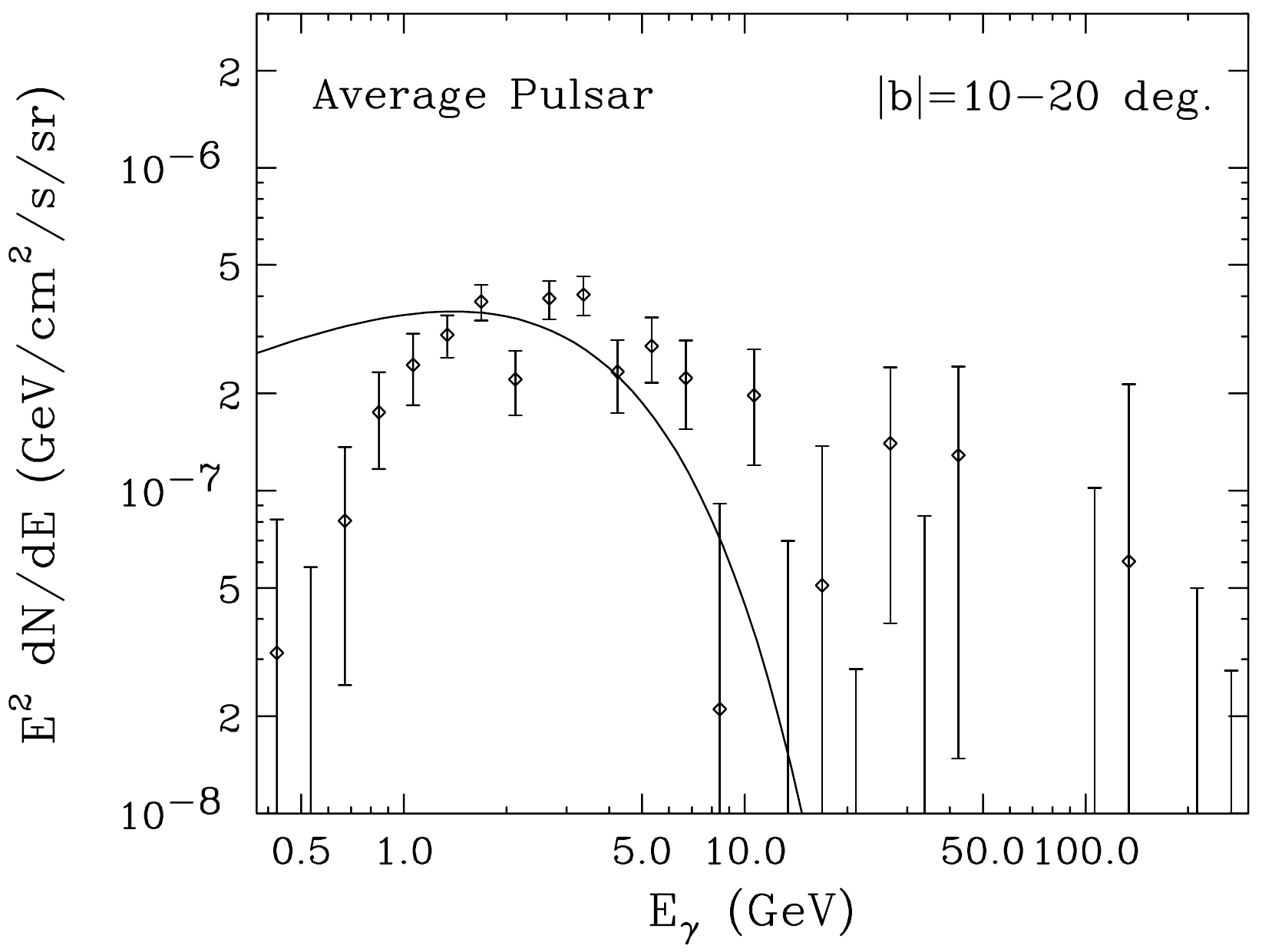}
\includegraphics[width=0.45\textwidth ]{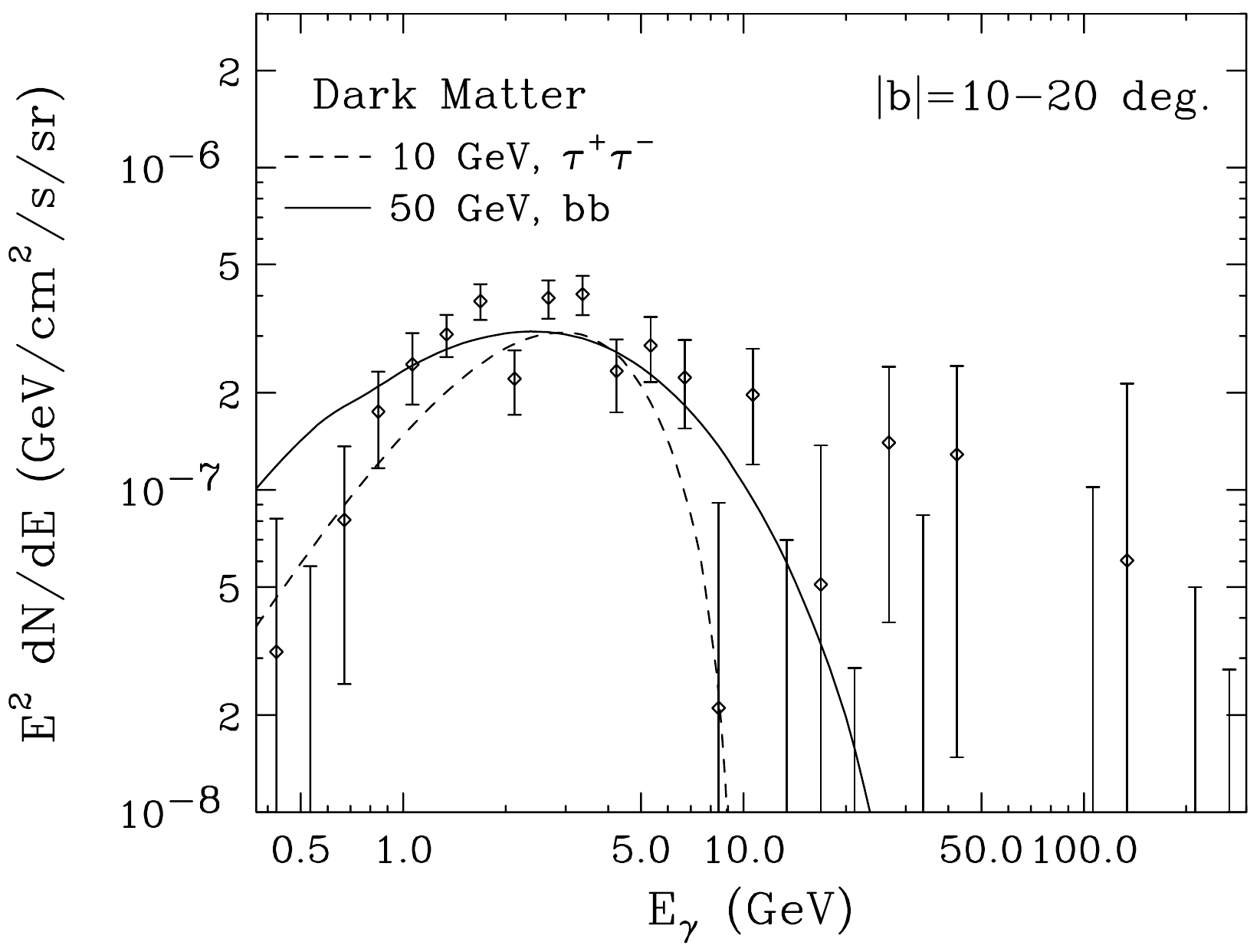}
\caption{Comparisons of the observed gamma-ray spectrum of the low-latitude emission, after subtracting the contribution from inverse Compton scattering (see Fig.~\ref{nonIC}) to that predicted from the scattering of cosmic ray protons with gas (\emph{upper left}), millisecond pulsars (\emph{upper right}), and dark matter annihilations (\emph{lower}). For proton-gas collisions, the solid and dashed lines denote cosmic ray proton spectra which take the form of a delta function at 25 GeV or a broken power-law following $E_p^{-2}$ below 25 GeV and $E_p^{-3}$ at higher energies, respectively. To accommodate the spectral shape of the observed gamma-ray emission, the cosmic ray proton spectrum throughout the inner several kiloparces of the \emph{Fermi} Bubbles must peak very strongly at approximately 25 GeV. The spectrum shown for pulsars is that corresponding to the average millisecond pulsar observed by the \emph{Fermi} collaboration~\cite{Abdo:2009,Malyshev:2010xc}. For annihilating dark matter, we show results for two models: 10 GeV particles annihilating to tau leptons (dashed) and 50 GeV particles annihilating to $b\bar{b}$. In each case, we have adopted a generalized NFW profile with an inner slope of $\gamma=1.2$, and normalized the signal to a local density of 0.4 GeV/cm$^3$ and an annihilation cross section of $\sigma v=2\times 10^{-27}$ cm$^3$/s ($\tau^+ \tau^-$) or $\sigma v=8\times 10^{-27}$ cm$^3$/s ($b\bar{b}$).}
\label{pion}
\end{figure*}

In the existing literature, a number of possibilities have been discussed for the origin of the extended gamma-ray emission observed from the region of the Galactic Center, including annihilating dark matter~\cite{Hooper:2010mq,Hooper:2011ti, Goodenough:2009gk,Abazajian:2012pn}, a population of millisecond pulsars~\cite{Hooper:2010mq,Hooper:2011ti,Abazajian:2010zy,Wharton:2011dv,Abazajian:2012pn}, and cosmic ray interactions with gas~\cite{Hooper:2010mq,Hooper:2011ti,Linden:2012iv,Linden:2012bp,Abazajian:2012pn}. In this section, we discuss these possibilities in light of the new information provided in this study, including the evidence that this emission is not confined to the inner Galaxy, but instead extends out to at least 2-3 kpc from the Galactic plane.

\subsection{Diffuse Emission Mechanisms}

As we have previously argued in Sec.~\ref{low}, no realistic spectrum of cosmic ray electrons can produce the observed spectral features of the GeV emission identified in this study. Cosmic ray protons scattering with gas also fail to provide a reasonable explanation for the observed gamma-rays. Firstly, the morphology of the observed emission is not highly correlated with the distribution of gas in the Milky Way, as would be predicted in such a scenario. Secondly, the spectral shape of the observed gamma-ray emission requires a peculiar spectrum of cosmic ray protons, sharply peaked at approximately 25 GeV. In the upper left frame of Fig.~\ref{pion}, we compare the observed spectrum of this emission to the spectral shape predicted from proton collisions with gas for a proton spectrum given by a delta function at 25 GeV (\emph{solid}), or by a broken power-law, following $E_p^{-2}$ below 25 GeV and $E_p^{-3}$ at higher energies (\emph{dashed}). While neither provides a particularly good fit, it is clear that from this comparison that a very sharply peaked spectrum of protons would be required to potentially account for this emission. And while we cannot absolutely rule out the existence of such a strongly peaked cosmic ray proton spectrum, this rather extreme requirement further disfavors cosmic ray origins for the observed low-latitude emission.

Note that in each frame of Fig.~\ref{pion}, we make comparisons to the observed gamma-ray spectrum (after subtracting the contribution from inverse Compton), as extracted using the diffuse model template and from the $|b|=10^{\circ}-20^{\circ}$ regions of the Bubbles. We have chosen to compare to this region over that extracted from lower latitudes, or from the NFW-template, as this region provides a spectrum that is more robust to contamination from the Galactic Disk. In particular, a comparison of the low-energy spectrum extracted from the $|b|<10^{\circ}$ Bubbles template or (especially) the NFW templates varies considerably depending on how the disk is masked, and on whether we consider both hemispheres, or only the southern sky. On the other hand, the spectral shape extracted at higher latitudes depends more strongly on the subtraction of the inverse Compton component, which is negligible compared to the bump for $|b| < 10^\circ$, but increasingly substantial as one moves farther from the Galactic plane. However, the consistency in the spectrum of the bump between the $10-20^\circ$ band and even higher latitudes supports the simple picture where the electron spectrum generating the ICS component (or the proton spectrum, in a hadronic scenario) is essentially invariant.

\subsection{Millisecond Pulsars}

Instead of relying on diffuse emission mechanisms, a large population of unresolved gamma-ray point sources could, in principle, be responsible for the observed emission. In particular, the spectrum of gamma-ray pulsars has been observed to peak at GeV energies, and it has been suggested that a collection of $\sim$$10^3$ such objects may be able to account for the extended gamma-ray emission observed from the Galactic Center~\cite{Abazajian:2010zy,Wharton:2011dv,Abazajian:2012pn}. Of particular interest are millisecond pulsars, which are thought to be formed as parts of low-mass X-ray binary systems. This could help to accommodate the very concentrated distribution of the gamma-ray emission observed around the Galactic Center (the stellar distribution in the inner Galaxy scales as $n\propto r^{-1.2}$, but objects formed through interactions between stars could plausibly be distributed as steeply as the square of this distribution)~\cite{Abazajian:2012pn,Voss:2006az}. The binary companions of such pulsars could also act as a tether, explaining why they do not free-stream out of the Galactic Center as a result of velocity kicks. Furthermore, millisecond pulsars are spun up through accretion, and can thus produce high luminosities of gamma-ray emission over much longer timescales than other types of pulsars.

Unfortunately, relatively little is known about the gamma-ray spectrum of millisecond pulsars. \emph{Fermi} has reported spectra from only eight millisecond pulsars, which together yield an average spectrum that is well fit by $dN_{\gamma}/dE_{\gamma} \propto E^{-1.5}_{\gamma} \exp(-E_{\gamma}/2.8 \, \rm{GeV})$~\cite{Abdo:2009,Malyshev:2010xc}. In the right frame of Fig.~\ref{pion}, we show this spectral shape (with arbitrary normalization) compared to the observed emission from the low-latitude Bubbles. This does not provide a good fit, especially at low energies (although this is also where the potential systematic errors are most significant). One could imagine, however, that the eight millisecond pulsars reported by \emph{Fermi} may not be representative, perhaps being biased toward the brightest or most locally common examples of such objects. To address this issue, it has been suggested that the gamma-ray spectra of globular clusters (which are thought to contain large numbers of millisecond pulsars) could provide a more reliable determination of the average spectrum from millisecond pulsars~\cite{Abazajian:2010zy}. At present, however, the error bars on the gamma-ray spectra of globular clusters are quite large, and (on average) do not appear to favor a much harder spectral index than is observed from individual resolved pulsars~\cite{collaboration:2010bb} (see also Fig.~9 of Ref.~\cite{Hooper:2011ti}). Future data from \emph{Fermi} could potentially be useful in further testing the possibility that millisecond pulsars produce a gamma-ray spectrum compatible with the low-latitude, GeV emission under consideration here. 

In addition to their gamma-ray spectra, the distribution of millisecond pulsars within the Milky Way is not well constrained empirically. That being said, one expects the formation of millisecond pulsars to roughly follow that of stars, possibly with an additional preference for regions of very high stellar density, such as in globular clusters. To account for the observed morphology of this signal, however, there would have to be a significant number of millisecond pulsars in the halo, at distances of at least a few kpc outside the Galactic plane. If this emission is generated by such objects, they would require a distribution that is very different from that observed among other stellar populations. And while such distributions have been proposed~\cite{FaucherGiguere:2009df}, such distributions are constrained by the small number of millisecond pulsars resolved by \emph{Fermi} and by the high degree of isotropy observed in the gamma-ray background~\cite{Ackermann:2012uf}.

\subsection{Annihilating Dark Matter}

In the previous two sub-sections, we have described some of the possibilities and challenges involved in explaining this gamma-ray signal with astrophysical sources or mechanisms. Annihilating dark matter can provide a simple explanation for the sharply peaked spectrum and distinctive morphology of this emission. In the lower frame of Fig.~\ref{pion}, we compare the gamma-ray spectrum predicted in two dark matter scenarios to the observed spectrum. First, we consider a 10 GeV dark matter particle species annihilating to tau leptons. We also show the spectrum resulting from a 50 GeV particle annihilating to $b\bar{b}$ which, given the systematic uncertainties in the extraction of the spectrum, we also consider to be a viable possibility. For a generalized NFW profile ($\gamma=1.2$) normalized to a local density of 0.4 GeV/cm$^3$, we require an annihilation cross section of $\sigma v=2\times 10^{-27}$ cm$^3$/s in the 10 GeV $\tau^+ \tau^-$ case, or $\sigma v=8\times 10^{-27}$ cm$^3$/s in the case of a 50 GeV particle annihilating to $b\bar{b}$. As annihilations to electrons, muons, or neutrinos do not significantly contribute to the gamma-ray spectrum, the total annihilation cross section in the leptonic case may be larger by a factor of a few or several.

We note that the dark matter distribution required to fit the observed signal is well supported by the results of numerical simulations. As has been known for some time, N-body simulations which model the evolution of cold dark matter without baryons typically find halos of the NFW form, with inner slopes of $\rho \propto r^{-1}$ ($\gamma=1$)~\cite{Navarro:1995iw,Navarro:1996gj}. Hydrodynamical simulations, which include the effects of baryonic processes involved in galaxy formation and evolution, in many cases predict the steepening of this inner slope, typically from $\gamma=1$ to $\gamma=1.2$ to 1.5 (see Ref.~\cite{Gnedin:2011uj} and references therein). Supporting this picture further are observations of the Milky Way's rotation curve and microlensing optical depth, which together find that the dark matter distribution is best fit by $\gamma=1.3$ (although with large uncertainties)~\cite{Iocco:2011jz}.\footnote{Note, however, that the inclusion of baryonic feedback processes may flatten cusps in dark matter halos (e.g. \cite{Zolotov:2012xd}), and there is observational evidence for profiles shallower than NFW in some dwarf galaxies~\cite{Walker:2011zu}.}

The annihilation cross section required to normalize the signal in question is also an attractive value from a theoretical perspective. To be produced thermally in the early universe in an abundance equal to the measured dark matter density, a ($\sim$10-50 GeV) dark matter particle must have an annihilation cross section of $\langle \sigma v \rangle \approx 2 \times 10^{-26}$ cm$^3$/s, when thermally averaged over the process of freeze-out~\cite{Steigman:2012nb}. In many typical dark matter models, however, the annihilation cross section in the low velocity limit (as is relevant for annihilation in the halo) is smaller than this value by a factor on the order of a few. Neutralinos, for example, when annihilating to leptons through the $t$-channel exchange of sleptons typically exhibit $\langle \sigma v\rangle_{\rm FO}/\sigma v_{v=0} \sim 5$. While neutralino annihilations through $p$-wave amplitudes (which are suppressed at low velocities) significantly contribute to the relic abundance calculation, they do not factor into the current annihilation rate. We also note that annihilation cross sections in the range required here ($\sigma v \sim (2-8)\times 10^{-27}$ cm$^3$/s) are currently consistent with all observational and experimental constraints, including those derived from dwarf galaxies~\cite{GeringerSameth:2011iw,Ackermann:2011wa}, the cosmic microwave background~\cite{Finkbeiner:2011dx,Slatyer:2009yq,Hutsi:2011vx}, the spectrum of cosmic-ray antiprotons~\cite{Kappl:2011jw,Cerdeno:2011tf}, and from searches for monophoton-plus-missing energy events at LEP~\cite{Fox:2011fx}. 

From a dark matter model building perspective, there are a number of possibilities one could consider. In the case of light dark matter particles which annihilate primarily to leptons, the annihilations could proceed through the $t$-channel exchange of new charged particles which carry lepton number (such as sleptons), or thorough the exchange of a leptophillic $Z'$~\cite{Buckley:2010ve} or a leptophillic higgs~\cite{Logan:2010nw}. Alternatively, one could imagine a scenario in which the dark matter itself carries lepton number. The dark matter could also annihilate to a pair of light force carriers which interact with the Standard Model only through kinetic mixing the with photon. The decays of the light force carrier then yield combinations of mesons and charged leptons, leading to a gamma-ray spectrum that is very similar to that predicted from tau decays~\cite{Hooper:2012cw}.

One additional consideration is that if the $\sim$10 GeV dark matter particles also annihilate to $e^+ e^-$ at a rate similar to $\tau^+ \tau^-$, then they would also be capable of explaining the anomalous synchrotron emission observed from the Milky Way's radio filaments~\cite{Linden:2011au}, and possibly a significant fraction of the isotropic radio background~\cite{Hooper:2012jc,Fornengo:2011cn}. Such annihilations could also lead to a feature in the cosmic ray positron spectrum, potentially identifiable by AMS~\cite{Hooper:2012gq}.

\section{Discussion and Conclusions}
\label{conclusions}

In this study, we have identified a new component of gamma-ray emission from the low-latitude regions of the \emph{Fermi} Bubbles. This emission does not appear to be compatible with originating from either inverse Compton scattering or proton collisions with gas (for any physically realistic spectrum of cosmic rays). The spectrum of this emission peaks at energies of a few GeV and is distributed with an approximately spherically symmetric morphology about the Galactic Center, with a luminosity per volume that is consistent with a squared (generalized) NFW profile with an inner slope of $\rho \propto r^{-1.2}$ (although the best-fit slope depends somewhat on the region chosen for the fit). The broad features of this signal are robust to the various details of our analysis, such as the choice of templates and the degree of masking around the disk. 

The diffuse gamma-ray signal we have identified here is consistent both in spectrum and morphology with being the more spatially extended counterpart of the emission previously observed from the Galactic Center~\cite{Hooper:2010mq,Hooper:2011ti, Goodenough:2009gk, Boyarsky:2010dr, Abazajian:2012pn}. In addition to further confirming the existence of this gamma-ray source, the results presented here are important in that they confirm that the morphology of this signal extends to at least $\sim2-3$ kpc from the Galactic Center (whereas previous studies of the Galactic Center were sensitive only to the inner few hundreds of parsecs). This provides us with valuable information with which to test various interpretations of the observed emission. Millisecond pulsars, in particular, have been proposed as a source of the emission observed from the Galactic Center. In such a scenario, the significant high-latitude component of the signal reported in this paper would seem to require a large high-latitude population of faint unresolved millisecond pulsars that does not trace known stellar populations. In contrast, the morphology of the signal \emph{is} consistent with that predicted a priori from annihilating dark matter.

If annihilating dark matter is responsible for the emission reported here, we find that a halo profile of the generalized NFW form, with an inner slope of $\rho \propto r^{-1.2}$, is preferred. The spectral shape of the emission is well described by an approximately 10 GeV particle annihilating to tau leptons, with a normalization corresponding to an annihilation cross section of $\sigma v \sim 2 \times 10^{-27}$ cm$^3$/s (in addition to annihilations proceeding to electrons, muons and/or neutrinos, each of which provide comparatively few gamma-rays). We note that dark matter in this mass range may also be capable of accommodating a number of other anomalous signals reported from direct and indirect detection experiments (for a summary, see Ref.~\cite{Hooper:2012ft}). Alternatively, the annihilations of a $\sim$40-50 GeV dark matter particle to quarks can also provide a reasonable fit, requiring a normalization of $\sigma v \sim 8 \times 10^{-27}$ cm$^3$/s.


\bigskip

\acknowledgments{We would like to thank Kev Abazajian, Roland Crocker, Doug Finkbeiner, Manoj Kaplinghat, Tim Linden, Simona Murgia, Annika Peter and Neal Weiner for helpful discussions. We thank Ettore Carretti, Roland Crocker and Greg Dobler for providing data from their own work. This work has been supported by the US Department of Energy. TRS is supported by the National Science Foundation under grants PHY-0907744 and AST-0807444. We would like to thank the Aspen Center for Physics, where this project originated, for their hospitality.}

\bibliography{hazebubbles}

\begin{figure*}[!]
\includegraphics[width=\textwidth]{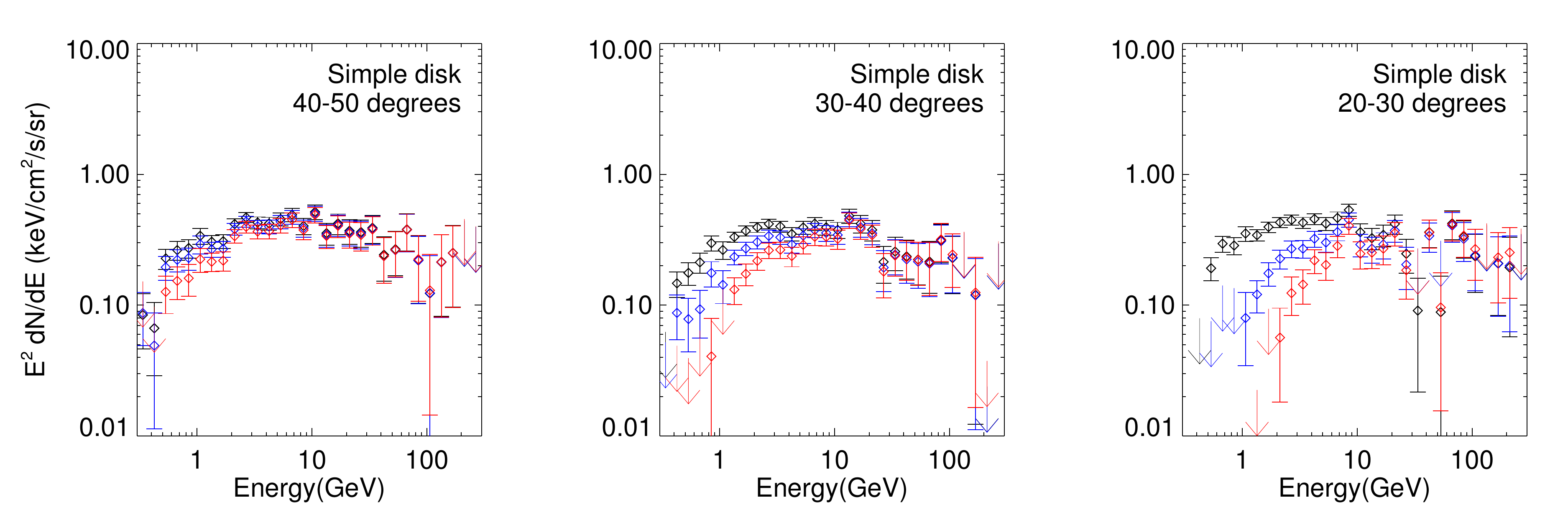} \\
\includegraphics[width=0.66\textwidth]{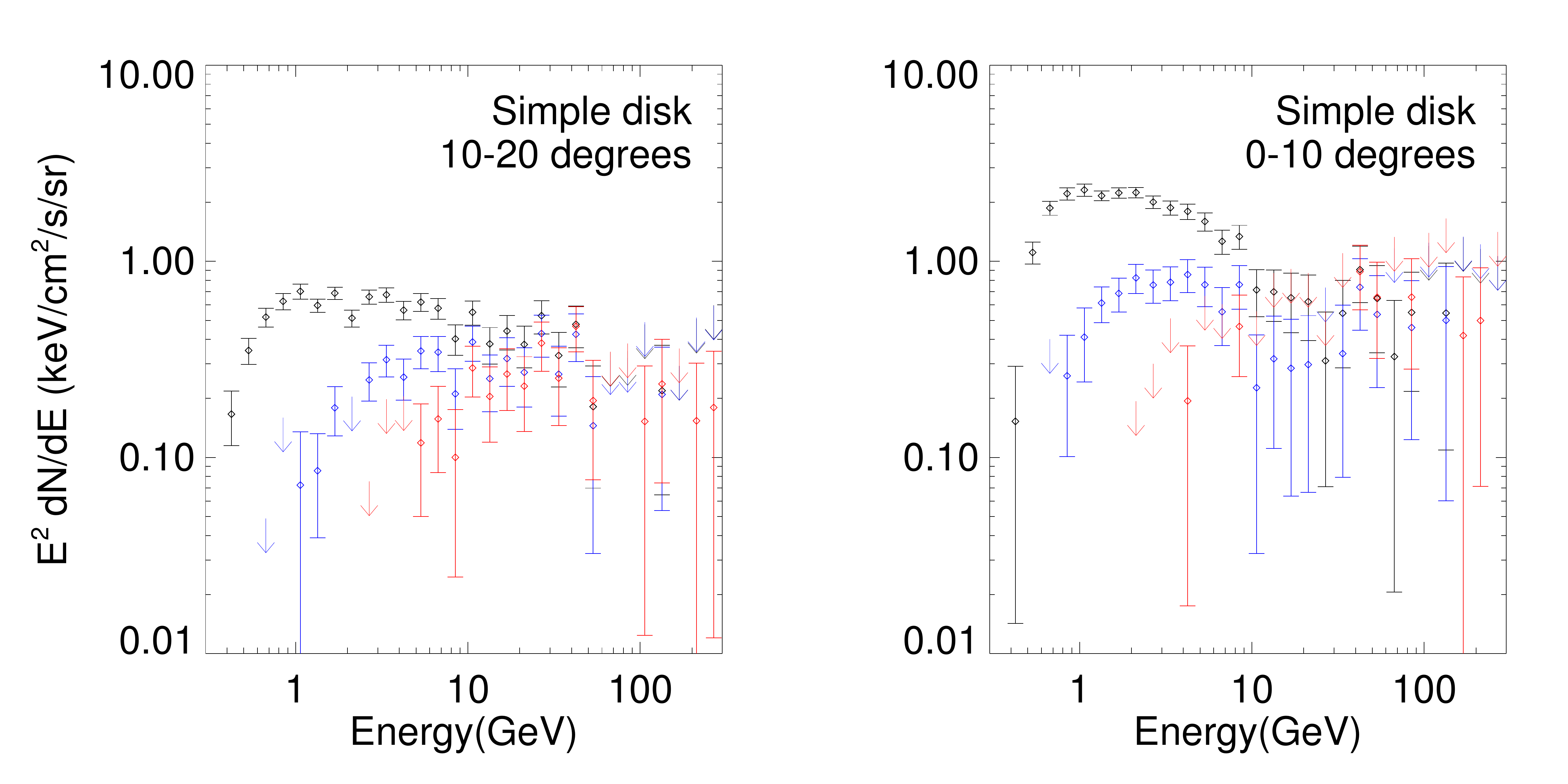}
\includegraphics[width=0.33\textwidth]{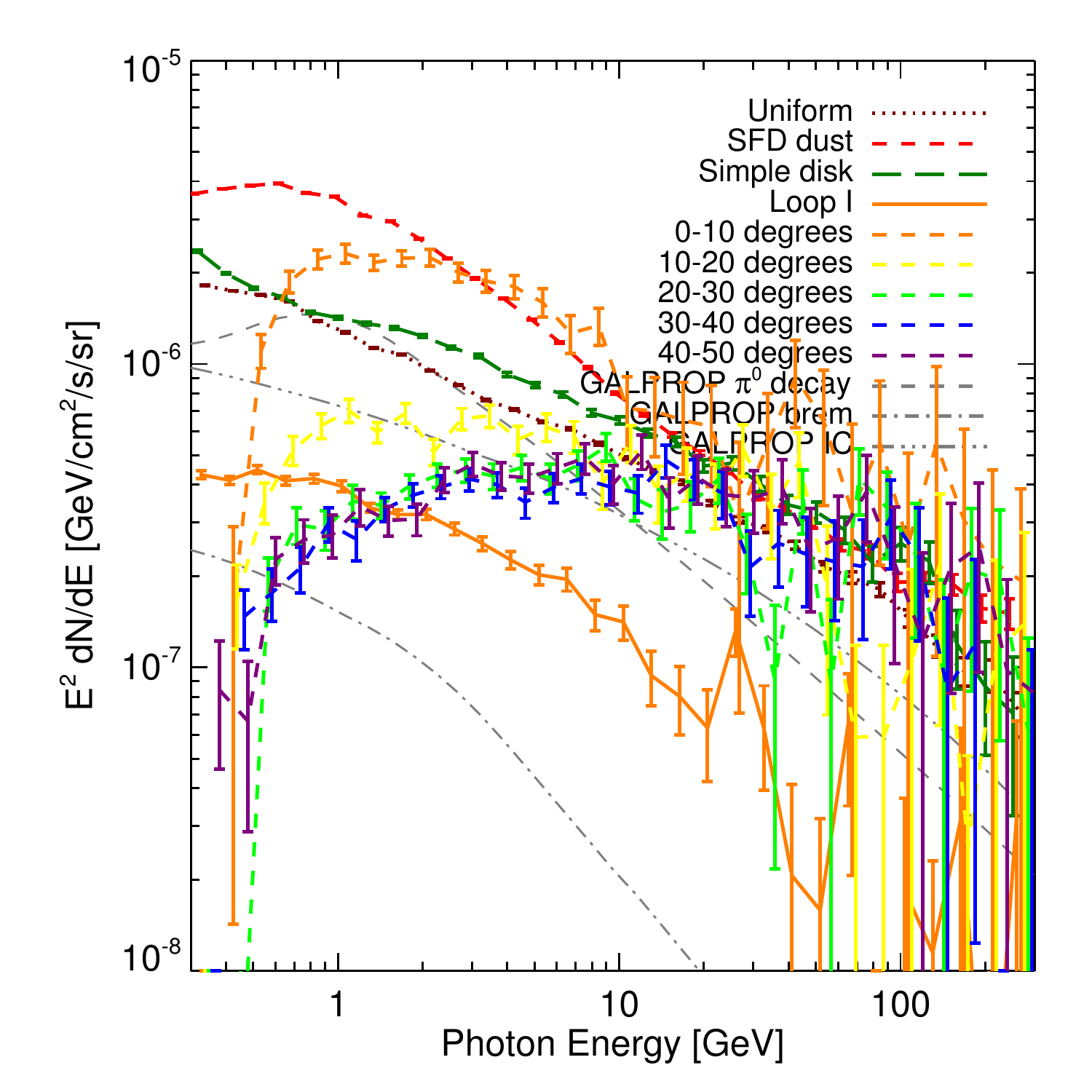}
\caption{\label{fig:simpledisk}
The spectrum extracted for the gamma-ray Bubbles in ten-degree latitude bands: in order from the top left, $40^\circ < |b| < 50^\circ$, $30^\circ < |b| < 40^\circ$, $20^\circ < |b| < 30^\circ$, $10^\circ < |b| < 20^\circ$, $|b| < 10^\circ$. The fit is performed using the simple disk model as a template for ICS, the SFD dust map, a uniform map, a template for Loop I, and the five latitudinal slices through the Bubbles templates. The different colors show different choices for the latitude cut to remove the Galactic Disk: $|b| < 1^\circ$ (black), $|b| < 2^\circ$ (blue), $|b| < 5^\circ$ (red). Where the $1 \sigma$ error bars overlap with zero, we instead plot downward-pointing arrows corresponding to the $3 \sigma$ upper limits on the emission. The \emph{bottom right} panel shows the spectrum of the various templates in the fit for the $|b| < 1^\circ$ mask.}
\end{figure*}

\begin{appendix}

\section{The Simple Disk model}
\label{app:simpledisk}

The primary template subtraction technique initially used in Ref.~\cite{Su:2010qj} to reveal the Bubbles employed a simple geometric template to model ICS emission from the disk. The functional form of this template is $\csc|b| - 1$ in latitude and Gaussian with $\sigma_l = 30^\circ$ in longitude. Such a model, while oversimplified relative to the actual emission associated with the Galactic Disk, is smooth and cannot mimic sharp features, and is demonstrably effective in subtracting ICS emission at high latitudes.

In the current work, we attempted to fit the data using this template to model ICS, along with the SFD dust map to trace emission from pion decay, the isotropic background model, a flat template for Loop 1, and the latitude-sliced Bubbles templates. This is identical to the ``low-energy template'' fit described in the main text except that the dust-subtracted $0.5-1$ GeV map is replaced with the simple disk. The results of this approach are shown in Fig.~\ref{fig:simpledisk}. This model turns out to be poorly suited to fits at low latitudes, with the fit being highly unstable to the masking of the disk; we suspect that the steep rise of the template at small $|b|$ causes the fit to be largely driven by the lowest-latitude data, where the data are known to be contaminated by unsubtracted point sources.

\section{Consistency of results using the diffuse model and low energy template}
\label{app:dmle}

At first appearance, the gamma-ray spectra as extracted using the diffuse model and low-energy template methods are quite different (see the left and right frames of Fig.~\ref{fig:specbylatitude}). This comparison can be misleading, however, as the spectra extracted using the low-energy template do \emph{not} describe the emission present in the Bubbles region after subtraction of some physical model -- rather, they describe the degree to which the Bubbles-correlated emission is harder than the ordinary ICS emission associated with the disk (which dominates the 0.5-1 GeV skymap after removal of the dust-correlated emission), with a pivot point at 0.5-1 GeV. In contrast, the diffuse model template fit extracts the excess over the diffuse model spatially correlated with the \emph{Fermi} Bubbles. Of course, the diffuse model is fitted to the data, and so may also soak up emission physically associated with the Bubbles.

In this Appendix, we provide a direct comparison of these techniques. Comparing the gamma-ray spectra extracted using the two methods, we see that there are discrepancies at both low and high energy (with the latter being particularly pronounced at low latitudes). These discrepancies have different origins; we will discuss them in turn.

At high energies and low latitudes, the low-energy template fit returns a roughly flat (in $E^2 dN/dE$) Bubbles-correlated spectrum, similar to the spectrum obtained at high latitudes. In contrast, the diffuse model fit detects virtually no emission. This indicates that the diffuse model is ``soaking up'' this hard-spectrum emission in some way. The diffuse model has, of course, been adjusted to fit the data, and has the potential to over-subtract Bubbles-correlated emission in trying to fit the data without the freedom to include the Bubbles explicitly. Given the information to hand, it is not possible to distinguish between this scenario and one where the \emph{Fermi} diffuse model is adequately capturing the physics of this spectral component, which should be assigned to the ordinary Galactic emission even though it is harder than the norm. The amplitude of this hard spectral component, at low latitudes, is not of great interest for the purposes of this study; to facilitate checking agreement of the few-GeV spectral feature between the two methods, we subtract a component with $dN_\gamma/dE_\gamma \propto E_\gamma^{-2}$ from the results for the low-energy template fit, designed to remove this high-energy emission.

The low-energy discrepancy is simpler to deal with, as it is an entirely natural and expected consequence of the different methodologies being employed. Since the low-energy template already includes some contribution from the Bubbles, to obtain the true Bubbles spectrum we should re-add a component with the same spectrum as the low-energy template ($dN_{\gamma}/dE_{\gamma} \propto E_{\gamma}^{-2.65}$, to a good approximation). The normalization of this component is a priori unknown, but to check consistency between the two methods, we can normalize it to match the diffuse model result at 0.5-1 GeV.

\begin{figure*}[!]
\includegraphics[width=0.45\textwidth ]{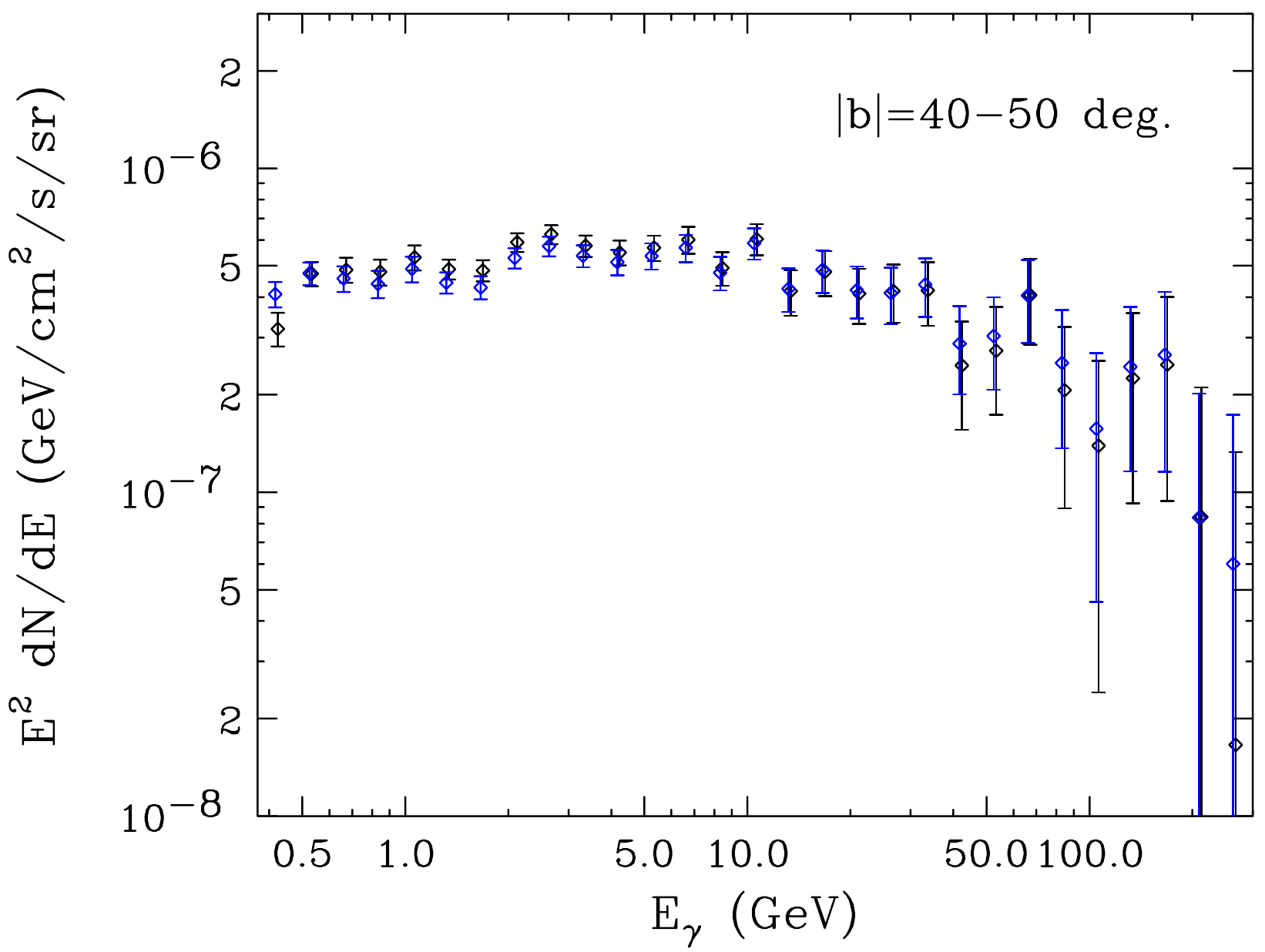}
\hspace{0.4cm}
\includegraphics[width=0.45\textwidth ]{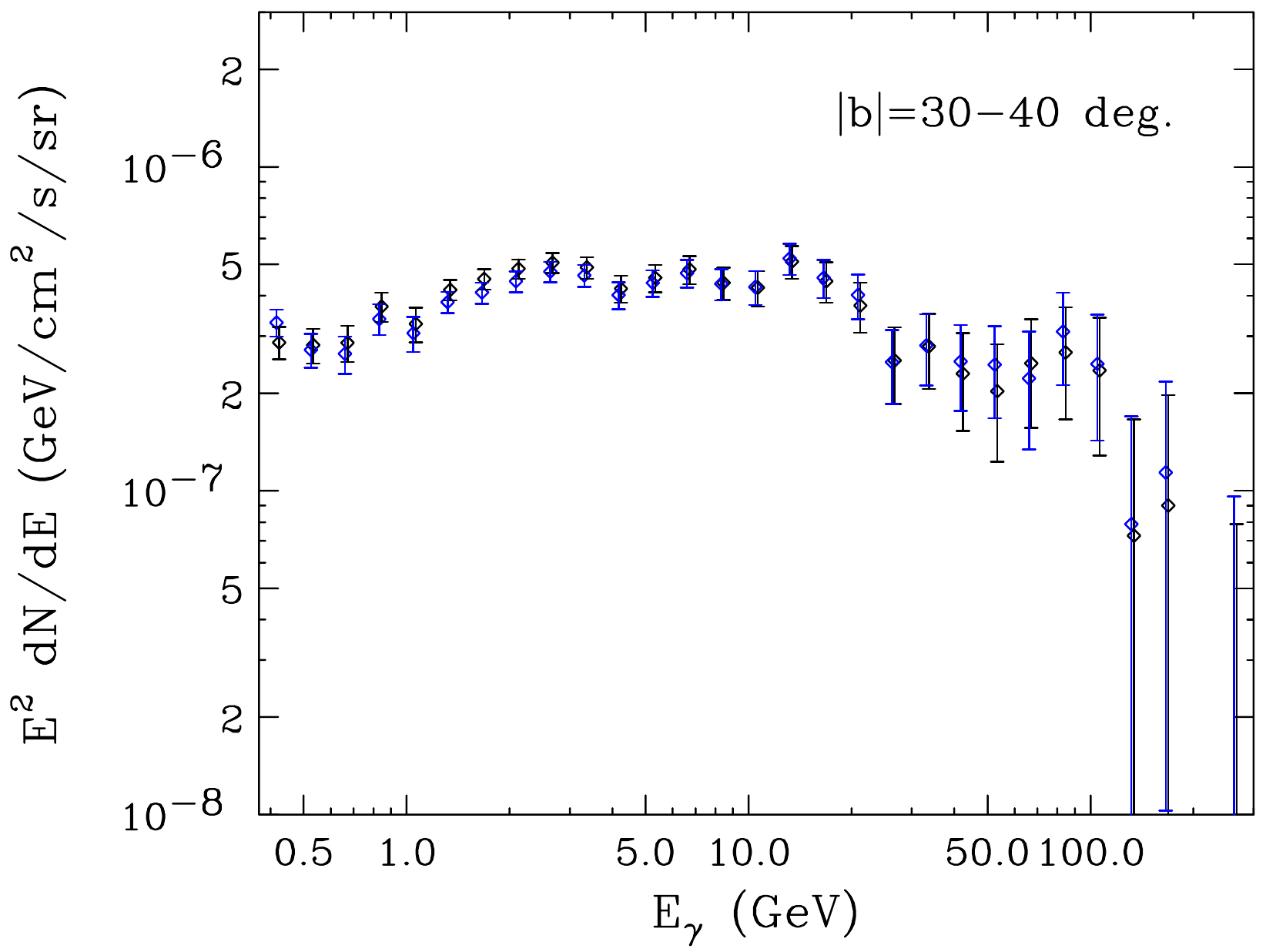}
\hspace{0.4cm}
\includegraphics[width=0.45\textwidth ]{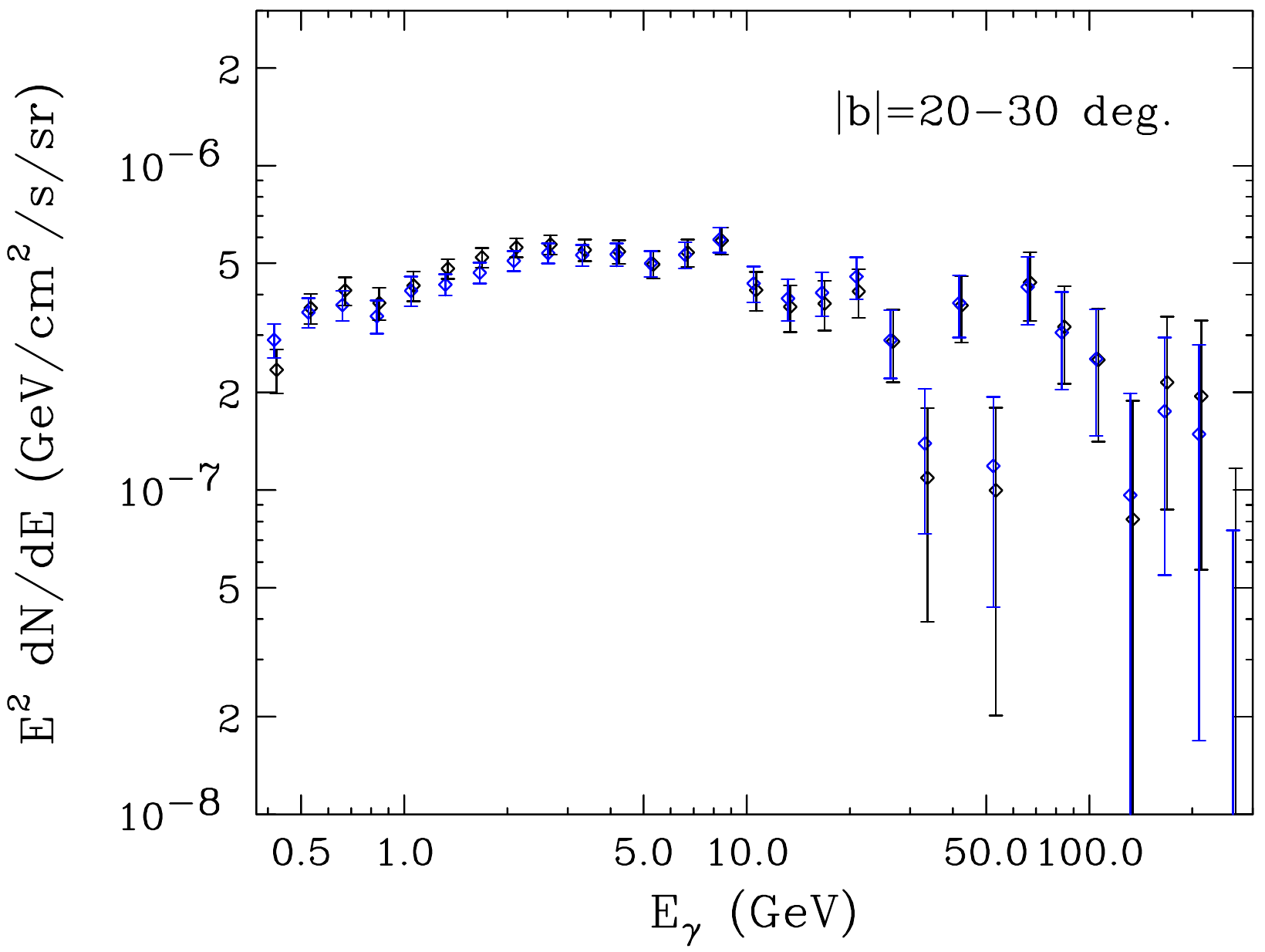}
\hspace{0.4cm}
\includegraphics[width=0.45\textwidth ]{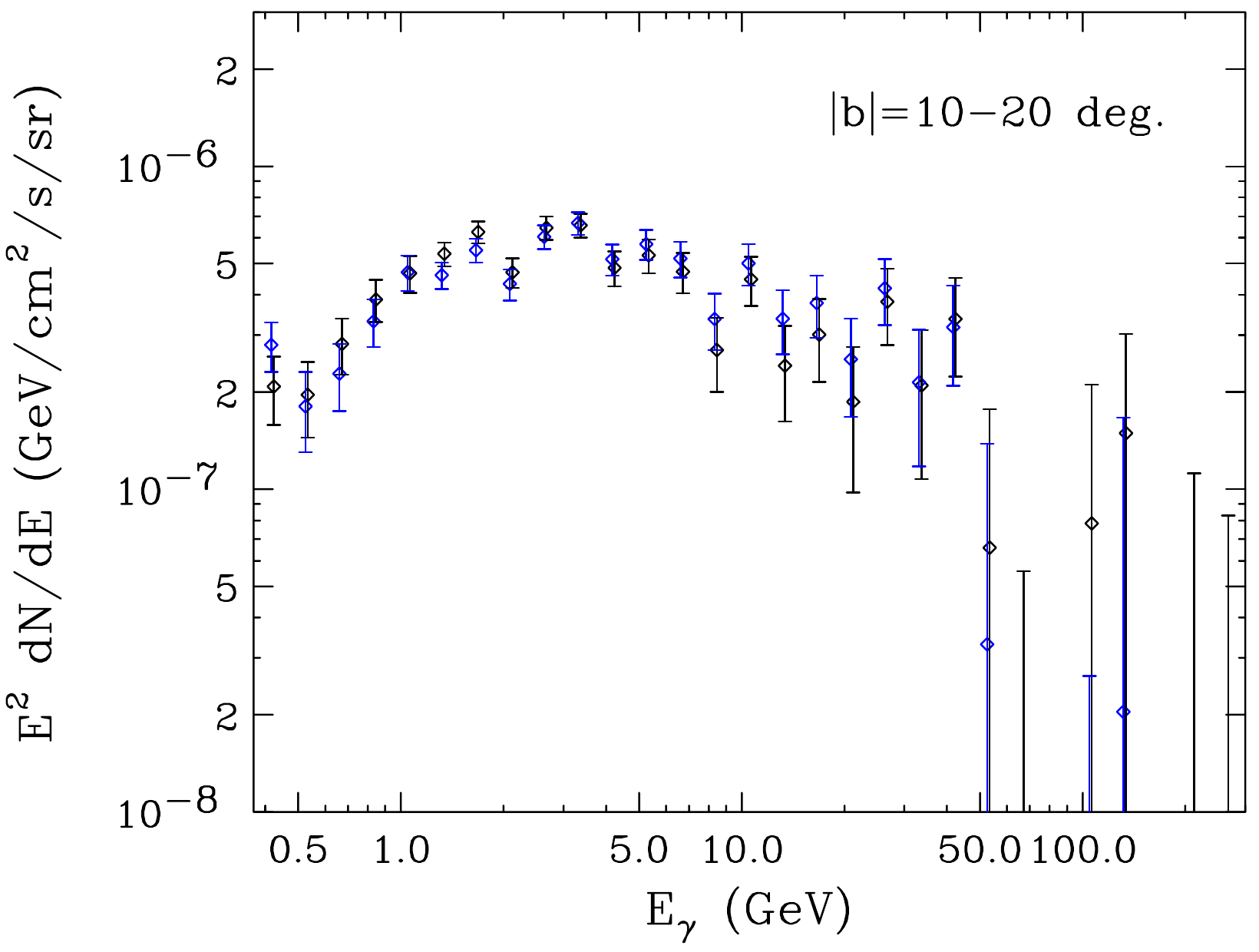}
\hspace{0.4cm}
\includegraphics[width=0.45\textwidth ]{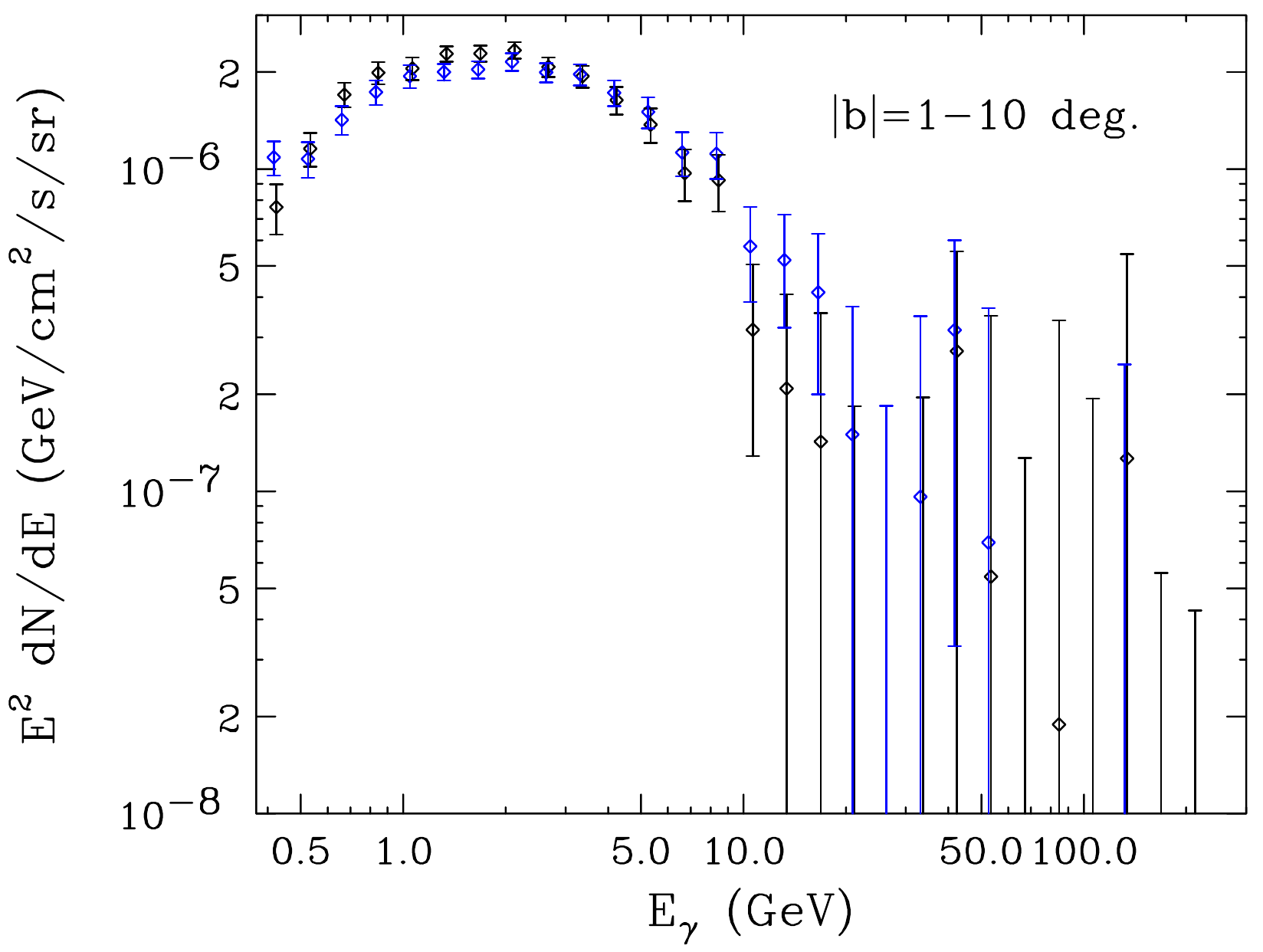}
\caption{A comparison of the gamma-ray spectra extracted from various latitude regions of the \emph{Fermi} Bubbles using the diffuse model template (black) to that extracted using the low-energy template, after performing the corrections described in the text (blue). After accounting for these corrections, the low-energy template results are almost identical to those derived using the diffuse model template.}
\label{dmlecompare}
\end{figure*}

Having chosen these two free parameters (the amplitude of the high-energy spectrum and the amplitude of the correction due to the low-energy template), we compare in Fig.~\ref{dmlecompare} the spectra extracted using the diffuse model template (black) to that extracted using the low-energy template, after performing the adjustments described in this Appendix (blue). After these corrections, the low-energy template results are almost identical to those derived using the diffuse model template.

\begin{figure*}
\includegraphics[width=.4\textwidth]{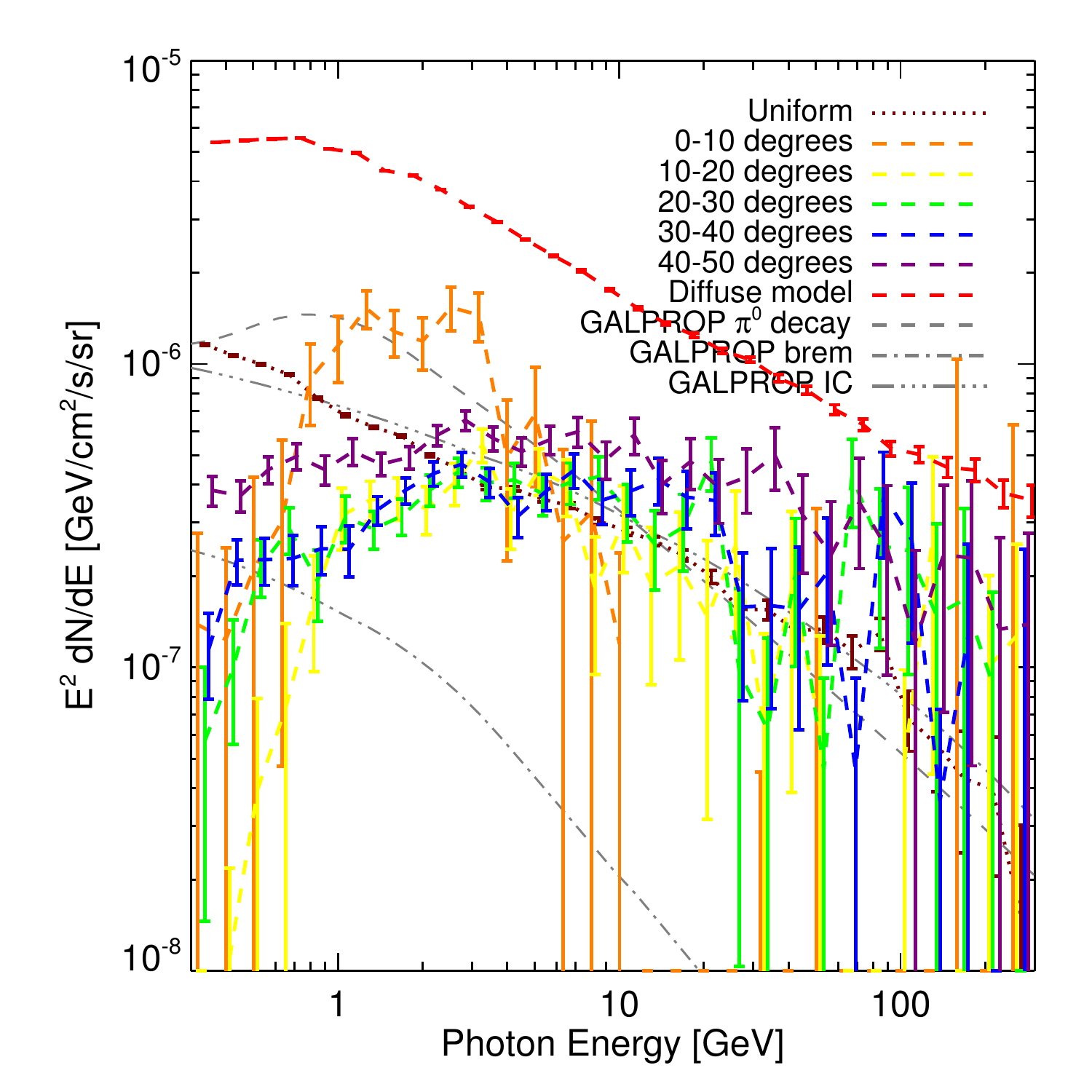}
\includegraphics[width=.4\textwidth]{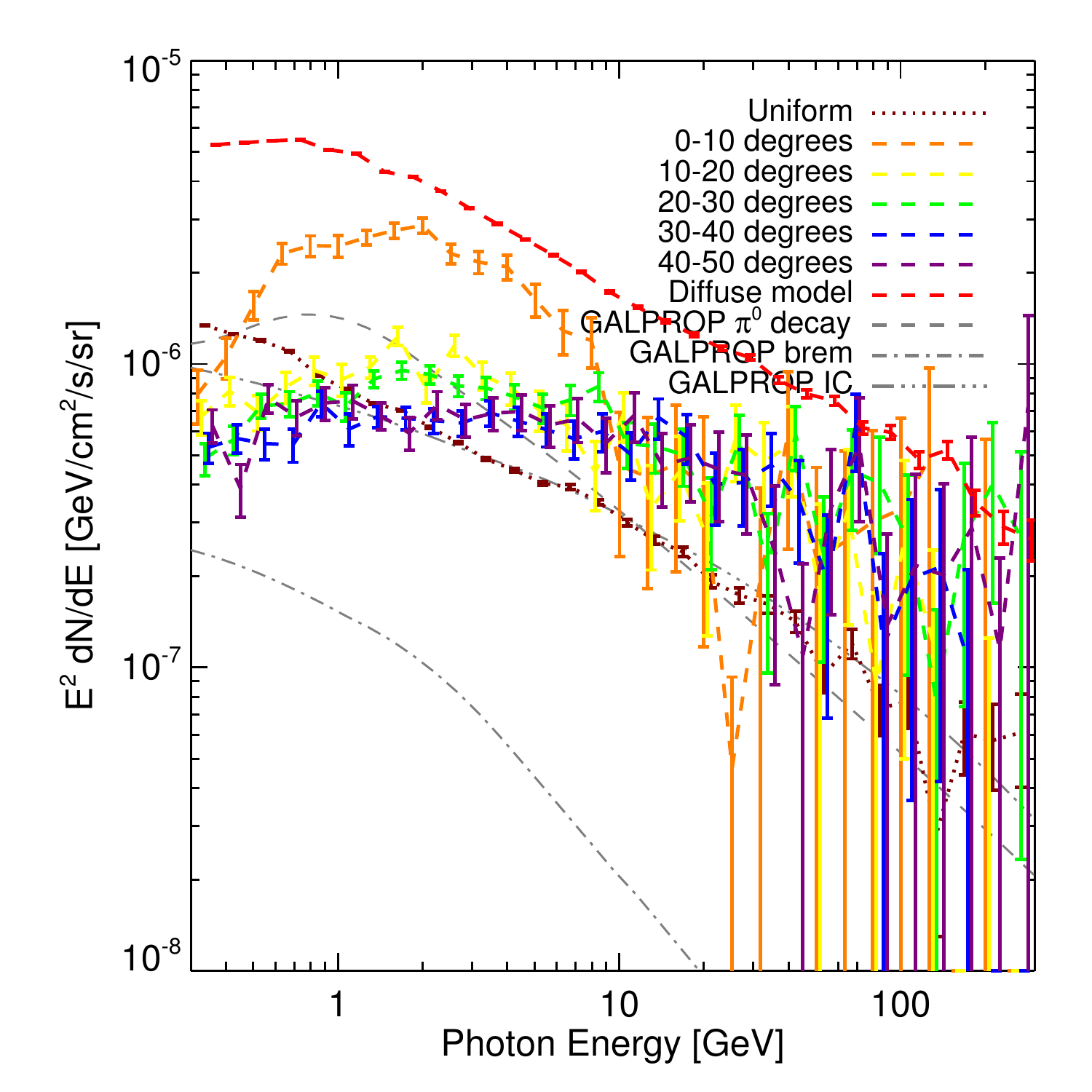} \\
\includegraphics[width=.4\textwidth]{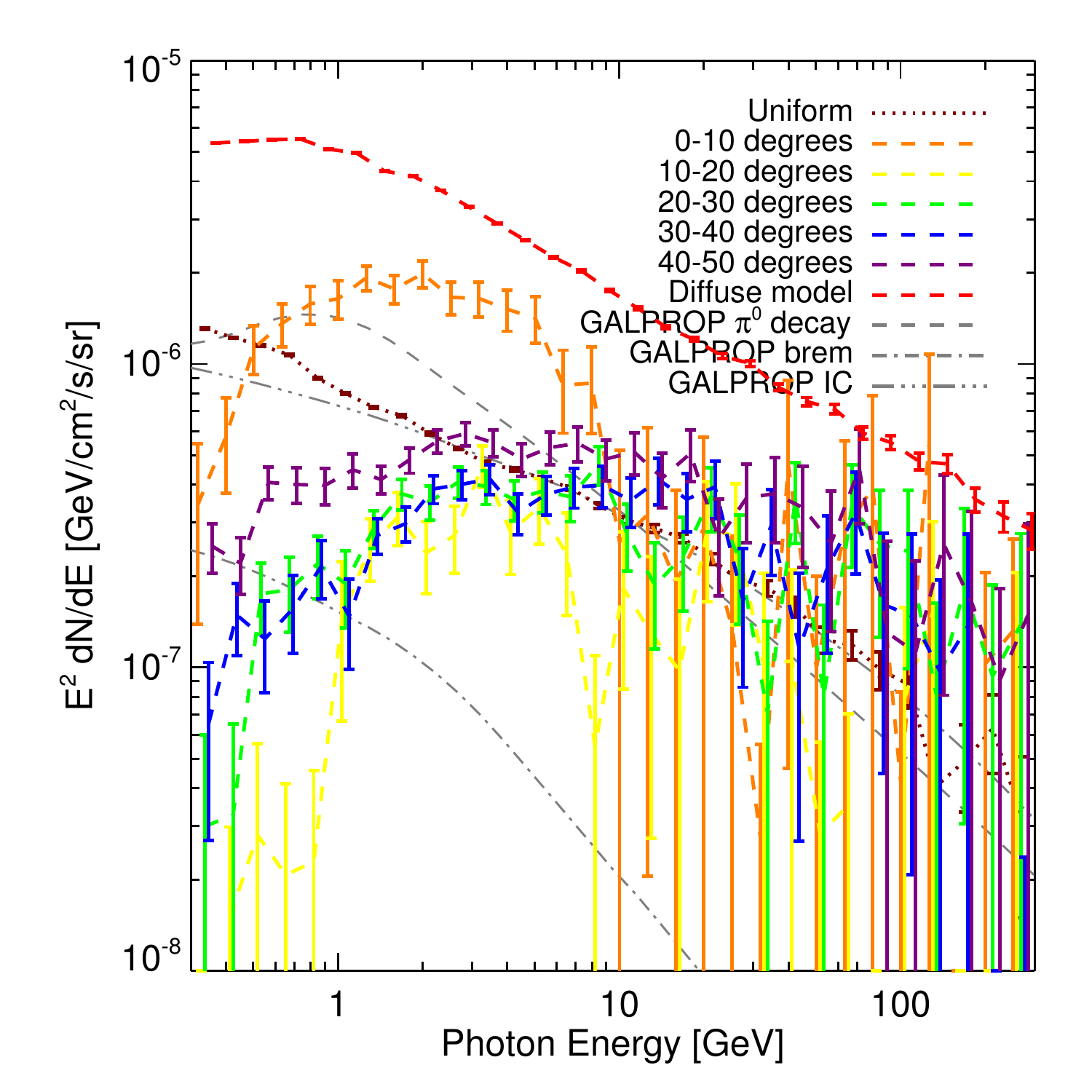}
\includegraphics[width=.4\textwidth]{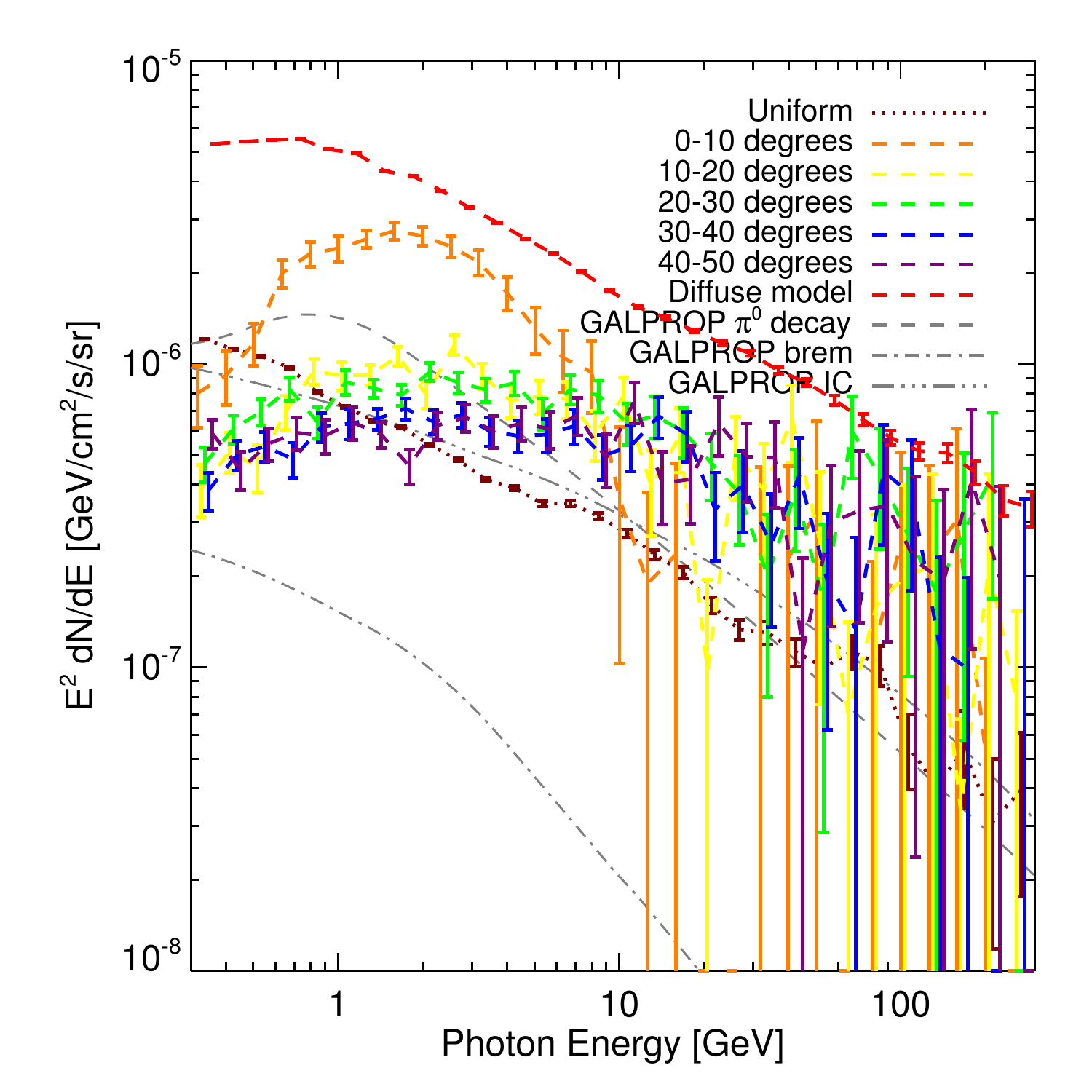}
\caption{\label{fig:leftrightnorthsouthchecks}
The spectra of the various fit components, including five separate latitude-sliced templates for the Bubbles (see text), for the diffuse model foreground template, restricting the fit to $b < 0$ (\emph{top left}), $b > 0$ (\emph{top right}), $l < 0$ (\emph{bottom left}), $l > 0$ (\emph{bottom right}). The Galactic Disk is masked for $|b| < 1^\circ$ in all cases.}
\end{figure*}

\section{The symmetry of the signal}

One might ask if the signal possesses north-south symmetry, as well as left-right symmetry (negative vs positive $l$). In Fig.~\ref{fig:leftrightnorthsouthchecks} we show the variation in the results when the fit is performed only in the north, only in the south, only for $l < 0$, or only for $l > 0$. In all cases, the spectral feature is apparent in the lowest latitude bin. The largest difference is in the relatively flat-spectrum (in $E^2 dN/dE$) emission attributed to the high-latitude Bubbles; this is larger in the north and for $l > 0$. This is to be expected, as Loop I and the associated gamma-ray arcs occur in the north quadrant with $l > 0$.

\section{Searching outside the Bubbles}
\label{app:complement}

In Sec.~\ref{nfw}, we showed that the low-latitude GeV-scale spectral feature is considerably better fit by a spherically symmetric NFW profile than by the Bubbles templates. In this Appendix, we perform a cross check of this conclusion by explicitly asking whether this spectral component is observed from the regions of the sky not associated with the Bubbles. The absence of such emission from the ``bubble complement'' region would strongly disfavor any dark matter interpretation of the signal. 

\begin{figure*}
\includegraphics[width=.3\textwidth]{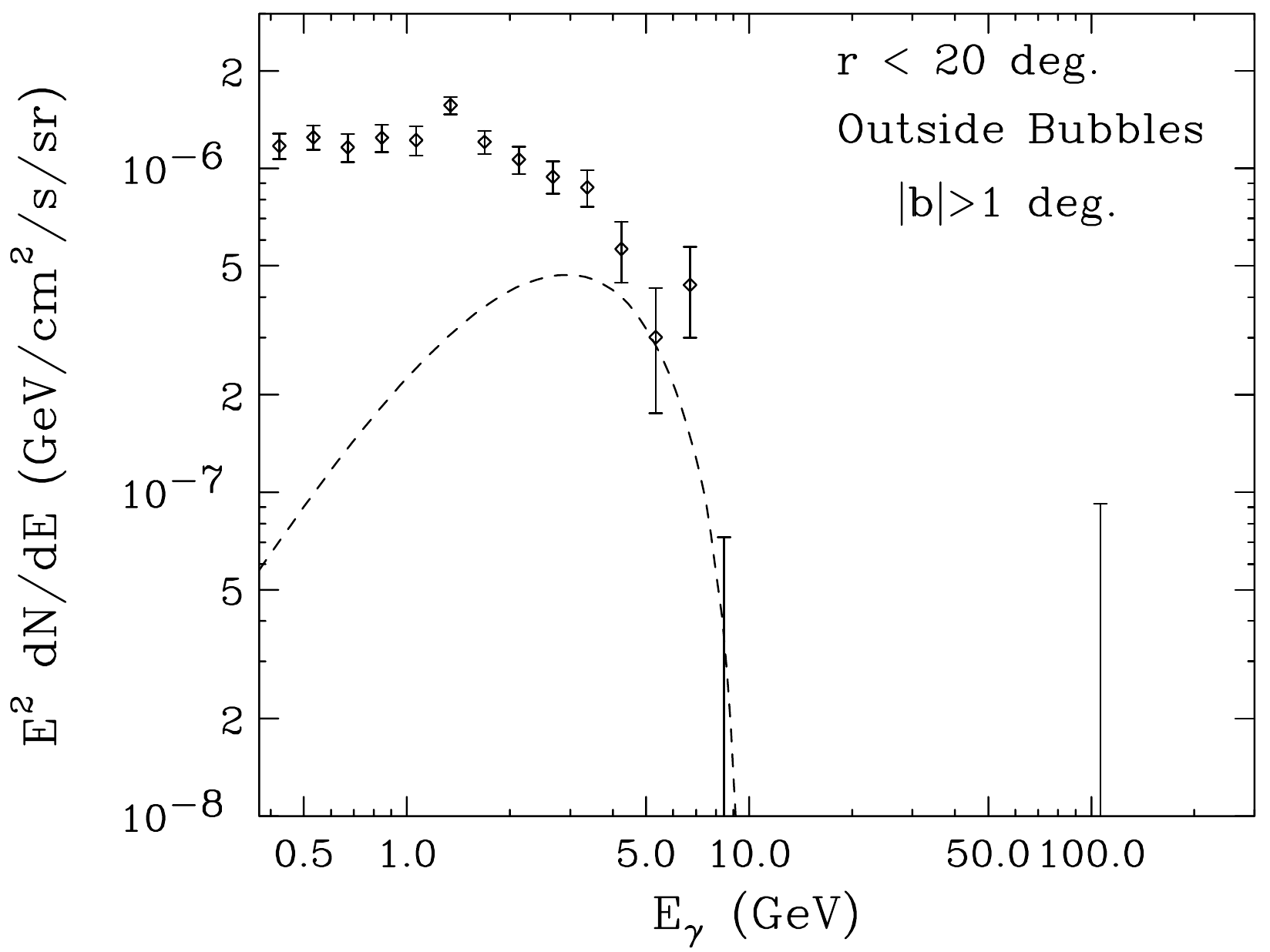}
\includegraphics[width=.3\textwidth]{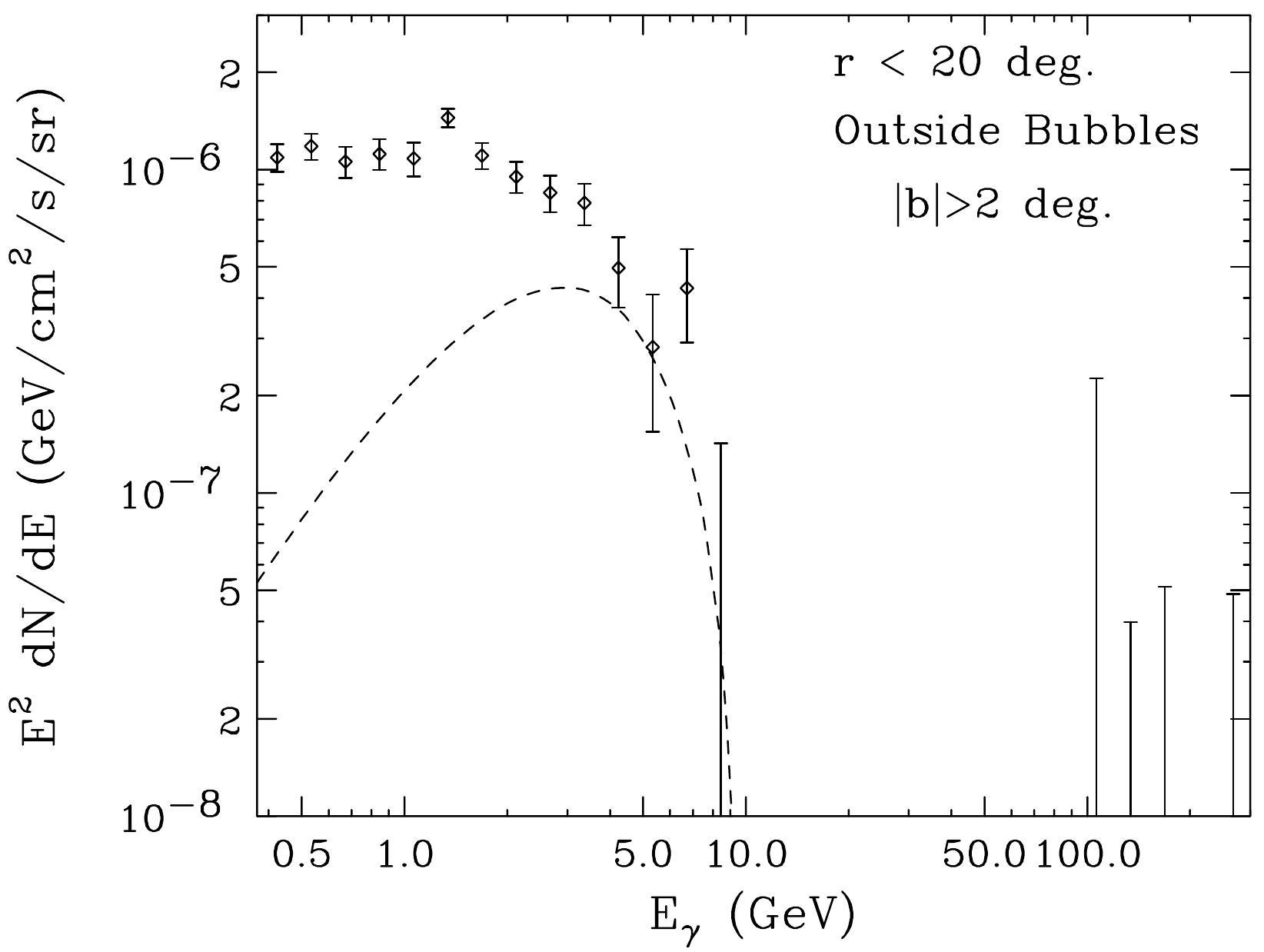}
\includegraphics[width=.3\textwidth]{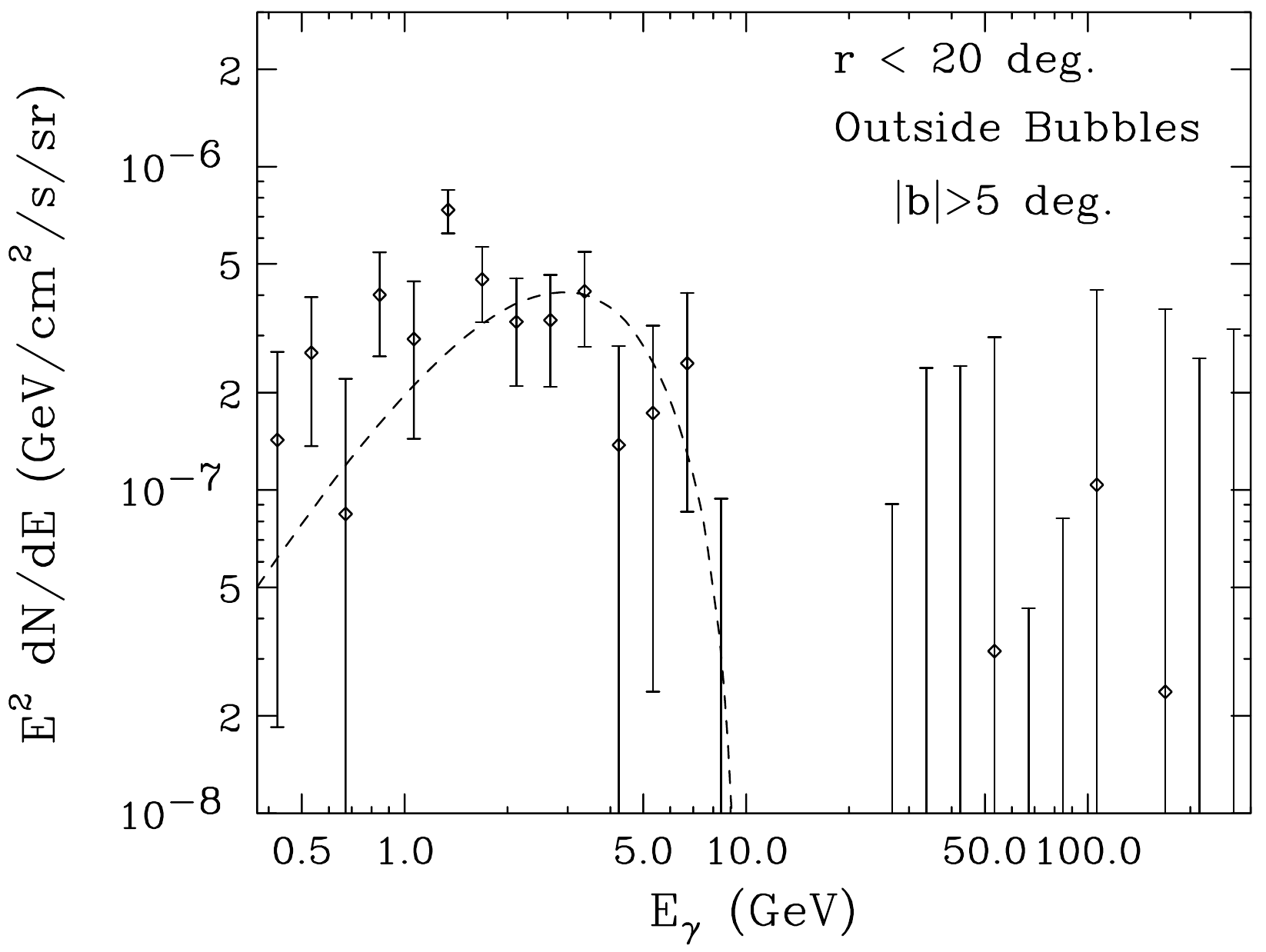}
\caption{\label{fig:complementspec}
The spectrum of the ``bubble complement'' region, defined as $\sqrt{b^2 + l^2} < 20^\circ$ and that does \emph{not} lie within the Bubbles templates, as extracted from a fit including the complement template, the five latitudinally sliced component Bubbles templates, and background templates as described in Fig.~\ref{fig:specbylatitude}. Here, we have used the \emph{Fermi} diffuse model. The different frames correspond to different choices for the latitude cut to remove the Galactic disk: $|b| < 1^\circ$, $|b| < 2^\circ$, $|b| < 5^\circ$. (Note that this substantially changes the region over which the complement spectrum is averaged, and thus may truly modify the result.) The dashed line in each frame denotes the average spectrum expected in this region from the dark matter model and distribution as shown in Fig.~\ref{nonIC}}
\end{figure*}

To address this question, we re-fit the data with an additional template, corresponding to the region lying within $20^\circ$ of the Galactic Center, but \emph{not} within the Bubbles. This region is indicated by dotted lines in Fig.~\ref{sky}. Fig.~\ref{fig:complementspec} shows the results for the ``diffuse model'' and ``low-energy template'' fits with this component added.

In each frame (corresponding to differing degrees of masking of the disk), the spectrum of the Bubbles complement template exceeds or equals the predicted gamma-rays from dark matter (for the same dark matter model and distribution as shown by dashed lines in Fig.~\ref{nonIC}). In contrast, little or no such emission is associated with this template at energies above 10 GeV. 

At low energies, the spectrum and amplitude of the Bubbles complement template is quite dependent on the degree to which we mask the Galactic Disk, and has resemblances in shape to the spectra associated with the NFW template. This supports our earlier hypothesis that the low-energy emission associated with the NFW template in these cases is being largely driven by the structures associated with Loop I and the Galactic disk.

\section{Residuals}
\label{app:residuals}

\begin{figure*}
\includegraphics[width=.6\textwidth]{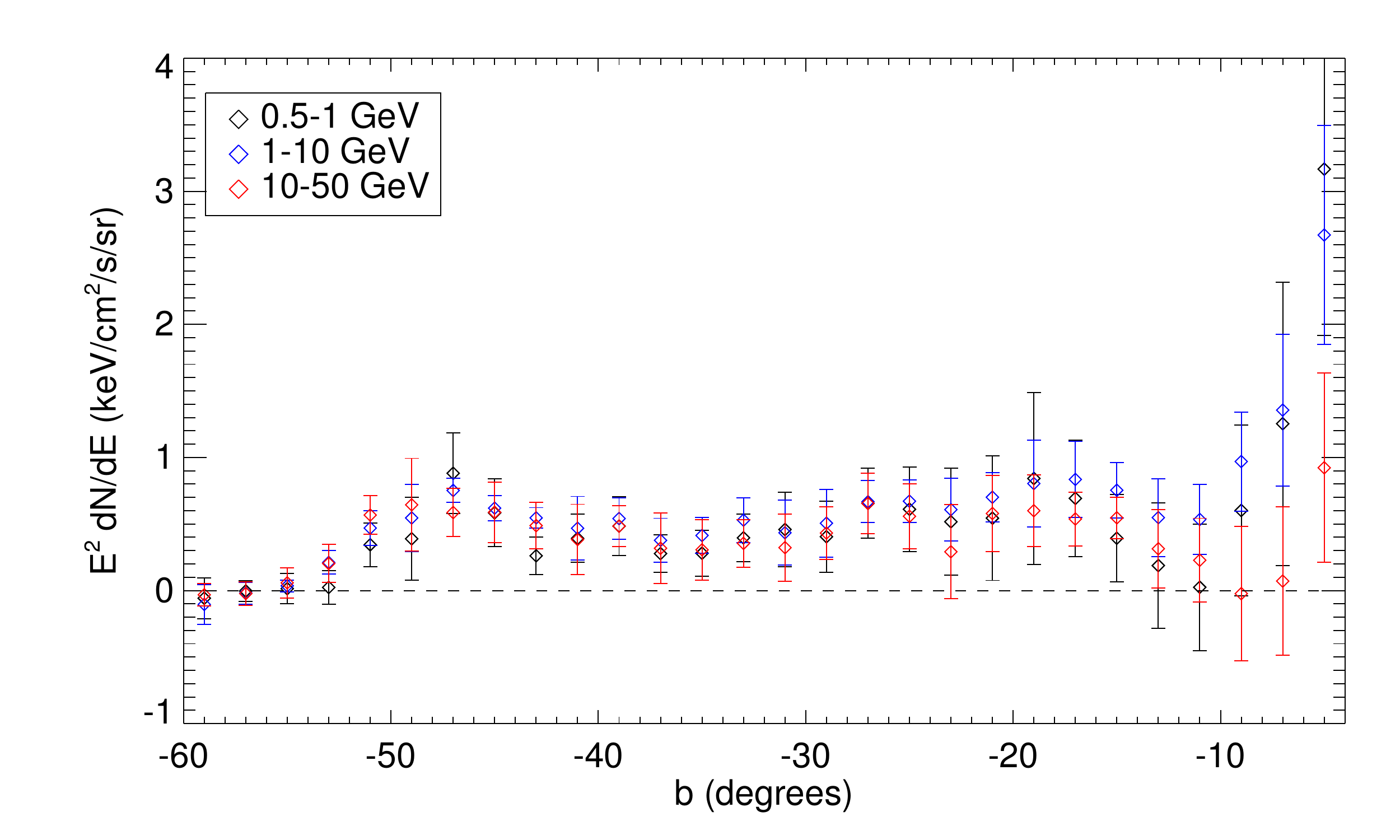} \\
\includegraphics[width=.6\textwidth]{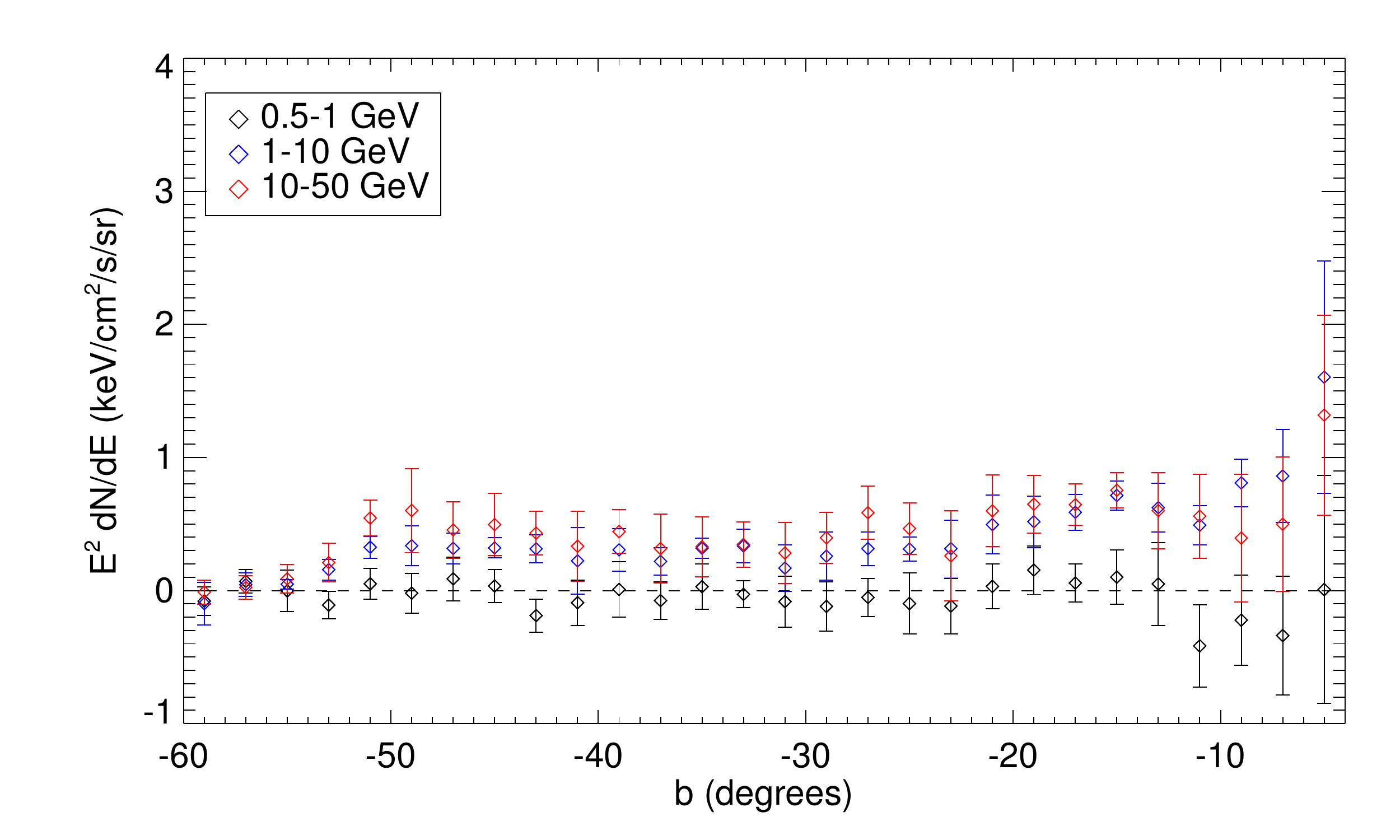}
\caption{\label{fig:residualbyenergy}
The residual emission after re-adding the latitude-sliced Bubbles templates with their best-fit coefficients, as a function of energy. Emission is $E^2 dN/dE$ averaged over $|l| < 5^\circ$ and in bins of width $\Delta b = 2^\circ$; the error bars describe the pixel-to-pixel scatter within each bin. The upper panel shows the fit using the diffuse model, the lower the fit using the low-energy template, both masked at $5^\circ$ from the plane.}
\end{figure*}

As discussed in Sec.~\ref{morphology}, residual maps with the latitude-sliced Bubbles re-added provide a characterization of the excess emission which is less dependent on the choice of signal templates. As in the main text, we restrict ourselves to $|l| < 5^\circ$, and bin in latitude in two degree steps, $\Delta b = 2^\circ$. In Fig.~\ref{fig:residualbyenergy}, we show the resulting ``Bubbles residuals'' for energy bands from 0.5-1 GeV, 1-10 GeV, and 10-50 GeV. The residuals are relatively flat in $b$ for $|b|>10^{\circ}$, and similar (in $E^2 dN/dE$) for the three energy bands. At smaller $|b|$, in contrast, the results for low energies diverge from those for the high-energy band.

\begin{figure*}
\includegraphics[width=.6\textwidth]{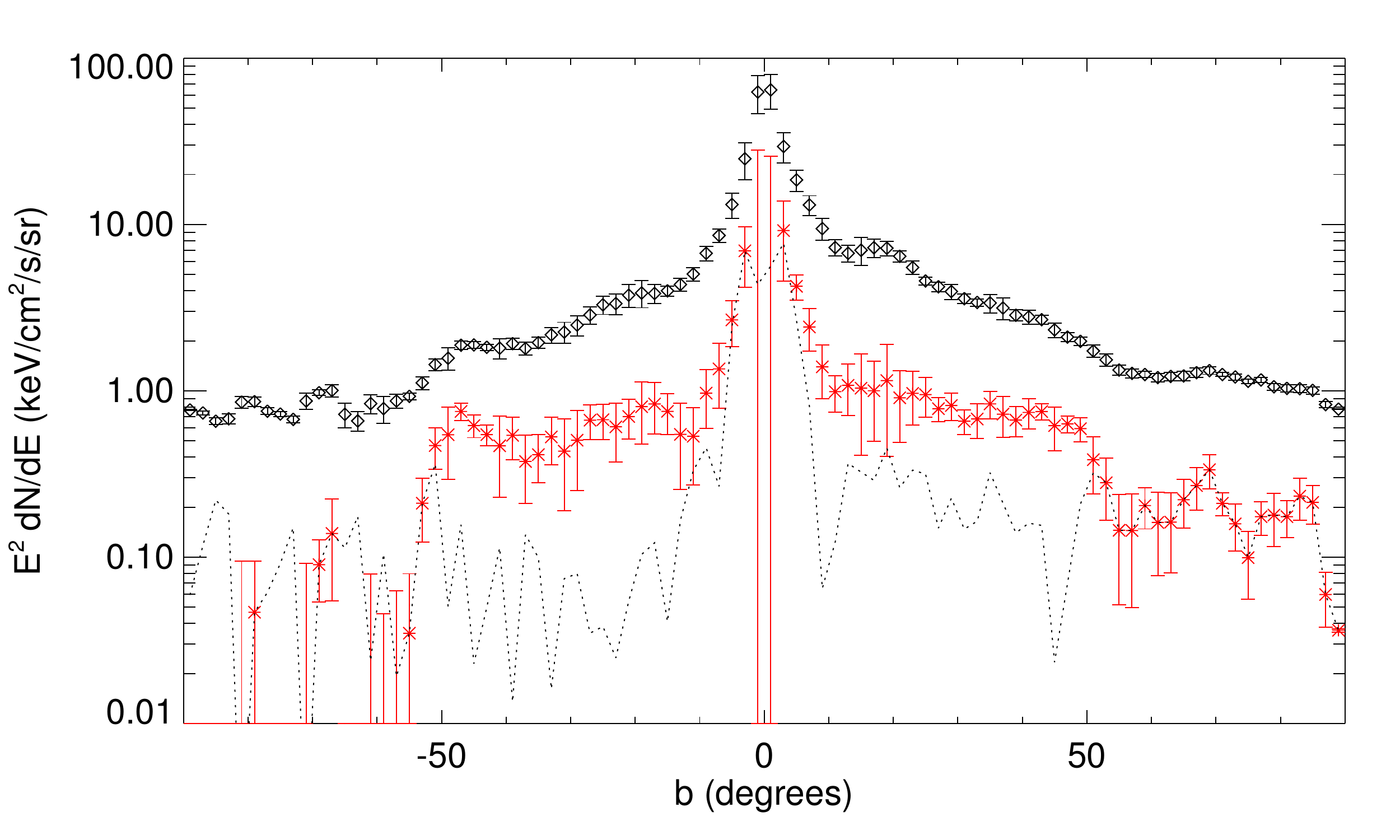} \\
\includegraphics[width=.6\textwidth]{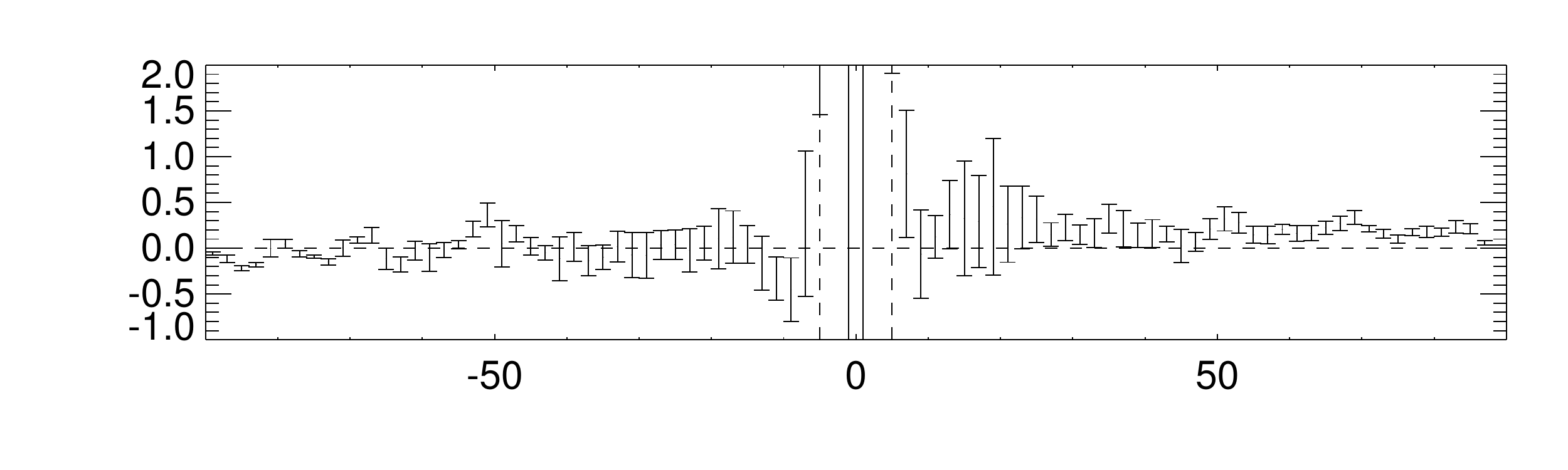} \\
\includegraphics[width=.6\textwidth]{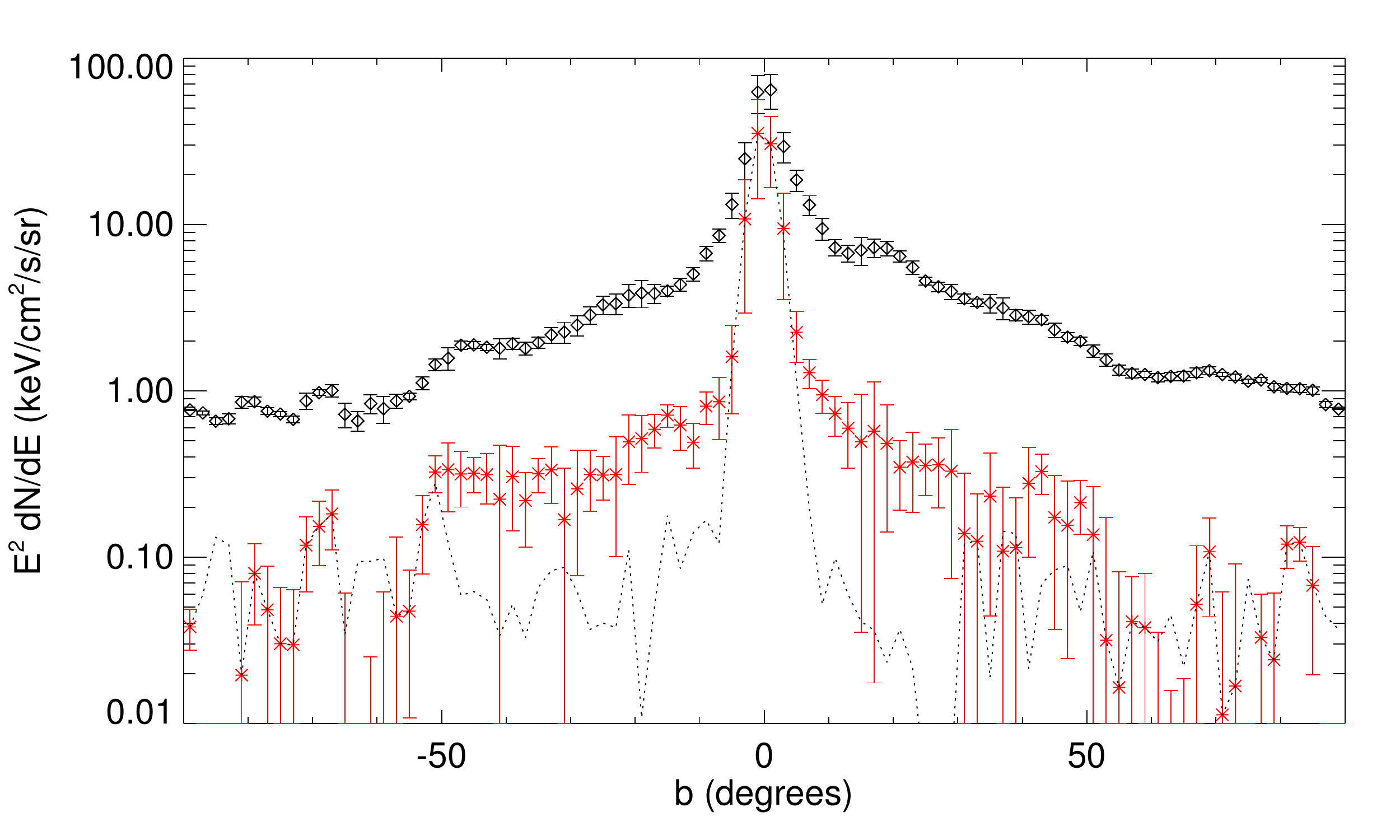} \\
\includegraphics[width=.6\textwidth]{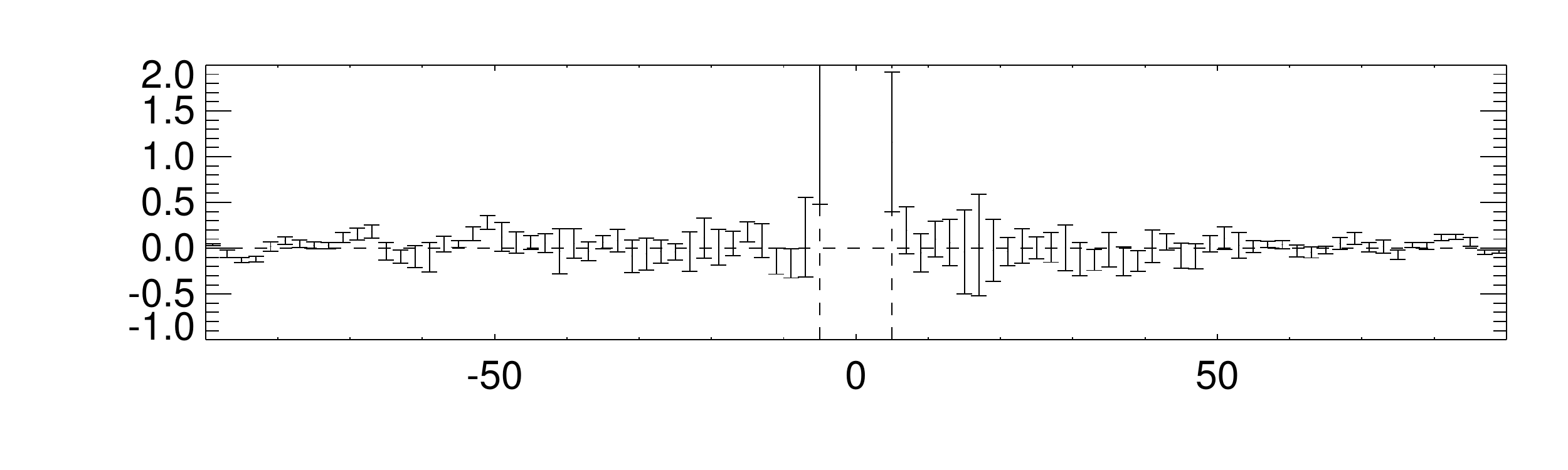} \\
\caption{\label{fig:residualhists}
The absolute value of the residual emission after subtraction of all templates (dotted line) and the emission after re-adding the latitude-sliced Bubbles templates with their best-fit coefficients (red stars), for 1-10 GeV. The total emission is shown in black diamonds. Emission is $E^2 dN/dE$ averaged over $|l| < 5^\circ$ and in bins of width $\Delta b = 2^\circ$; the error bars describe the pixel-to-pixel scatter within each bin. The second and fourth panels show the residual emission after subtraction of all templates (not its absolute value). The upper two panels show the fit using the diffuse model, the lower the fit using the low-energy template; in both cases the fit was performed for $|b| > 5^\circ$.}
\end{figure*}

In Fig.~\ref{fig:residualhists}, we look more closely at the $1-10$ GeV energy band, where the signal is present. We display the total emission, the Bubbles residual (i.e. the residual of the fit + the best-fit Bubbles templates), and the fit residual. We see that, with the exception of the region close to the Galactic plane where the Bubbles templates vanish, the absolute value of the fit residual is generally a factor of a few lower than the best-fit Bubbles-correlated emission, and fluctuates between positive and negative values; the Bubbles-correlated emission is another factor of several below the total emission in this energy band, depending on latitude. Thus, except for the region closest to the Galactic plane, this approach appears to provide an adequate model of the features in the data, and unaccounted-for residuals are unlikely to skew the signal. Close to the Galactic plane, there are large residuals, as the small extent of the Bubbles means they are unable to absorb an extended signal; given the uncertainty in the shape of the Bubbles close to the plane, this is not unexpected, and motivates the addition of a template to absorb this additional emission.

\end{appendix}

\end{document}